\documentclass[twocolumn,superscriptaddress,
 amsmath,amssymb,
 aps,
 prl,
]{revtex4-1}
\usepackage{dcolumn}% Align table columns on decimal point
\usepackage{bm}% bold math
\usepackage{graphicx}
\usepackage{comment}
\usepackage[caption=false]{subfig}
\usepackage{color}
\usepackage{etoolbox}
\usepackage{float}
\usepackage{hyperref}

\hypersetup{
    colorlinks=true,       % false: boxed links; true: colored links
    linkcolor=red,          % color of internal links (change box color with linkbordercolor)
    citecolor=blue,        % color of links to bibliography
    filecolor=magenta,      % color of file links
    urlcolor=cyan           % color of external links
}

\newcommand{\blue}{\textcolor{blue}}

\newcommand{\bea}{\begin{eqnarray}}

\newcommand{\eea}{\end{eqnarray}}

\begin{document}

%\preprint{}
\title{Large Deviations in Switching Diffusion: from Free Cumulants to Dynamical Transitions}% Force line breaks with 
\author{Mathis Gu\'eneau}
\affiliation{Sorbonne Universit\'e, Laboratoire de Physique Th\'eorique et Hautes Energies, CNRS UMR 7589, 4 Place Jussieu, 75252 Paris Cedex 05, France}
\author{Satya N. Majumdar}
\affiliation{LPTMS, CNRS, Univ.  Paris-Sud,  Universit\'e Paris-Saclay,  91405 Orsay,  France}
\author{Gr\'egory Schehr}
\affiliation{Sorbonne Universit\'e, Laboratoire de Physique Th\'eorique et Hautes Energies, CNRS UMR 7589, 4 Place Jussieu, 75252 Paris Cedex 05, France}

%\date{\today}

\begin{abstract}
{We study the diffusion of a particle with a time-dependent diffusion coefficient $D(t)$ that switches between random values drawn from a distribution $W(D)$ at a fixed rate $r$. Using a renewal approach, we compute exactly the moments of the position of the particle $\langle x^{2n}(t) \rangle$ at any finite time $t$, and for any $W(D)$ with finite moments $\langle D^n \rangle$. For $t \gg 1$, we demonstrate that the cumulants $\langle x^{2n}(t) \rangle_c$ grow linearly with $t$ and are proportional to the free cumulants of a random variable distributed according to $W(D)$. For specific forms of $W(D)$, we compute the large deviations of the position of the particle, uncovering rich behaviors and dynamical transitions of the rate function $I(y=x/t)$. Our analytical predictions are validated numerically with high precision, achieving accuracy up~to~$10^{-2000}$.} %in some specific case.} 
%for the rate function in a two-state model -- $D(t)=D_1$ or $D_2$.}
\end{abstract}

\maketitle

%\tableofcontents

\newpage
%{\color{red}
%- Motiver le besoin d'un modèle à coefficient de diffusion dynamique
%- Motiver la pertinence d'un modèle qui change de coefficient de manière discrète
%- motiver la nécessité physique d'une distribution à support borné
%- Motiver la pertinence d'un modèle qui a des phases d'immobilisme alternées avec de la diffusion: modèle à 2 états avec un état sans diffusion
%}

\noindent {\it Introduction.} Anomalous diffusion processes have attracted significant interest across diverse scientific fields, including complex and disordered systems \cite{BG90,MK00}, soft materials such as colloids \cite{Golest09} or living cells \cite{WGLFGH2019}, movement ecology \cite{Viswa96}, or financial markets \cite{MS99}. Typically, anomalous diffusion refers to deviations from standard Brownian scaling, where the mean squared displacement (MSD) of the particle position ${x}(t)$ behaves with time $t$ as ${\rm MSD}[{x}(t)]\propto t^{2\alpha}$ with $\alpha \neq 1/2$. However, recent studies have revealed numerous cases displaying standard Brownian scaling ($\alpha = 1/2$) accompanied by distinctly non-Gaussian fluctuations \cite{Wang2012}, contradicting the standard kinetic theory of normal diffusion. For instance, experiments on colloids \cite{WABG2009} have demonstrated a crossover in the position distribution from Gaussian behavior at short distances to an exponential tail at larger distances. %These observations indicate that even in situations where the overall diffusion appears normal, underlying processes---such as heterogeneity, intermittency, or complex interactions---can induce significant deviations from Gaussian statistics in the displacement distributions.

To theoretically capture and describe these ``diffusive yet non-Brownian'' behaviors, a broad spectrum of models has been proposed. These include continuous-time random walks and their variants \cite{Barkai1,Sokolov,Burov1,Burov2,BB}, as well as random diffusivity models \cite{Chubynsky2014,Sposini2018,Sposini2} -- which have also been studied in finance, for instance in the Heston model~\cite{Heston_DY}. In the latter models, a key feature is the incorporation of stochasticity or randomness into the time evolution of the diffusion coefficient $D(t)$. In the context of disordered systems, this random diffusion coefficient effectively accounts for the spatial heterogeneities present in the system \cite{BG90}.
For such models in the simple one-dimensional setting, the MSD, which is the second cumulant (or variance) of the particle position, typically behaves as $\text{MSD}[{x}(t)] = \text{Var}[{x}(t)] \approx 2 \, D_{\text{eff}} \, t$, 
where $D_{\text{eff}}$ is an effective diffusion coefficient that has been computed for various models. 
%According to basic probability theory,
The non-Gaussian fluctuations of ${x}(t)$ are usually captured by the higher-order cumulants of ${x}(t)$, like the skewness and kurtosis (respectively the third and fourth cumulants). 
%Indeed, the presence of a non-vanishing cumulant of order strictly higher than two serves as a signature of the non-Gaussianity of ${x}(t)$. 
Understanding these higher-order cumulants is thus crucial for characterizing non-Gaussianities of ${x}(t)$. Cumulants are also interesting because they carry information on the large deviations of ${x}(t)$ that characterize its atypical large~fluctuations.
%, which, in turn, allow us to characterize the non-Gaussian tails of the probability density function (PDF) of ${x}(t)$.

However, calculating higher-order cumulants is often quite challenging, as it requires evaluating higher-order correlation functions of $x(t)$. Consequently, there are very few results in the literature concerning these cumulants 
%(see however \cite{Chubynsky2014} in the context of diffusing diffusivity models) 
or the large deviations of the position distribution in random diffusivity models. The aim of this paper is to present a detailed analytical study of these important observables for a broad class of such models, specifically focusing on stochastically switching diffusion models.

\begin{figure}[t]
    \centering
\includegraphics[width=1\linewidth]{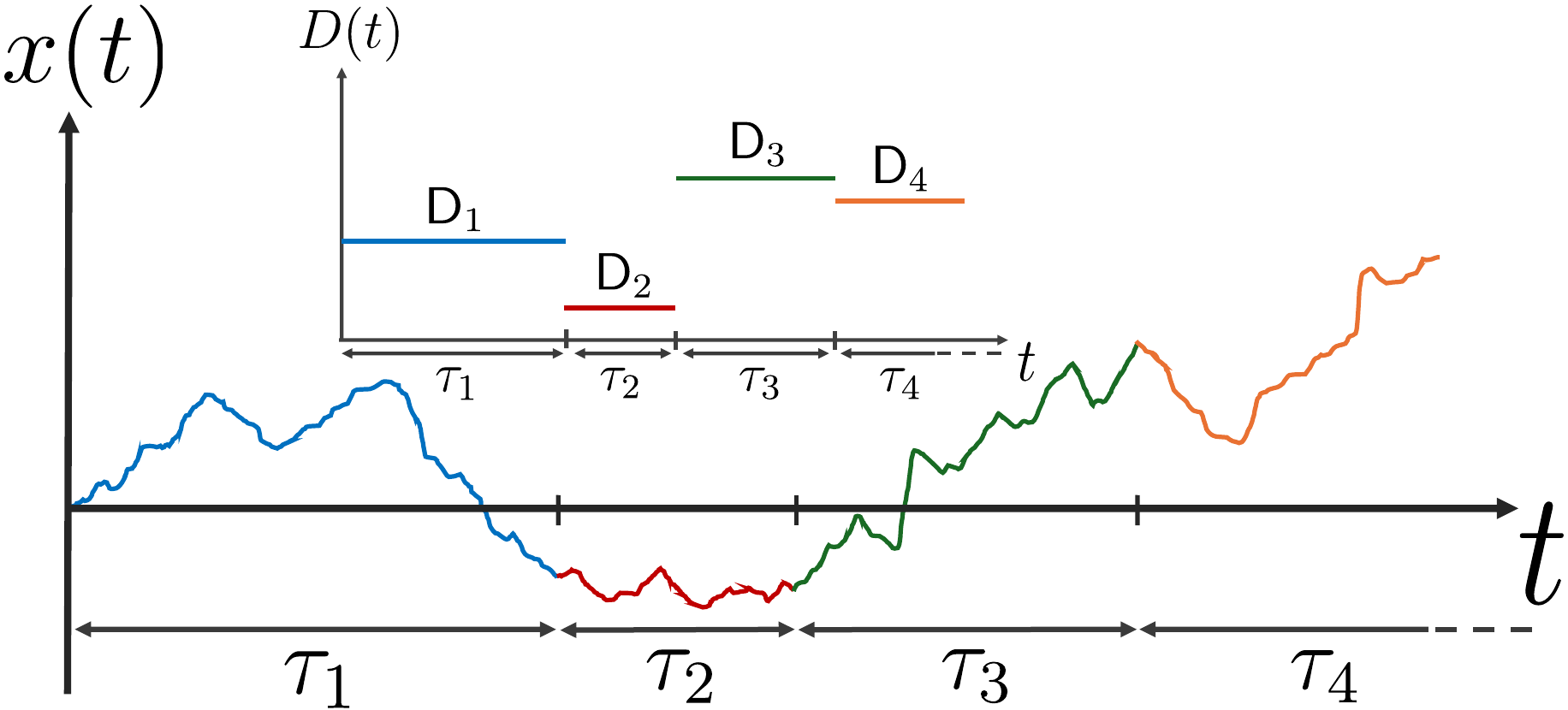}
  \caption{Trajectory of a switching diffusion process in one-dimension. During each time interval $\tau_i$, the particle performs an independent Brownian motion with a diffusion coefficient ${\sf D}_i$. In the model studied here, the $\tau_i$'s are independent exponential random variables, while the ${\sf D}_i$'s, which are also independent, are drawn from an arbitrary distribution $W(D)$.}
  \label{drawing} 
\end{figure}
In this Letter, we consider 
%a stochastic switching diffusion model, sometimes called the annealed transit time model (ATTM) \cite{ATTM}, 
a model in which a particle, starting from the origin, performs a standard one-dimensional Brownian motion with a diffusion coefficient ${\sf D}_1$ over a time $\tau_1$. Both ${\sf D}_1$ and $\tau_1$ are random variables drawn from a joint distribution $P_{\rm joint}(D, \tau)$. After this time $\tau_1$, the particle resumes its motion from its current position, now performing a new Brownian motion with diffusion coefficient ${\sf D}_2$ for a duration $\tau_2$, which are drawn independently from the same distribution $P_{\rm joint}(D, \tau)$ as ${\sf D}_1$ and $\tau_1$. This process continues iteratively for a fixed period of time $t$ (see Fig. \ref{drawing} for an illustration of this process). Such models have been used to model recent experiments on cytoplasmic membranes (which control the movement of substances in and out of a cell)
showing patches of strongly varying diffusivity \cite{Serge2008,English2011,Masson2014,Weron2017}. Here we will mainly consider a simpler version of this model where ${\sf D}_i$'s and $\tau_i$'s are independent, that is, $P_{\rm joint}(D,\tau) = W(D)\, p(\tau)$. More specifically, we will study the case where the $\tau_i$'s are exponential random variables with a rate $r$, i.e., $p(\tau) = r\, e^{-r \tau}$, while $W(D)$ is an arbitrary probability distribution function (PDF). A well-known example is the case where
$W(D)$ is a superposition of Dirac delta peaks, i.e. $W(D)=\sum_{i=1}^N\, p_i\,  \delta(D-D_i)$, with $D_1>D_{2}>\cdots>D_{N}$ and $\sum_{i=1}^N\, p_i=1$. This model, sometimes called
``composite Markov process''~\cite{VanKampen}, has been studied in various contexts ranging from disordered systems \cite{Luczka1, BarkaiBurov}, biophysics \cite{Holcman2009,Bressloff2017,Metzler2023, natureGprotein, Singha}, nuclear magnetic resonance \cite{Karger1985}, finance \cite{Naik} or movement ecology \cite{Morales2004,Fagan2020}. In the latter, mixtures
of random walks with switching dynamics between them are
widely used to model intermittent
searches where an animal/a particle can employ different motion modes \cite{Morales2004,Benichou2011}. In the case $N=2$ (referred to as the two-state model), the mode with $D=D_2 < D_1$ would then model local search,
while the one with $D=D_{1}$  corresponds to an exploratory motion with larger displacements. Incidentally, this model with $N=2$ recently appeared in the context of stochastic resetting with two resetting points~\cite{Boyer2024}. 
%An interesting limit corresponds to the case where the slow mode is totally immobile, i.e., $D_2 = 0$, which can be used to model molecular caging effect \cite{Grebenkov2019}. This limit was also proposed as an effective model of autonomous ratcheting by stochastic resetting in \cite{Ghosh2023}. 
Besides the case of discrete diffusion modes, various studies, both theoretical \cite{ATTM, Metzler2023, Sposini2018} and experimental \cite{Kuhn2011,Roichman2020, Chubynsky2014, Manzo2015}, have considered a continuous distribution for $W(D)$~including exponential and gamma distribution 
\cite{ATTM, Manzo2015,Roichman2020,Sposini2018} but also distribution with a finite support~\cite{Metzler2023, Chubynsky2014}.  

%%\begin{figure}[t]
%%    \centering
%\includegraphics[width=0.49\linewidth]{cumu_twostates.pdf}
%\includegraphics[width=0.49\linewidth]{cumu_wigner.pdf}
%  \caption{two states: $r = 1$, $p = 1/3$, $D_2 = 1$,$D_1 = 2$. Wigner: $r = 1$, $D_{\text{max}}=1$ {\color{blue} M: je crois qu'il manque $O(1)$ dans la legénde à droite pour $rt\gg 1$ et le $\propto$ n'est pas approprié}. \textcolor{red}{Cette figure peut aller dans le SM.}}
%  \label{Fig_cumul} 
%\end{figure}

\noindent {\it Summary of our main results.} 
%It is useful to summarize our main results. 
First, for this class of models illustrated in Fig. \ref{drawing}, 
we have obtained an exact analytical expression for the moments of the positions $\langle x^{2n}(t) \rangle $, for any integer $n$~\cite{footnote1} and arbitrary $t$ and for any distribution $W(D)$ (with all its moments well defined). %which we assume to have all its moments well defined~\footnote{note that the odd moments of $x(t)$ vanish by symmetry $x \to -x$.}. 
Their explicit expression is given in~Eq.~(\ref{finite_t}). Here the notation $\langle \cdots \rangle$ means a simultaneous average over all the sources of randomness on the same footing (in the language of disordered systems, 
we consider here an ``annealed'' average). Of course, the $2n$-th cumulant, denoted as $\langle x^{2n}(t) \rangle_c$, can be formally obtained from~(\ref{finite_t}). 
%This expression (\ref{finite_t}) is indeed convenient to extract the small time behavior of $\langle x^{2n}(t) \rangle_c$. 
However, their large time behavior is more conveniently extracted from the cumulant generating function, which, as shown below, can be computed explicitly. Their asymptotic behaviors at small and large time read
\bea \label{summary_cum}
\langle x^{2n}(t)\rangle_c \simeq
\begin{cases}
& \frac{(2n)!}{n!}\langle D^n\rangle_c \,t^n \quad , \quad r\,t \ll 1 \, , \\
& \frac{(2n)!}{r^{n-1}}\, \kappa_n(D)\,t \quad , \quad \,r\,t \gg 1 \;.
\end{cases}
\eea
In the first line, $\langle D^n \rangle_c$ denotes the (standard) cumulant of $D$, while the coefficients $\kappa_n(D) \neq \langle D^n \rangle_c$ also depend in a nontrivial way on the moments of $D$. This result~(\ref{summary_cum}) clearly shows that, at large times $t$, the higher cumulants of $x(t)$ grow linearly with time, revealing the presence of non-Gaussian fluctuations in this model. 

But what are these nontrivial coefficients $\kappa_n(D)$ that characterize this linear growth? As we will show, they are none other than the {\it free cumulants of $D$}, a class of combinatorial objects central to the field of free probability theory. Free probability theory is a mathematical framework developed to study non-commutative random variables~\cite{Voiculescu}, where the classical notion of independence is replaced by a new concept called {\it freeness}. Analogous to classical cumulants, which encode statistical independence, free cumulants capture the structure of freeness and play a central role in this theory. Free probability has found applications in various fields, in particular in random matrix theory (RMT)~\cite{BIPZ, VivoBook, PottersBouchaud, GuionnetHouches, Voiculescu}, and has sparked significant interest in both mathematics~\cite{BianeBook, BianeQSSEP, Speicher_book, Speicher_book2} and physics~\cite{Burda}, notably in quantum mechanics~\cite{Bernard1, ETHfree}.

While such free cumulants appeared before in more complicated classical models of {\it interacting} particles \cite{Bernard2,BB24}, their appearance in such a simple {\it single} particle
model here is highly surprising and intriguing. Similar to the classical case, where conventional cumulants $\langle D^n \rangle_c$ relate polynomially to the moments $\langle D^p \rangle$, with $p=1,\dots, n$, via Eq.~(\ref{cumulantclassic}), free cumulants also have a fairly explicit expression in terms of these moments~\eqref{free_norm}.%{\color{red}Although conventional cumulants $\langle D^n \rangle_c$ are related to the moments $\langle D^p \rangle$, with $p=1,\dots, n$ via Bell polynomials~\eqref{cumulantclassic}, free cumulants have a fairly explicit expression in terms of these moments~\eqref{free_norm}.}
 This enables us to compute them explicitly for various distributions of interest~\cite{SM}. For instance, for the two-state model $W(D) = p\delta(D-D_{1}) + (1-p) \delta(D-D_2)$ with $0\le p\le 1$, one has
$\langle x^2(t)\rangle_c \sim 2 (p D_1 + (1-p)D_2) t$, while for $n \geq 2$ the higher cumulants are also explicit and linear in time~\cite{SM}.
%\begin{equation} \label{moment_dble}
%\langle x^{2n}(t) \rangle_c \approx - (\Delta D)^n \frac{(2n)!}{r^{n-1}} \frac{\sqrt{p(1-p)}}{n(n-1)} P_{n-1}^{1}(1-2p)\, t, 
%\end{equation}
%where $\Delta D = (D_{1}-D_{2})$ and $P_{n}^m(z)$ denotes the associated Legendre polynomial of degree $n$ and parameter $m$. 
It is also interesting to study the case where $W(D)$ is a continuous PDF with a finite support, as discussed e.g. in \cite{Metzler2023, Chubynsky2014}. For example, we consider the case where $W(D)$ is given by the Wigner semi-circle on $[0,D_{\text{max}}]$,~i.e., $W(D)=8\sqrt{D(D_{\max}-D)}/(\pi D_{\max})$
for which it is well known, from RMT, that the corresponding free cumulants are quite simple \cite{SM}, i.e., $\kappa_n(D) = 0$ for $n \geq 3$. In this case, one finds 
\bea
\langle x^2(t)\rangle_c \approx D_{\text{max}} t \quad , \quad \langle x^4(t)\rangle_c \approx \frac{3}{2}\, \frac{D^2_{\text{max}}}{r} t \;,
\eea
while higher order cumulants vanish to leading order in~$t$ [see Eq. (\ref{summary_cum})]. 
In fact, for $n \geq 3$, $\langle x^{2n}(t)\rangle_c = O(1)$ can also be computed \cite{SM}. 
%For such a distribution, the non-Gaussianities are somewhat weaker since only the kurtosis is growing with time, while higher order cumulants are simply constant. 
%\blue{GS: the case of a uniform distribution can be computed in terms of Bernoulli numbers. Give it in SM.}        

What about the full probability distribution $p_r(x,t)$ of $x(t)$, both at short and large times?
%One may naturally wonder: what can be inferred from these asymptotic behaviors of the cumulants (\ref{summary_cum}) about the full distribution of $x(t)$, both at short and large times? 
At short time $rt \ll 1$, {the particle does not have enough time to switch states and hence diffuses freely with a propagator $e^{-x^2/(4{\sf D}_1\tau)}/\sqrt{4\pi {\sf D}_1 \tau}$. Averaging over ${\sf D}_1$ leads to
%${\sf D}_1$ the free propagator $e^{-x^2/(4{\sf D}_1\tau)}/\sqrt{4\pi {\sf D}_1 \tau}$ and it takes the scaling form}%it is easy to see that the PDF $p_r(x,t)$ takes the scaling form
\begin{equation} \label{diff_diff}
p_r(x,t) \approx   \int_{0}^{+\infty} dD\, W(D) \frac{e^{-\frac{x^2}{4Dt}}}{\sqrt{4 \pi D t}}\;.
\end{equation}Not surprisingly, this PDF (\ref{diff_diff}) has exactly the form found for diffusing diffusivity model \cite{Chubynsky2014,Wang2012}. On the other hand, at large time $r\,t \gg 1$, one finds that the PDF of the position takes a large deviation form
\bea\label{large_dev_form}
p_r(x,t) \approx e^{- t \, I\left(y= {x}/{t}\right)} \;,
\eea
where $I(y)$ is a large deviation function (LDF), whose precise shape depends on $W(D)$. However, its asymptotic behaviors for small and large arguments are universal and are given by
\bea \label{asympt_I}
I(y) \approx
\begin{cases}
& \frac{y^2}{4 \langle D \rangle} \quad \quad \quad \;, \quad y \to 0 \;,
 \\
& r + \frac{y^2}{4 D_{\max}} \quad, \quad y \to \infty \;.
 \end{cases}
\eea
Here, $D_{\max}$ denotes the right edge of the support of $W(D)$ \cite{footnote2}. %\cite{Note1}. %. 
These two asymptotic behaviors can be physically understood as follows. When $y \to 0$, i.e.  $x\ll t$, the Gaussian behavior near
the center of the PDF picks up the average $\langle D \rangle$ (since there are many switchings, the particle samples the average of $D$). On the other hand, for $y\to \infty$, i.e., $x \gg t$, this behavior is due to very rare trajectories where the particle diffuses with the largest diffusion coefficient $D_{\rm max}$ without undergoing any switch, which occurs with a
probability $e^{-rt}$. 
%Hence, $x\sim O(t)$ acts like a separatrix between typical and atypical trajectories. A similar scenario also occurs in the case of the resetting Brownian motion \cite{MSS, resettingReview} or for Brownian particle with a nonzero death rate~\cite{YMS}. As we show below, for some distributions $W(D)$, these two asymptotic behaviors (\ref{asympt_I}) are separated by a transition point where the function $I(y)$ displays a singularity, signaling the presence of a dynamical transition. In some cases, like in the two-state model, there may even be two transitions~(see~Fig. \ref{Fig_twostates}).  

\noindent {\it Renewal approach.} Our approach is based on a renewal argument, which enables us to derive an exact equation for $p_r(x,t)$, the PDF of the particle’s position at time $t$. The details are given in Appendix B of the End Matter.
%It reads \cite{SM}
%\begin{eqnarray}
%&&  p_r(x,t) = e^{-r\,t} \int_0^\infty dD\, W(D)\, \frac{e^{-\frac{x^2}{4D t}}}{\sqrt{4\pi D t}}\,  \label{renewal_1} \\
%&&\hspace*{-0.5cm}+ r\hspace*{-0.1cm}\int_0^{t} \hspace*{-0.2cm} d\tau\hspace*{-0.1 cm}\int_0^\infty\hspace*{-0.3cm} dD\,e^{-r \tau}\,W(D) \int_{-\infty}^{+\infty}\hspace*{-0.2cm}dz\, p_r(z,t-\tau) \frac{e^{-\frac{(x-z)^2}{4D \tau}}}{\sqrt{4\pi D \tau}} \;. \nonumber
%\end{eqnarray}
%In Eq.~(\ref{renewal_1}), the first term represents trajectories where no switching of the diffusion coefficient occurs. The second term corresponds to the case where there is at least one switching
%event in $[0,t]$.
%Suppose that the last switching before $t$ takes place at $t-\tau$, and let $z$ be the position
%of the walker just before this last switching. Then $p_r(z,t-\tau)$ is the propagator
%until $t-\tau$. After a switching to a new diffusion coefficient $D$ drawn from $W(D)$ at $t-\tau$, the particle propagates freely during the interval $[t-\tau, t]$ with a Gaussian propagator $e^{-(x-z)^2/(4D\tau)}/\sqrt{4\pi D \tau}$. Multiplying these two propagators over $[0,t-\tau]$ and $[t-\tau, t]$, integrating over $z$ and averaging over $D$ drawn from $W(D)$, gives the second term in Eq. (\ref{renewal_1}).
%In the following, we restrict our analysis to the case where $W(D)$ has a finite support $[0,D_{\text{max}}]$ and refer to \cite{SM} for more details when the support extends over the full real axis. 
%
\begin{figure}[t]
    \centering
\includegraphics[width=0.49\linewidth]{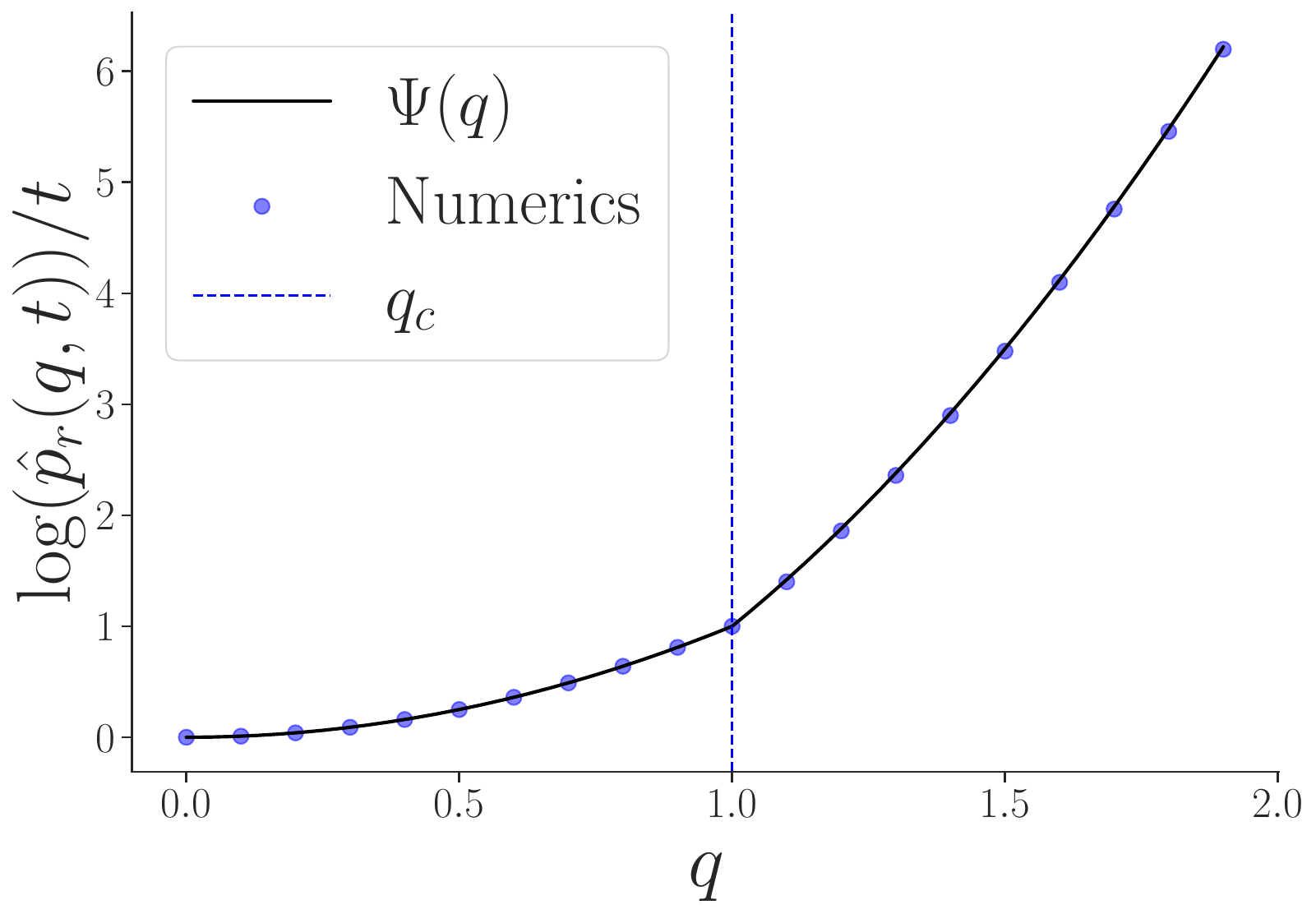}
\includegraphics[width=0.49\linewidth]{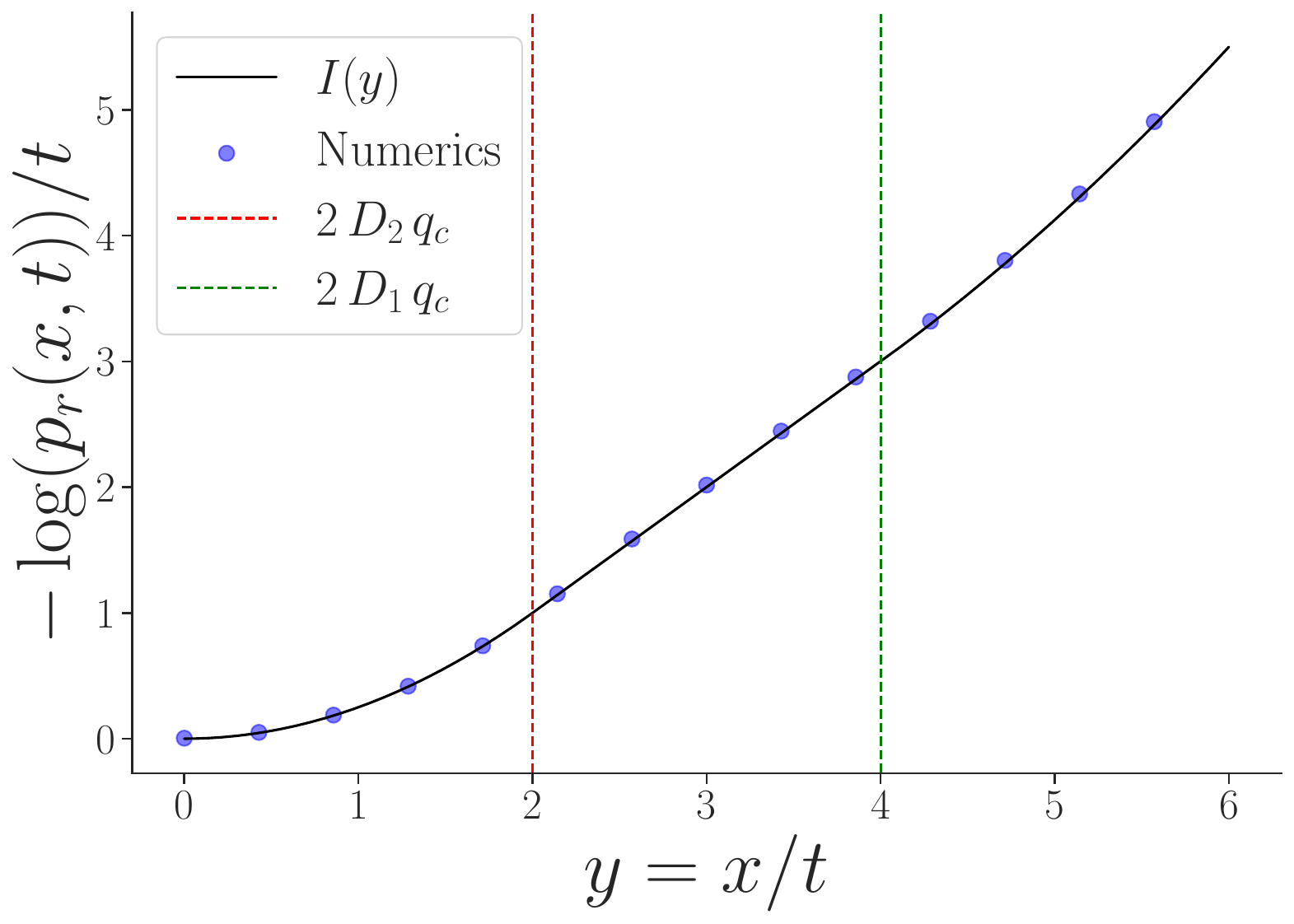}
  \caption{Plot of $\log(\hat{p}_r(q,t))/t$ vs $q$ (left panel) and of $-\log(p_r(x,t))/t$ vs $y=x/t$ (right panel) for the two-state model in the limit $p\to0^+$. The symbols correspond to numerical results (see \cite{SM} for details on numerical methods). \textbf{Left:} the solid line shows the exact analytical result $\Psi(q)$ -- see Eq.~(\ref
  {pzeroPsitwostates}). \textbf{Right:} the solid line shows the rate function $I(y)$ given in Eq.~(\ref{expr_I_2st}), displaying two transition points indicated by the dotted lines.
%  for the two-state model. The dots correspond to numerical results (see \cite{SM} for the details on the method) while the the solid line shows the exact analytical result $\Psi(q)$ -- see Eq.~(\ref
 % {pzeroPsitwostates}). \textbf{Right:} Plot of $-\log(p_r(x,t))/t$ vs $y=x/t$. The dots correspond to numerical results while 
%  the solid line shows the rate function $I(y)$ given in Eq.~(\ref{expr_I_2st}), displaying two transition points indicated by the dotted lines. 
The corresponding values of the probabilities are as small as $10^{-2000}$. Here we used $r = 1$, $p = 10^{-10}$, $D_1 = 2$, $D_2 = 1$ and $t=1000$.}
  \label{Fig_twostates} 
\end{figure}
This renewal equation (\ref{renewal_1}) has a convolution structure, both in time and space variables. It is thus natural to introduce the generating function of $x$ together with its Laplace transform (with respect to $t$)
\begin{eqnarray}
\hat p_r(q,t) = \langle e^{q\,x} \rangle \quad , \quad
\tilde p_r(q,s) = \int_0^\infty dt \, e^{-st} \hat p_r(q,t) \;,\label{def_dble_Laplace}
\end{eqnarray}
where $\langle e^{qx} \rangle = \int_{-\infty}^{\infty} \hspace*{-0.1cm} dx\, e^{qx}  p_r(x,t)$. In this paper, we restrict our analysis to the case where $W(D)$ has a finite support $[0,D_{\text{max}}]$ and refer to \cite{SM} for more details when the support extends over the full real axis. Using the aforementioned convolution structure of (\ref{renewal_1}), $\tilde p_r(q,s)$ can be computed explicitly, leading to the exact expression~\cite{SM}
\begin{equation}
  \tilde{p}_r(q,s) = \frac{J_r(q,s)}{1-r\, J_r(q,s)} \;,\; J_r(q,s) = \int_0^{D_{\text{max}}}\hspace*{-0.3cm}dD\, \frac{W(D)}{r+s-D q^2}. \label{exact_dbleLaplace}
\end{equation}
%Note that these functions in the $(q,s)$-plane are defined in the region $r+s > D_{\text{max}} q^2$, such that the integrand defining $J_r(s,q)$ in (\ref{exact_dbleLaplace}) is free of any singularity. 
These formulae (\ref{def_dble_Laplace})-(\ref{exact_dbleLaplace}) provide an exact representation of the generating function, allowing the computation of the moments given in Eq.~(\ref{finite_t}) -- see~\cite{SM}. Carrying out explicitly the double inversion with respect to $q$ and $s$ to recover  $p_r(x,t)$ remains a formidable challenge. However, analytical progress can be made to extract the large-time behavior of $p_r(x,t)$.  In this regime, the behavior of $\hat{p}_r(q,t)$ is governed by the singularities of $\tilde{p}_r(q,s)$ in the complex $s$-plane. Indeed, we show that, for large $t$, the generating function $\hat{p}_r(q,t)$ reads
\bea \label{def_Psi}
\hat p(q,t) \approx e^{t \, \Psi(q)} \;, \; t \to \infty \;,
\eea
where $\Psi(q) = s^*(q)$ is the singularity of $\tilde{p}_r(q,s)$ in Eq.~(\ref{exact_dbleLaplace}) with the largest real part in the complex $s$-plane. The function $\Psi(q)$ is a central object since this is the scaled cumulant generating function (SCGF). Indeed, this form in (\ref{def_Psi}) already shows that all the cumulants of $x(t)$ are a priori of order $O(t)$ for large $t$ and given by the behaviors of $\Psi(q)$ near $q=0$, namely $\langle x^n(t)\rangle_c \approx t \, \partial_q^n \Psi(q) \vert_{q=0}$.
%More precisely, they are given by
%\bea \label{expr_cumul}
%\langle x^n(t)\rangle_c \approx t \, \frac{\partial^n \Psi(q)}{\partial q^n} \Bigg \vert_{q=0} \;, \; t \to \infty \;.
%\eea

Since $\Psi(q)$ is symmetric, we only study it for $q \geq 0$. For sufficiently small $q$, the leading singularity of $\tilde{p}_r(q,s)$ in the complex $s$-plane that determines $\Psi(q)$
is a pole, namely a root of the denominator in Eq. (\ref{exact_dbleLaplace}) \cite{SM}. Hence, for $q$ small enough, $\Psi(q) = s^*(q)$ is given implicitly by the root with the largest real part of the equation 
\bea \label{implicit}
1 = r \int_0^{D_{\text{max}}} dD \frac{W(D)}{r+\Psi(q) - D q^2} \, . 
\eea
%Of course, for general $W(D)$, it seems complicated to compute the derivatives of $\Psi(q)$ explicitly, and get the cumulants from the implicit relation (\ref{implicit}). 
Remarkably, Eq.~(\ref{implicit}) has a quite familiar structure which is well known in the context of {\it free probability} and its application to RMT~\cite{VivoBook, PottersBouchaud, GuionnetHouches, Voiculescu, MaidaGuionnet}. More precisely the SCGF $\Psi(q)$ is given by (at least in a neighborhood of $q=0$) 
\bea \label{psi_R}
\Psi(q) = q^2 \, R\left({q^2}/{r}\right) \;,
\eea
where $R(z)$ is the so-called {\it R-transform} of $W(D)$. Given the PDF $W(D)$, its {\it R-transform} is the generating function of the
free cumulants $R(z) = \sum_{n \geq 1} z^{n-1} \kappa_n(D)$ and it can be obtained from the Cauchy-Stieljes transform of $W(D)$ [see Eq. (\ref{free_relation})].
%which by definition is the generating function of the free-cumulants of the random variable $D$, namely $R(z) = \sum_{n \geq 1} z^{n-1} \kappa_n(D)$. 
This result (\ref{psi_R}) thus leads to the second line of Eq. (\ref{summary_cum})~\cite{footnote3}.
%which reads \cite{explicit_free_cumu1,explicit_free_cumu2}  
%\begin{eqnarray} \label{free_norm}
%    \kappa_n(D)= \sum_{j=1}^{n} \frac{(-1)^{j-1}}{j}\binom{n+j-2}{j-1}\widetilde{\sum_{\vec{q}}} \prod_{k=1}^j \langle D^{q_k}\rangle\, ,
%\end{eqnarray}  
%where $\widetilde{\sum_{\vec{q}}}$, with $\vec q = (q_1, \cdots, q_j)$ denotes a constrained sum such that  $q_1+q_2+\ldots +q_j=n$ with integers $q_k \geq 1$ \footnote{Note that we have corrected a typo compared to \cite{explicit_free_cumu1,explicit_free_cumu2}, where instead $q_k \geq 0$.}. When specified to the two-state model, for which $\langle D^q \rangle = p D_1^q + (1-p) D_2^q$ one obtains, after some manipulations, the formula given in (\ref{moment_dble}). 

For any distribution $W(D)$ with a finite support on $[0,D_{\max}]$, the asymptotic behaviors of the SCGF are \cite{SM}
\bea
\Psi(q) = 
\begin{cases}
\langle D \rangle \, q^2 &\;,  \quad q \to 0 \;, \\%[10pt]
D_{\text{max}} q^2 - r &\;,  \quad q \to \infty \;.
\end{cases}
\label{asymptpsi}
\eea
What happens between these two limits depends essentially on the behavior of $W(D)$ near $D_{\max}$, as in the extreme value statistics in the Weibull
universality class \cite{EVSBook}. Let us 
assume that $W(D)$ behaves as $W(D) \sim (D_{\text{max}} - D)^{\nu}$ when $D \to D_{\text{max}}$ with $\nu >-1$. For $-1<\nu \leq 0$, $\Psi(q)$ is given by Eq.~(\ref{psi_R}) for all $q$ and it is an analytic function of all $q \in \mathbb{R}$. Instead, for $\nu > 0$, (\ref{psi_R}) only holds for small $q$,~i.e., 
\bea
\Psi(q) = 
\begin{cases}
q^2 \, R\left( q^2/r\right)&\;,  \quad q <q_c \;, \\%[10pt]
D_{\text{max}} q^2 - r &\;,  \quad q >q_c \;,
\end{cases}
\label{psinupositive}
\eea
where the SCGF undergoes a transition at $q=q_c$, with $q_c^2 = r g(D_{\max})$, $g(x)$ being the Cauchy-Stieltjes transform of $W(D)$ -- see Eq.~(\ref{psi_cauchy}). While $\Psi(q)$ is continuous, its higher derivatives display singularities at $q=q_c$~(see~\cite{SM} for details). In particular, for $\nu > 1$ (as well as for the two-sate model in the limit $p\to 0^+$), the first derivative of $\Psi(q)$ is discontinuous -- see the left panel of Fig.~\ref{Fig_twostates}. Interestingly, a very similar transition occurs in the study of Harish-Chandra-Itzykson-Zuber matrix integrals (or spherical integrals) in large dimensions \cite{MaidaGuionnet} although these two problems are seemingly unrelated.

\noindent {\it The LDF $I(y=x/t)$.} From the standard theory of large deviations \cite{LD_Touchette,MS17}, the exponential form of the SCGF in (\ref{def_Psi}) implies the large deviation form of $p_r(x,t)$ in Eq. (\ref{large_dev_form}) where the LDF $I(y)$ is given by the Legendre transform of $\Psi(q)$, namely
\bea \label{legendre}
I(y) = \max_{q \in \mathbb{R}} (q\,y - \Psi(q)) \;.
\eea
Using this formula and the asymptotics of $\Psi(q)$ from Eq.~(\ref{asymptpsi}), we find that $I(y)$ behaves as in Eq.~(\ref{asympt_I}). Since $I(y)$ is symmetric, we study it only for $y\geq 0$.

For a distribution $W(D)$ with a finite support $[0,D_{\max}]$ as discussed above with $-1 <\nu \leq 0$, the LDF $I(y)$ is regular and crosses over smoothly between the two asymptotic behaviors given in~(\ref{asympt_I}). This is, for instance, the case of a uniform distribution~\cite{SM}. However, for $0<\nu<1$, the LDF exhibits a dynamical transition of the form
\bea 
\label{ratenu01}
I(y) = 
\begin{cases}
\phi_\nu(y) &\quad, \quad y \leq y_c = 2 D_{\max} q_c \, ,\\
 r + \frac{y^2}{4 D_{\text{max}}} &\quad, \quad y \geq y_c \;,
\end{cases}
\eea
where $\phi_\nu(y)$ is the Legendre transform of Eq.~(\ref{psi_R}) -- which we can compute explicitly in the case of the Wigner semi-circle law ($\nu = 1/2$) \cite{SM}. 
This transition for the case when $W(D)$ vanishes as $D \to D_{\max}$ has an interesting physical implication. The sharp dynamical transition at $y=y_c$ implies the existence of a ``light cone'' $x = \pm y_c t$ in the space-time plane (see the left panel of Fig.~\ref{transitions_diagrams}). This light cone acts like a separatrix between rare atypical trajectories and the typical trajectories, as seen in models of diffusion with resetting \cite{resettingPRL, MSS, YMS, resettingReview, Kundureset}. Trajectories that stay outside the light cone up to time $t$ are the ones which undergo very few switchings in time $t$, while those inside the light cone are the typical trajectories that experience a large number of switching events. However this sharp light cone and its associated sharp transition disappear when $W(D)$ does not vanish as $D \to D_{\max}$ (i.e., when $-1 < \nu \leq 0$). This is because, in that case, there is a nonzero probability for realizing many switching events but with a large fraction of them close to $D_{\max}$. As $x$ decreases, for a fixed $t$, such trajectories smoothly interpolate between atypical and typical trajectories, leading to the disappearance of the sharp transition.

Finally, when $\nu >1$, the LDF $I(y)$ exhibits two singular points between which its behavior is linear in $y$, namely
\bea
\label{nugeq1}
I(y) = 
\begin{cases}
\phi_\nu(y) &, \quad 0 < y < 2 D_{\rm eff} q_c \;, \\
q_c y -\gamma &, \quad 2 D_{\rm eff} q_c < y < y_c \;, \\
r + \frac{y^2}{4 D_{\text{max}}} &, \quad y > y_c \;,
\end{cases}
\eea
where $D_{\rm eff} < D_{\max}$ and $\gamma =D_{\max}q_c^2 - r >0$ can be computed explicitly~\cite{SM}. In \cite{SM}, we show that the two-state model exhibits the same transitions in the limit $p\to0^+$ -- see Appendix E and the right panel of Fig.~\ref{Fig_twostates}. Thus in this case, there are two transitions as a function of the scaled distance $y$, with a new intermediate phase for $2 D_{\rm eff} q_c < y < y_c = 2D_{\max}q_c$, sandwiched between the atypical and typical regimes. In this new intermediate phase, the PDF takes the form $p_r(x,t) \sim e^{- q_c(x-v\,t)}$ where $v = y_c - r D_{\max}/y_c > 0$. Thus, in this intermediate phase, the position distribution has the shape of a traveling front, with a nontrivial velocity $v$ \cite{saarloos}. Hence, in the space-time plane, we now have two light cones respectively with slopes $2 D_{\rm eff} q_c$ and $y_c$ that separate three regimes of trajectories \cite{SM} (see the right panel of Fig.~\ref{transitions_diagrams}). Note that while $I(y)$ and $I'(y)$ are continuous across the two transitions, the second derivative $I''(y)$ is generically discontinuous at these two points -- and similarly at $y=y_c$ in Eq.~(\ref{ratenu01}) (see \cite{SM} for more details). This type of change of behaviors in the position distribution was also found in some models of CTRW \cite{Barkai1,Sokolov,Burov1,Burov2,BB}.

%Hence, $x\sim O(t)$ acts like a separatrix between typical and atypical trajectories. A similar scenario also occurs in the case of the resetting Brownian motion \cite{MSS, resettingReview} or for Brownian particle with a nonzero death rate~\cite{YMS}. As we show below, for some distributions $W(D)$, these two asymptotic behaviors (\ref{asympt_I}) are separated by a transition point where the function $I(y)$ displays a singularity, signaling the presence of a dynamical transition. In some cases, like in the two-state model, there may even be two transitions~(see~Fig. \ref{Fig_twostates}).  

%It is easy to see that, at $y=y_2$, the first derivative of $I(y)$ is continuous while the second is not. However, at $y_1$ the situation is a bit more complicated and depends on $1<\nu<2$ or $\nu >2$
%$y_1$ and $y_2$, between which its behavior is linear. When $y<y_1$, $I(y)$ is a non-trivial function, and for $y>y_2$, $I(y)$ is given by the second line of Eq.~(\ref{ratenu01}) \cite{SM}. 

\noindent {\it Conclusion.} We have investigated the dynamics of a Brownian particle with a switching diffusion coefficient, obtaining the exact expression of the moments at any finite time $t$ and for any $W(D)$ with finite moments. 
At large times, our analysis of the cumulants and the large deviation function reveals significant deviations from Gaussian behavior in the position distribution of the particle, with intermediate exponential decay emerging in certain cases (\ref{nugeq1}).
%At large time, our analysis of the cumulants and the large deviations, reveals non-Gaussian behaviors in the distribution of the position of the particle, displaying exponential tails in some cases. Notably, we unveiled a remarkable connection between switching diffusion and free probability theory, which is surprising for a such a classical diffusion model. It remains challenging to understand the origin of this connection. Besides, one may naturally wonder what happens in higher dimensions $N>1$ or, equivalently, to $N$-particles subjected to simultaneous switching, extending the recent works in the context of simultaneous resetting~\cite{Marco1, Marco2, Marco3}. Finally, it will be interesting to study the effect of an external confining potential, which would lead to a non-equilibrium stationary state~\cite{Luczka2,Hanggi88}.
Remarkably, we uncovered a surprising connection between switching diffusion and free probability theory, an unexpected link in such a classical single particle diffusion model. The origin of this connection remains a challenging and intriguing question for further investigation. Another unexpected connection has recently been noticed between switching diffusion and a random multiplicative growth model~\cite{BBLD2025}. As shown there, the growth rate in that model is analogous to the SCGF $\Psi(q)$ of our switching diffusion model, and thus shares similar transitions and relations to free cumulants. In~\cite{SM}, we specify the mapping between the~two~models. 

Our work opens several natural extensions. A key question is the generalization to %higher dimensions ($N~>~1$) or, equivalently to 
$N$ particles subjected to simultaneous switching dynamics. This direction could build upon recent studies in the context 
of simultaneous resetting~\cite{Marco1, Marco2}. Similar questions were recently studied for $N$ particles in a harmonic trap in the presence of switching stiffnesses \cite{Marco3} and switching centers~\cite{SabhapanditMajumdar, KulkarniMajumdar}. 
%Finally, it will be interesting to study the effect of an external confining potential, which would lead to a non-equilibrium stationary state~\cite{Luczka2,Hanggi88}.
The extension to higher dimensions is also natural. In dimension $d > 1$, the result (\ref{exact_dbleLaplace}) generalizes straightforwardly by replacing $q$ with its norm~\cite{SM}. Thus, the distance to the origin exhibits the same properties as the one-dimensional case. Moreover, switching events introduce nontrivial correlations between the components $x_i$'s with $i = 1,\ldots, d$. For instance, one can show that, for $i \neq j$, $\langle x_i^2 x_j^2\rangle - \langle x_i^2 \rangle \langle x_j^2\rangle \propto \kappa_2(D)\,t$ in dimension $d>1$ -- see~\cite{SM}. It would be very interesting to probe experimentally these higher order correlations -- as well as higher order cumulants -- and compare with the linear growth $\propto t$ predicted here [see Eq. (\ref{summary_cum})], which is a clear indication of non-Gaussian diffusion.  
%Finally, measuring higher-order cumulants -- which we can compute exactly -- in experimental systems, and in particular, verifying whether they grow linearly in time, would represent an important step toward validating switching diffusion models.

\vspace*{0.5cm}
\noindent {\it Acknowledgments.}{ We thank O. Arizmendi, D. Bernard, M. Bernard, P. Biane, J.-P. Bouchaud, A. Guionnet, A. Hartmann, P. Le Doussal, M. Maïda, R. Speicher, H. Touchette, L. Touzo and J. B. Zuber for useful discussions. We also acknowledge support from ANR Grant No. ANR-
23-CE30-0020-01 EDIPS.}

\newpage

\newpage
\begin{widetext}
\newpage
{\color{black}
\begin{center}
{\bf End Matter}
\end{center}

\vspace*{0.0cm}
\begin{center}
{\bf Appendix A: Exact expressions of the first three moments}
\end{center}
\vspace*{0.0cm}

We provide the exact expression for the moments of the position in the switching diffusion process, valid at any finite time $t$ and for any distribution $W(D)$ with finite moments, as derived in the supplementary materials~\cite{SM}. It reads
\begin{eqnarray}
    && \langle x^{2n}(t) \rangle = \frac{(2n)!}{r^n} \, \sum_{m=1}^n\, \frac{(r t)^{m+n-1}}{(m+n-1)!}\, M(n-1, m + n, -r t)\, \hat{B}_{n,m}\left(\langle D \rangle, \ldots ,\langle D^{n-m+1} \rangle \right)\, ,
    \label{finite_t}
\end{eqnarray}
where $\hat{B}_{n,m}$ is the ordinary Bell polynomial of $n-m$ variables and of homogeneous degree $m$~\cite{ComtetL, Riordan_book, SM}. In Eq.~(\ref{finite_t}), the function $M(a,b,x)$ denotes Kummer's function. We also give explicitly in~\cite{SM} the first three non-zero moments.

\begin{center}
{\bf Appendix B: Renewal approach}
\end{center}
\vspace*{0.0cm}

We present in this appendix the renewal equation satisfied by $p_r(x,t)$. It reads (more details are provided in~\cite{SM})
\begin{eqnarray}
p_r(x,t) = e^{-rt}\, \int_0^{+\infty}dD\, W(D)\, \frac{e^{-\frac{x^2}{4Dt}}}{\sqrt{4\pi D t}}\,  + \int_0^{t} d\tau\, r\, e^{-r \tau} \int_{-\infty}^{+\infty}dz\, p_r(z,t-\tau) \, \int_0^{+\infty}dD\, W(D)\, \frac{e^{-\frac{(x-z)^2}{4D \tau}}}{\sqrt{4\pi D \tau}} \, .
\label{renewal_1}
\end{eqnarray}
In Eq.~(\ref{renewal_1}), the first term represents trajectories where no switching of the diffusion coefficient occurs. The second term corresponds to the case where there is at least one switching
event in $[0,t]$.
Suppose that the last switching before $t$ takes place at $t-\tau$ (with associated probability $rd\tau e^{-r\tau}$), and let $z$ be the position
of the walker just before this last switching. Then $p_r(z,t-\tau)$ is the propagator
until $t-\tau$. After a switching to a new diffusion coefficient $D$ drawn from $W(D)$ at $t-\tau$, the particle propagates freely during the interval $[t-\tau, t]$ with a Gaussian propagator $e^{-(x-z)^2/(4D\tau)}/\sqrt{4\pi D \tau}$. Multiplying these two propagators over $[0,t-\tau]$ and $[t-\tau, t]$, integrating over $z$ and $\tau$, and averaging over $D$ drawn from $W(D)$, gives the second term in Eq. (\ref{renewal_1}).

\vspace*{0.0cm}
\begin{center}
{\bf Appendix C: Cumulants and free cumulants}
\end{center}
\vspace*{0.0cm}

For a random variable $D$ with distribution $W(D)$, the classical cumulants are related to the moments via the following explicit formula
\begin{eqnarray}
    \langle D^{n} \rangle_c = \sum_{k=1}^{n}(-1)^{k-1}(k-1)!\, B_{n,k}\left(\langle D \rangle, \ldots,\langle D^{n-k+1} \rangle  \right)\, ,
    \label{cumulantclassic}
\end{eqnarray}    
where $B_{n,k}$ are the partial exponential Bell polynomials. We give below the first few classical cumulants
%\begin{eqnarray}
%\langle D \rangle_c&=& \langle D \rangle \, ,\\
%\langle D^2 \rangle_c&=& \langle D^2 \rangle - \langle D \rangle^2 \, ,\\
%\langle D^3 \rangle_c &=& \langle D^3 \rangle - 3\langle D^2 \rangle\langle D \rangle + 2\langle D \rangle^3 \, ,\\
%\langle D^4 \rangle_c &=& \langle D^4 \rangle - 4\langle D^3 \rangle\langle D \rangle - 3\langle D^2 \rangle^2 + 12\langle D^2 \rangle\langle D \rangle^2 - 6\langle D \rangle^4 \, ,\\
%\langle D^5 \rangle_c &=& \langle D^5 \rangle - 5\langle D^4 \rangle\langle D \rangle - 10\langle D^3 \rangle\langle D^2 \rangle + 20\langle D^3 \rangle\langle D \rangle^2 + 30\langle D^2 \rangle^2\langle D \rangle - 60\langle D^2 \rangle\langle D \rangle^3 + 24\langle D \rangle^5 \,.
%\end{eqnarray}
\begin{align}
\langle D \rangle_c &= \langle D \rangle, \quad
\langle D^2 \rangle_c = \langle D^2 \rangle - \langle D \rangle^2, \quad
\langle D^3 \rangle_c = \langle D^3 \rangle - 3\langle D^2 \rangle\langle D \rangle + 2\langle D \rangle^3, \\
\langle D^4 \rangle_c &= \langle D^4 \rangle - 4\langle D^3 \rangle\langle D \rangle - 3\langle D^2 \rangle^2 + 12\langle D^2 \rangle\langle D \rangle^2 - 6\langle D \rangle^4, \\
\langle D^5 \rangle_c &= \langle D^5 \rangle - 5\langle D^4 \rangle\langle D \rangle - 10\langle D^3 \rangle\langle D^2 \rangle + 20\langle D^3 \rangle\langle D \rangle^2 + 30\langle D^2 \rangle^2\langle D \rangle- 60\langle D^2 \rangle\langle D \rangle^3 + 24\langle D \rangle^5 \,.
\end{align}
The free cumulants, on the other hand, can be computed using the following explicit formula in terms of the moments of $D$, which reads~\cite{explicit_free_cumu1,explicit_free_cumu2}  
\begin{eqnarray} \label{free_norm}
    \kappa_n(D)= \sum_{j=1}^{n} \frac{(-1)^{j-1}}{j}\binom{n+j-2}{j-1}\widetilde{\sum_{\vec{q}}} \prod_{k=1}^j \langle D^{q_k}\rangle\, ,
\end{eqnarray}  
where $\widetilde{\sum_{\vec{q}}}$, with $\vec q = (q_1, \cdots, q_j)$ denotes a constrained sum such that  $q_1+q_2+\ldots +q_j=n$ with integers $q_k \geq 1$. Note that we have corrected a typo compared to \cite{explicit_free_cumu1,explicit_free_cumu2}, where instead $q_k \geq 0$. The first few free cumulants are given by
%\begin{eqnarray}
%\kappa_1(D)&=& \langle D \rangle \, ,\\
%\kappa_2(D)&=& \langle D^2 \rangle - \langle D \rangle^2 \, ,\\
%\kappa_3(D)&=& \langle D^3 \rangle - 3\langle D^2 \rangle\langle D \rangle + 2\langle D \rangle^3 \, ,\\
%\kappa_4(D) &=& \langle D^4 \rangle - 4\langle D^3 \rangle\langle D \rangle - 2\langle D^2 \rangle^2 + 10\langle D^2 \rangle\langle D \rangle^2 - 5\langle D \rangle^4 \, ,\\
%\kappa_5(D) &=& \langle D^5 \rangle - 5\langle D^4 \rangle\langle D \rangle - 5\langle D^3 \rangle\langle D^2 \rangle + 15\langle D^3 \rangle\langle D \rangle^2 + 15\langle D^2 \rangle^2\langle D \rangle - 35\langle D^2 \rangle\langle D \rangle^3 + 14\langle D \rangle^5 \,.
%\end{eqnarray}
\begin{align}
\kappa_1(D) &= \langle D \rangle, \quad 
\kappa_2(D) = \langle D^2 \rangle - \langle D \rangle^2, \quad 
\kappa_3(D) = \langle D^3 \rangle - 3\langle D^2 \rangle\langle D \rangle + 2\langle D \rangle^3, \\
\kappa_4(D) &= \langle D^4 \rangle - 4\langle D^3 \rangle\langle D \rangle - 2\langle D^2 \rangle^2 + 10\langle D^2 \rangle\langle D \rangle^2 - 5\langle D \rangle^4, \\
\kappa_5(D) &= \langle D^5 \rangle - 5\langle D^4 \rangle\langle D \rangle - 5\langle D^3 \rangle\langle D^2 \rangle + 15\langle D^3 \rangle\langle D \rangle^2 + 15\langle D^2 \rangle^2\langle D \rangle - 35\langle D^2 \rangle\langle D \rangle^3 + 1 4\langle D \rangle^5 \,.
\end{align}
%The expressions for classical and free cumulants differ starting from  $n > 3$. This is because the computation of the classical cumulant $\langle D^n\rangle_c$ usually involves a sum over all possible partitions of $\{1, 2, \dots, n\}$, including both crossing and non-crossing partitions. In contrast, free cumulants are computed using only non-crossing partitions of the indices~\cite{Speicher_book}.

%For  $n = 1$, $2$,  and  $3$, the possible partitions of the sets  $\{1\}$,  $\{1, 2\}$, and $\{1, 2, 3\}$  do not include any crossing partitions. Consequently, for  $n \leq 3$ , the set of all partitions coincides with the set of {\it non-crossing partitions} \cite{Speicher_book}, meaning the first three free cumulants are identical to the corresponding classical cumulants.

\vspace*{0.0cm}
\begin{center}
{\bf Appendix D: The SCGF $\Psi(q)$ in terms of the \textit{R-transform} of $W(D)$}
\end{center}
\vspace*{0.0cm}

For small values of $q$, we argued that the large time behavior of the generating function of $x$ is $e^{t\Psi(q)}$ where $\Psi(q)$ is solution of Eq.~(\ref{implicit}). This equation can also be written in terms of the Cauchy-Stieltjes transform $g(x)$ as 
\begin{eqnarray}
    \frac{q^2}{r} = g\left(\frac{r+\Psi(q)}{q^2}\right) \quad \text{where} \quad g(x) = \int_0^{D_{\max}}dD\, \frac{W(D)}{x - D}\, .
    \label{psi_cauchy}
\end{eqnarray}
For a real probability measure $W(D)$, the following relation holds
\begin{eqnarray}\label{free_relation}
    g\left(R(z) +\frac{1}{z}\right)=z\, ,
\end{eqnarray}
where $R(z)$ is the {\it R-transform} of the PDF $W(D)$ \cite{Speicher_book, Speicher_book2, Voiculescu, MingoSpeicher17,GuionnetHouches}. We recall that it can be written, at small $z$, as an expansion where the coefficients are the free cumulants of $W(D)$, denoted by $\kappa_n(D)$, namely $R(z) = \sum_{n \geq 1} z^{n-1} \kappa_n(D)$.
Therefore, by identifying terms in Eq.~(\ref{psi_cauchy}), we obtain the crucial relation
\begin{eqnarray}
    \Psi(q) =  \lim_{t \to \infty} \frac{\ln \hat p_r(q,t)}{t} = q^2 \, R\left( \frac{q^2}{r}\right) =r \sum_{n \geq 1} \left(\frac{q^2}{r} \right)^{n} \kappa_n(D)\, .
\end{eqnarray}
%\vspace*{0.3cm}
%\begin{center}
%{\textbf{Appendix D: Expansion of the cumulants at large time up to order $O(1)$}}
%\end{center}
%\vspace*{0.3cm}
%
%It is possible to derive the $O(1)$ term in the large-time expansion of the cumulants of the position of the particle. As demonstrated in \cite{SM}, this term can be expressed as a sum involving Bell polynomials, whose arguments are the free cumulants of the random variable $D$ distributed according to $W(D)$. The resulting expression is given by
%
%\bea \label{order1}
%\langle x^n(t)\rangle_c \isApproxTo{t\to\infty} \frac{(2n)!}{r^{n-1}}\, \kappa_n(D)\,t  - \frac{(2n)!}{r^n} \sum_{m=1}^n\frac{1}{m} \hat B_{n,m}\left(\tilde \kappa_1(D), \cdots, \tilde \kappa_{n-m+1}(D) \right) + O(e^{- r t}) \;, \; \tilde \kappa_n(D) = (n-1) \kappa_n(D)\, .
%\eea

\vspace*{0.0cm}
\begin{center}
{\bf Appendix E: {The SCGF $\Psi(q)$ and LDF $I(y)$ for the two-state model}}
\end{center}
\vspace*{0.0cm}

For the two-state model, the SCGF $\Psi(q)$ can be computed explicitly, leading  
\begin{eqnarray}\label{PsiqTwostates}
&&\Psi(q) =\frac{1}{2} \left((D_1 + D_2) q^2 - r + \Delta(q)\right) \quad , \quad \Delta(q) = \sqrt{\left((D_2 - D_1) q^2 + r\right)^2 + 4 (D_1 - D_2)p r q^2}\;.
\end{eqnarray}
Interestingly, in the limit $p\to 0^+$, $\Psi(q)$ exhibits a transition as $q$ crosses some value $q_c  = \sqrt{r/(D_1-D_2)}$. One also finds that $I(y)$ has a nontrivial limit $p \to 0^+$ (see \cite{SM} for details). They read
\noindent
\begin{minipage}[t]{0.48\linewidth}
  \begin{equation}\label{pzeroPsitwostates}
   \Psi (q) = 
\begin{cases}
D_2\, q^2 &\;,  \quad q < q_c \;, \\[10pt]
D_1\, q^2 - r &\;,  \quad q > q_c \;.
\end{cases}
  \end{equation}
\end{minipage}\hfill
\begin{minipage}[t]{0.48\linewidth}
  \begin{equation}\label{expr_I_2st}
I(y) = 
\begin{cases} 
\frac{y^2}{4 D_{2}} & , \,  |y| \leq 2 D_{2} q_c \, ,\\
q_c\, |y| - D_{2}\, q_c^2 & , \,  2 D_2 q_c \leq |y| \leq 2 D_1 q_c \, ,\\
r + \frac{y^2}{4 D_{1}} & , \,  |y| \geq 2 D_{1} q_c \;.
\end{cases}
  \end{equation}
\end{minipage}\vspace{0.3cm}
%\bea
%\lim_{p \to 0^+} \, \Psi (q) = 
%\begin{cases}
%D_2\, q^2 &\;,  \quad q < q_c \;, \\[10pt]
%D_1\, q^2 - r &\;,  \quad q > q_c \;.
%\end{cases}
%\label{pzeroPsitwostates}
%\eea
%\bea \label{expr_I_2st}
%I(y) = 
%\begin{cases} 
%\frac{y^2}{4 D_{2}} & , \quad |y| \leq 2 D_{2} q_c \, ,\\
%q_c\, |y| - D_{2}\, q_c^2 & , \quad 2 D_2 q_c \leq |y| \leq 2 D_1 q_c \, ,\\
%r + \frac{y^2}{4 D_{1}} & , \quad |y| \geq 2 D_{1} q_c \;.
%\end{cases}
%\eea
Interestingly, although $I(y)$ as well as its first derivative $I'(y)$ are continuous at $y=2 D_2 q_c$ and $y = 2 D_1 q_c$, the second derivative $I''(y)$ is discontinuous, signaling second order dynamical transitions at these two points.

\vspace*{0.0cm}
\begin{center}
{\bf Appendix F: {Space-time diagrams}}
\end{center}
\vspace*{0.0cm}

We consider a distribution $W(D)$ with finite support $[0,D_{\max}]$, and such that $W(D) \sim (D_{\max}-D)^\nu$ as $D \to D_{\max}$ (with $\nu>-1$). When $\nu \leq 0$ the rate function $I(y = x/t)$ smoothly interpolates between the two regimes described in Eq.~(\ref{asympt_I}). However, when $\nu > 0$, $I(y)$ exhibits transitions depending on the value of $\nu$. The nature of the transitions depend on the two different cases presented in the figure below:
\begin{figure}[H]
    \centering
\includegraphics[width=0.49\linewidth]{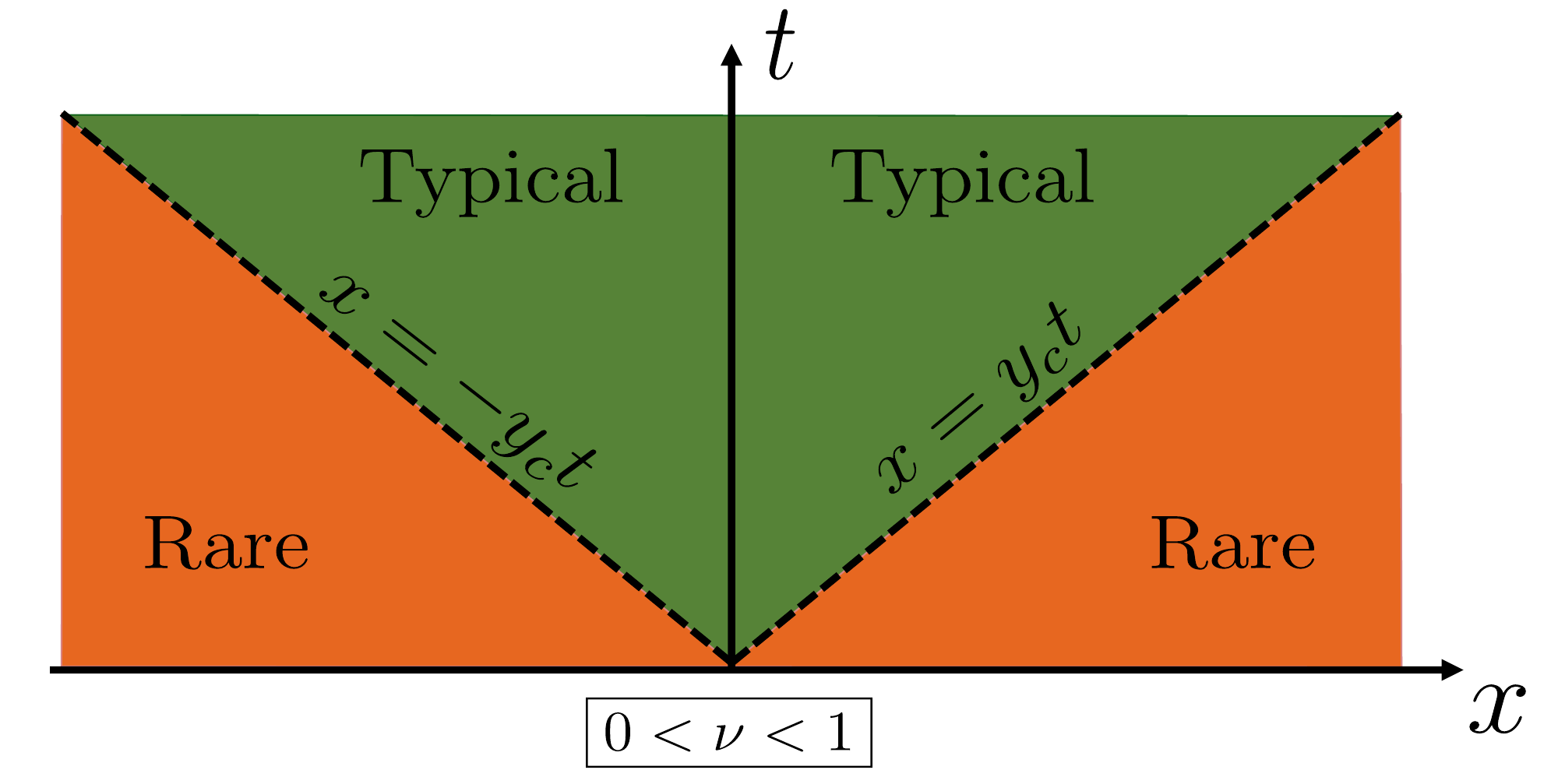}
\includegraphics[width=0.49\linewidth]{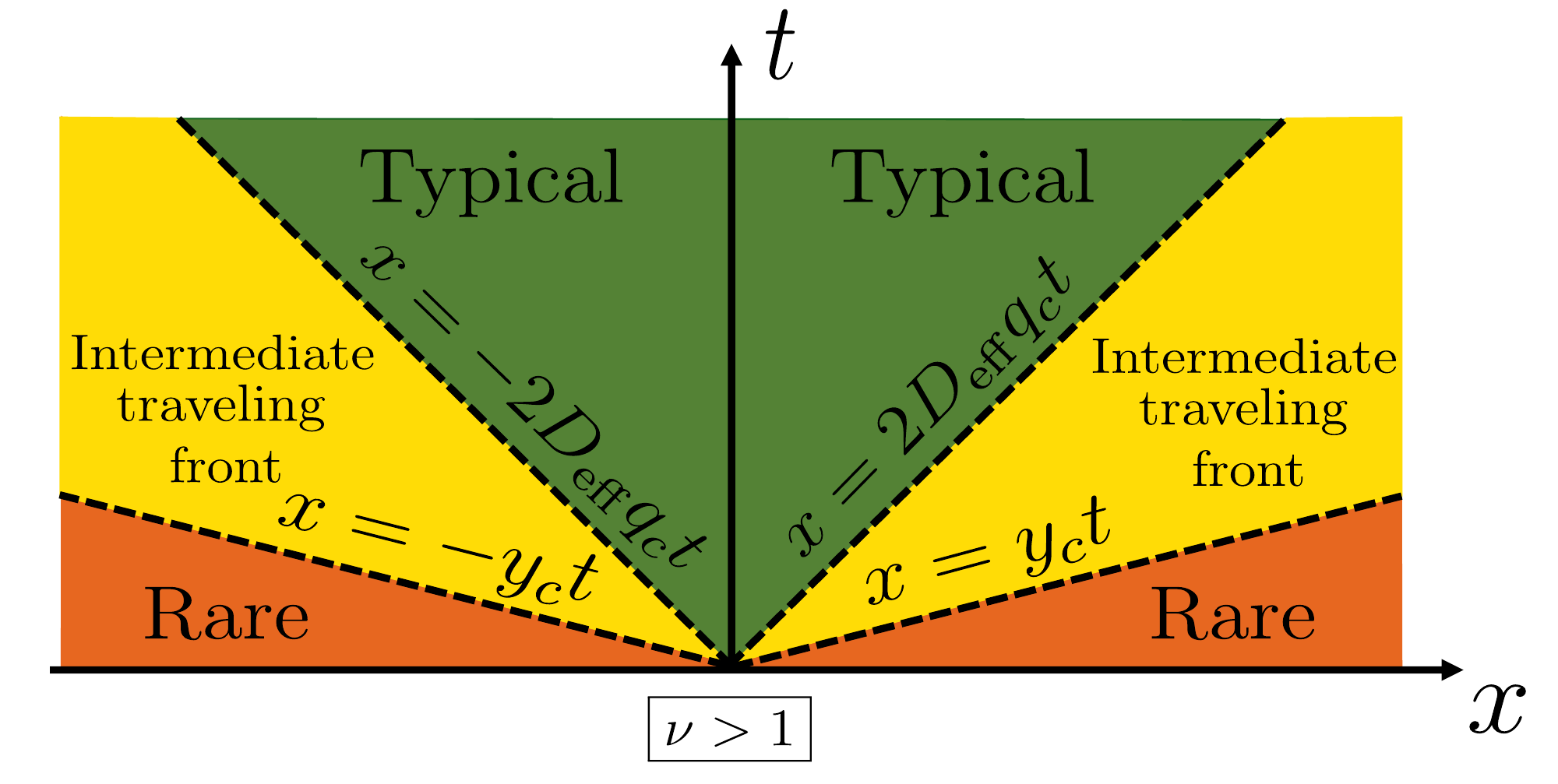}
\caption{
\textbf{Left:} For $0 < \nu < 1$, a light cone at $x = \pm y_c t$ separates two types of trajectories: typical ones, which switch frequently, and rare ones, which undergo almost no switches and spend most of the time in the $D_{\max}$ state. These rare trajectories dominate the large $x$ tail (see Eq.~(\ref{ratenu01})).
\textbf{Right:} For $\nu > 1$, a new exponential regime appears between the two existing for $0 < \nu < 1$. This exponential regime manifests as a traveling front (see Eq.~(\ref{nugeq1})). In both cases, the order of the transition depends on the specific value of $\nu$~\cite{SM}.}
  \label{transitions_diagrams} 
\end{figure}}

\newpage
\end{widetext}


\begin{thebibliography}{26}
\bibitem{footnote1}
Note that the odd moments of $x(t)$ vanish by symmetry $x \to -x$.
\bibitem{footnote2}
If the support of $W(D)$ is unbounded, then $D_{\max}$ is infinite.
\bibitem{footnote3}
Besides, the $O(1)$ correction to the cumulants can also be computed explicitly~\cite{SM}.
\end{thebibliography}
\end{document}

%\preprint{}
\title{Large Deviations in Switching Diffusion: from Free Cumulants to Dynamical Transitions: {\it Supplementary Material}}% Force line breaks with \\
\author{Mathis Gu\'eneau}
\affiliation{Sorbonne Universit\'e, Laboratoire de Physique Th\'eorique et Hautes Energies, CNRS UMR 7589, 4 Place Jussieu, 75252 Paris Cedex 05, France}
\author{Satya N. Majumdar}
\affiliation{LPTMS, CNRS, Univ.  Paris-Sud,  Universit\'e Paris-Saclay,  91405 Orsay,  France}
\author{Gr\'egory Schehr}
\affiliation{Sorbonne Universit\'e, Laboratoire de Physique Th\'eorique et Hautes Energies, CNRS UMR 7589, 4 Place Jussieu, 75252 Paris Cedex 05, France}

%\date{\today}

\begin{abstract}
We give the principal details of the calculations described in the main text of the Letter.
\end{abstract}

\maketitle

{
  \hypersetup{hidelinks}  
  \newpage
  \tableofcontents
}
\newpage

\section{Definition of the model}
{\color{black}
We consider a stochastic switching diffusion model whose dynamics is given by the following Langevin equation
\begin{equation}
    \dot{x}(t) = \sqrt{2 D(t)}\, \eta(t) \quad \, , \quad x(0)=x_0\quad \, ,
    \label{Langevin1}
\end{equation}
where $\eta(t)$ is a Gaussian white noise with zero mean $\langle \eta(t) \rangle = 0$ and unit variance $\langle \eta(t)\eta(t')\rangle = \delta(t-t')$. For a time $\tau_1$, the process $x(t)$ diffuses with a diffusion coefficient ${\sf D}_1$, where $({\sf D}_1, \tau_1)$ are drawn from a joint distribution $P_{\text{joint}}(D, \tau)$. After the duration $\tau_1$, a new pair $({\sf D}_2, \tau_2)$ is drawn independently from $P_{\text{joint}}$, and this selection process is repeated at each subsequent renewal time. In our case, we assume that the renewal times and diffusion coefficients are independently selected, such that $P_{\text{joint}}(D, \tau) = W(D)\, p(\tau)$, where $W(D)$ is an arbitrary distribution with finite moments, and the renewal times are exponentially distributed as $p(\tau) = r\, e^{-r \tau}$. As a result, all values ${\sf D}_i$'s and $\tau_i$'s are independent and identically distributed (i.i.d.) random variables. In the following section, we derive a renewal equation for the distribution of the position of the particle. For simplicity, we set the initial position to $x_0 = 0$.%, and we will eventually average over the initial diffusion coefficient ${\sf D}_1$.
\begin{figure}[t]
    \centering
\includegraphics[width=0.45\linewidth]{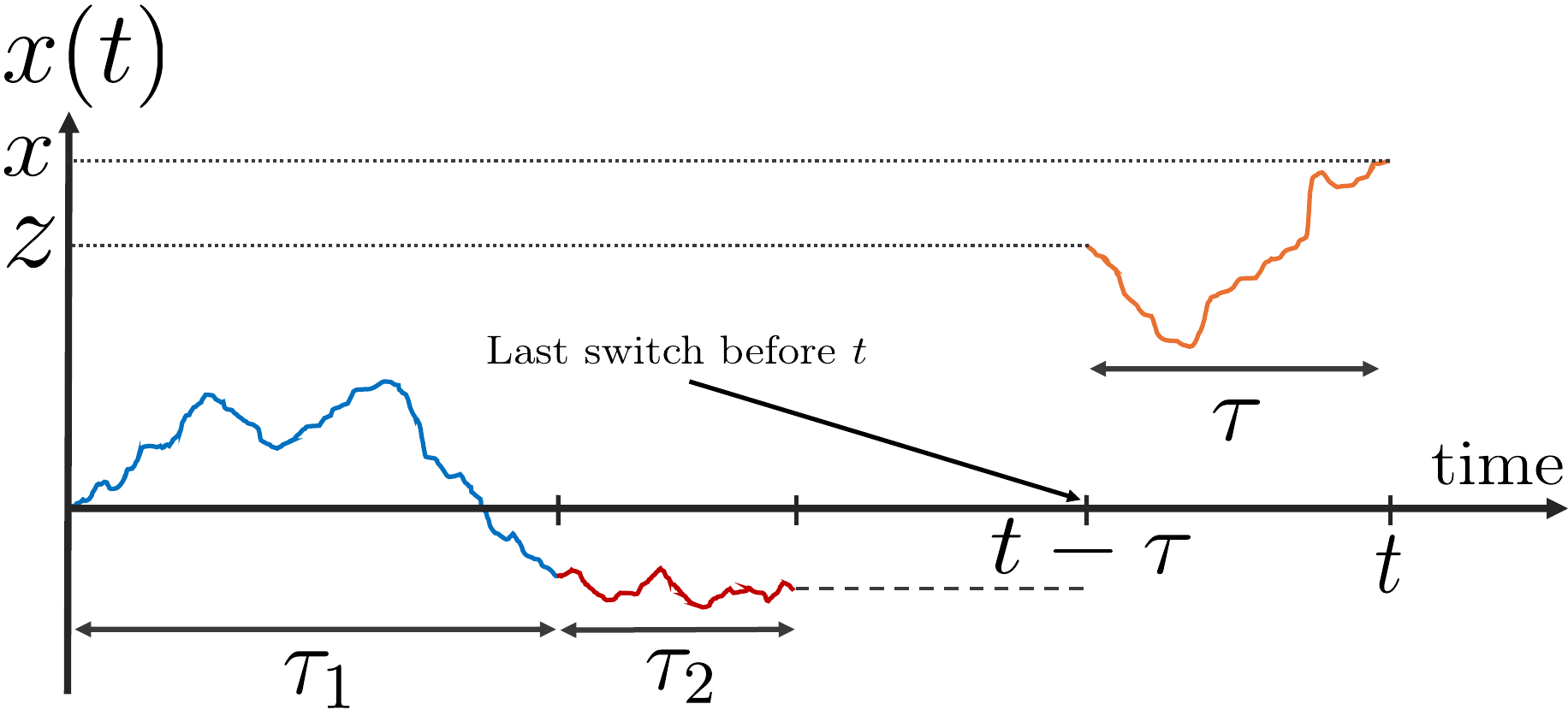}
\includegraphics[width=0.45\linewidth]{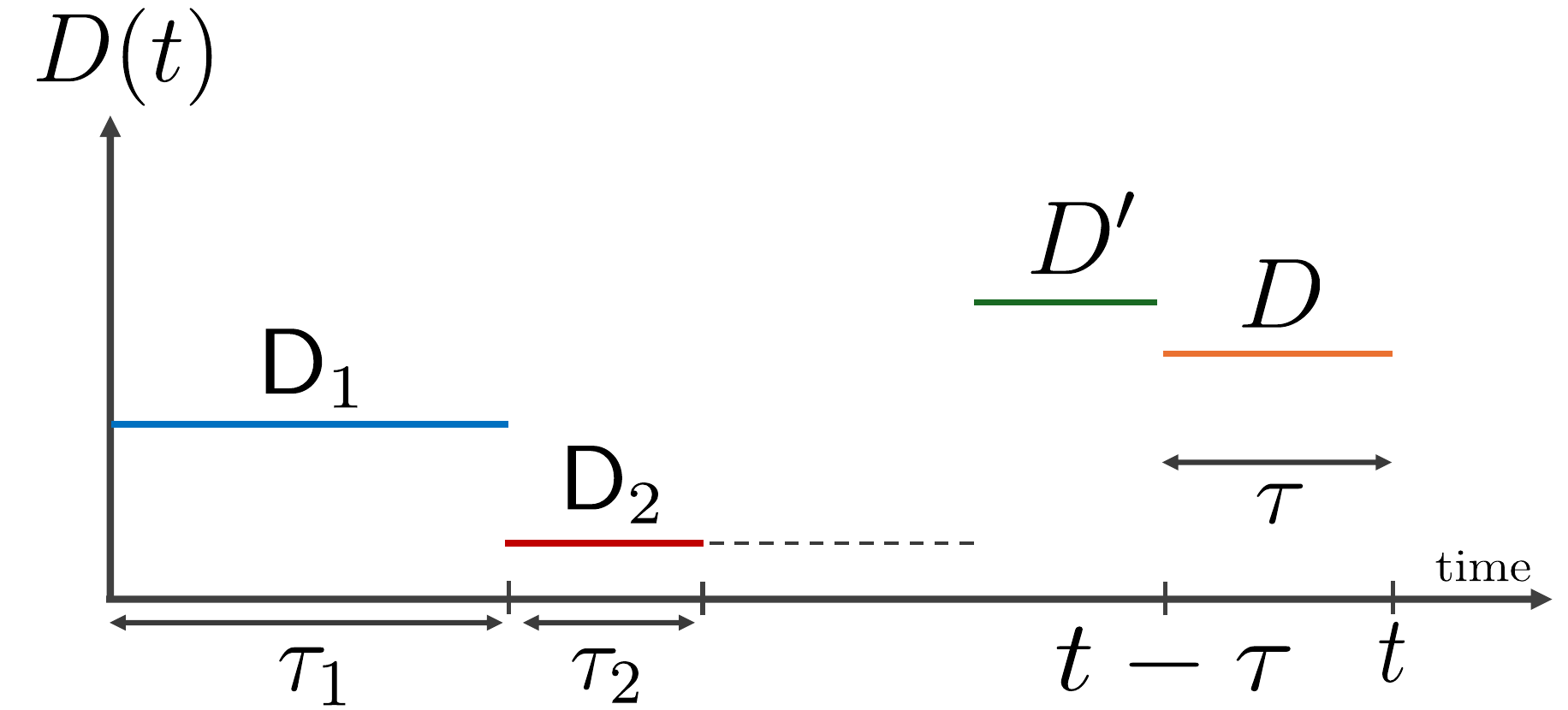}
  \caption{Schematic description of a typical time evolution of $x(t)$ and $D(t)$. In the model studied here, the $\tau_i$'s are independent exponential random variables, while the ${\sf D}_i$'s, which are also independent, are drawn from an arbitrary distribution $W(D)$. As considered in the renewal equation~(\ref{renewalpropa}), two situations are considered. The first one is when there is no switch, which happens with probability $e^{-rt}$, and we do not represent it on the figure. Here, we show the second situation where the last switch happens at time $t-\tau$. At this specific time, the particle is at position $z$,  and right before the switch, its diffusion coefficient is $D'$. At time $t$, the particle is at position $x$ with diffusion coefficient $D$.}
  \label{drawing_SM_renewal} 
\end{figure}
}

\section{Renewal equation and explicit solution}
We first derive a renewal equation for the joint distribution $p_r(x,D,t|{\sf D}_1)$ of the position $x(t)$ and the diffusing coefficient $D(t)$ at time $t$ conditioned on the first diffusion coefficient value (see Fig. \ref{drawing_SM_renewal} for an illustration)
\begin{eqnarray}
  p_r(x,D,t|{\sf D}_1) &&= e^{-rt}\, \frac{e^{-\frac{x^2}{4{\sf D}_1t}}}{\sqrt{4\pi {\sf D}_1 t}}\, \delta(D-{\sf D}_1)\,  + \nonumber \\
  &&\int_0^{t} d\tau\, r\, e^{-r \tau} \int_{-\infty}^{+\infty}dz\, \int_0^{+\infty}dD'\, p_r(z,D',t-\tau|{\sf D}_1) \, W(D)\, \frac{e^{-\frac{(x-z)^2}{4D \tau}}}{\sqrt{4\pi D \tau}} \, .
  \label{renewalpropa}
\end{eqnarray}
The first contribution comes from the event, that occurs with probability $e^{-r t}$, where there is no reset up to time $t$ and the dynamics follow a simple Brownian motion with diffusion coefficient ${\sf D}_1$. 
%This results in the standard Brownian propagator. Since there is no switch of the diffusion coefficient, ${\sf D}_1$ should match $D$, hence the dirac delta function. 
The second term accounts for the event where the last switch occurred at time $t - \tau$, at which point the particle was at position $z$ with diffusion coefficient $D'$. The probability that no switch occurred between $t - \tau$ and $t$ is $e^{-r \tau}$, while the probability of a switch occurring within the small time interval $[t - \tau, t - \tau + d\tau]$ is $r d\tau$. To account for all possible switch times, we integrate over $\tau$. Next, we integrate over $z$ and $D'$, taking into account the propagator $p_r(z, D', t - \tau)$ that describes the paths from the origin $x = 0$ at $t = 0$ to position $z$ at time $t - \tau$. We also include the Gaussian propagator that governs the motion from $z$ at time $t - \tau$ to $x$ at time $t$. Finally, we need to account for the transition probability of the diffusion coefficient changing from $D'$ (the value just before the switch at time $t - \tau$) to $D$ (the value immediately after the switch). Since the diffusion coefficients are i.i.d., this is simply given by the distribution $W(D)$.
If we integrate the renewal equation~(\ref{renewalpropa}) over all values of $D$ and perform the integral over $D'$ in (\ref{renewalpropa}), we obtain an integral equation for the propagator of $x(t)$ conditioned on the value ${\sf D}_1$. This is given by
\begin{eqnarray}
  p_r(x,t|{\sf D}_1) = e^{-rt}\, \frac{e^{-\frac{x^2}{4{\sf D}_1t}}}{\sqrt{4\pi {\sf D}_1 t}}\,  + \int_0^{t} d\tau\, r\, e^{-r \tau} \int_{-\infty}^{+\infty}dz\, p_r(z,t-\tau|{\sf D}_1) \, \int_0^{+\infty}dD\, W(D)\, \frac{e^{-\frac{(x-z)^2}{4D \tau}}}{\sqrt{4\pi D \tau}} \, .
  \label{renewalpropa2}
\end{eqnarray}
A more symmetrized expression can be obtained by averaging over the values of ${\sf D}_1$, resulting in
\begin{eqnarray}
  p_r(x,t) = e^{-rt}\, \int_0^{+\infty}dD\, W(D)\, \frac{e^{-\frac{x^2}{4Dt}}}{\sqrt{4\pi D t}}\,  + \int_0^{t} d\tau\, r\, e^{-r \tau} \int_{-\infty}^{+\infty}dz\, p_r(z,t-\tau) \, \int_0^{+\infty}dD\, W(D)\, \frac{e^{-\frac{(x-z)^2}{4D \tau}}}{\sqrt{4\pi D \tau}} \, .
  \label{renewalpropa3}
\end{eqnarray}
{We now assume in the remainder of this section that $W(D)$ has a finite support on $[0,D_{\text{max}}]$, while the case where $W(D)$ has support on $[0,+\infty)$ is discussed separately in section \ref{Winfinitesupport}. We first note that the second term exhibits a convolution structure in space. To leverage this property, instead of transitioning to Fourier space (which is used for $W(D)$ with infinite support instead -- see section \ref{Winfinitesupport}), we introduce the generating function of $x$ which is defined as the bilateral Laplace transform (BLT) of $p_r(x,t)$. This approach is better suited for the study of large deviations that will follow. It is given by
\begin{eqnarray}
    \hat{p}_r(q,t)= \langle e^{qx} \rangle =\int_{-\infty}^{+\infty}dx\, e^{q x}\, p_r(x,t) \;.
\end{eqnarray}
Note that the normalization of the PDF of $p_r(x,t)$ implies
\bea \label{normalisation}
\hat{p}_r(q=0,t) =\int_{-\infty}^{+\infty}dx\, p_r(x,t) = 1 \quad, \quad {\rm for \; all} \; t\;.
\eea
We also recall that the inversion formula is given by
\begin{eqnarray} \label{inverse_BLT}
    p_r(x,t) = \frac{1}{2i\pi}\, \int_{\gamma-i\infty}^{\gamma +i\infty}dq\, e^{-q x}\, \hat{p}_r(q,t) \;,
\end{eqnarray}
where the integral runs over the Bromwich contour which, in this case of a bilateral Laplace transform, lies within the region of convergence of $\hat p_r(q,t)$ in the complex $q$-plane. More precisely, the real $\gamma$ in (\ref{inverse_BLT}) is such that $\gamma \in ]s_0,s_1[$ where $]s_0,s_1[$ is the maximal real interval such that $\hat p_r(q,t)$ is an analytic function in the vertical strip delimited by $]s_0,s_1[$ in the complex $q$-plane. 
%where $\gamma$, which is real, is such that all the singularities of $\hat p_r(q,t)$ in the complex $q$-plane are to the left of the Bromwich contour of integration $(\gamma-i\infty, \gamma+i\infty)$.}
Taking the BLT of Eq.~(\ref{renewalpropa3}) yields
\begin{eqnarray}
  \hat{p}_r(q,t) = e^{-rt}\, \int_0^{D_{\text{max}}}dD\, W(D)\, e^{D q^2 t}  + \int_0^{D_{\text{max}}}dD\, W(D)\, \int_0^{t} d\tau\, r\, e^{-r \tau}\, e^{D q^2 \tau}\, \hat{p}_r(q,t-\tau)\, .\label{fourierRenewal}
\end{eqnarray}
Similarly, the convolution structure in time can be exploited to obtain a closed equation via the Laplace transformation with respect to the time variable $t$. It is thus useful to introduce the Laplace transform defined as
\begin{eqnarray}
  \tilde{p}_r(q,s) = \int_{0}^{+\infty}dt\, e^{-s t}\, \hat{p_r}(q,t)\, .
\end{eqnarray}
Taking the Laplace transform of Eq.~(\ref{fourierRenewal}) yields
\begin{eqnarray}
  &&\tilde{p}_r(q,s) = \int_0^{D_{\text{max}}}dD\, \frac{W(D)}{r + s - Dq^2}  + r\, \int_0^{D_{\text{max}}}dD\, \frac{W(D)}{r + s - Dq^2}\, \tilde{p}_r(q,s) \, .
\end{eqnarray}
Ultimately, the explicit solution takes the form
\begin{equation}
  \tilde{p}_r(q,s) = \frac{J_r(q,s)}{1-r\, J_r(q,s)} \;,\; J_r(q,s) = \int_0^{D_{\text{max}}}\hspace*{-0.3cm}dD\, \frac{W(D)}{r+s-D q^2}. \label{explicitLaplaceaveraged}
\end{equation}
When $q=0$, one has $J_r(0,s) = 1/(r+s)$ and we easily check from (\ref{explicitLaplaceaveraged}) that $\tilde{p}_r(0,s)= \frac{1}{s}$, which is consistent with the normalization condition (\ref{normalisation}). 

\section{Explicit formula for the moments}
In this section we derive an explicit formula for the moments $\langle x^{2n}(t) \rangle$ of the distribution $p_r(x,t)$ at any finite time~$t$. Since the distribution $p_r(x,t)$ is symmetric, i.e., $p_r(x,t) = p_r(-x,t)$, all odd moments vanish. Here, $\langle \ldots \rangle$ denotes an average over all sources of randomness present in the system, which are treated here on the same footing. To begin, we derive a recursive relation for the moments of the probability distribution function (PDF) $p_r(x,t)$. This relation can be obtained by expanding $\hat{p}_r(q,t)$ within Eq.~(\ref{fourierRenewal}), using the series expansion 
\begin{equation}
    \hat{p}_r(q,t)=\sum_{n=0}^{+\infty} \frac{q^{2n}}{(2n)!}\langle x^{2n}(t) \rangle\, .
\end{equation}
First, we substitute this power series expansion in Eq.~(\ref{fourierRenewal})
\begin{eqnarray}
    &&\sum_{n=0}^{+\infty} \frac{q^{2n}}{(2n)!}\langle x^{2n}(t) \rangle = e^{-rt}\, \sum_{n=0}^{+\infty}\, \langle D^n \rangle\, \frac{q^{2n} t^n}{n!}  + \int_0^{t} d\tau\, r\, e^{-r \tau}\, \sum_{p=0}^{+\infty}\, \langle D^p \rangle\, \frac{q^{2p} \tau^p}{p!}\, \sum_{l=0}^{+\infty} \frac{q^{2l}}{(2l)!}\langle x^{2l}(t-\tau) \rangle\, ,
\end{eqnarray}
and we select the term of order $q^{2n}$ on both sides of to get
\begin{eqnarray}
   \langle x^{2n}(t) \rangle= e^{-rt}\,  \langle D^n \rangle\, t^n\, \frac{(2n)!}{n!}  +\int_0^{t} d\tau\, r\, e^{-r \tau}\, \sum^{+\infty}_{\substack{p,l=0\\p+l=n}}\, \langle D^p \rangle\, \tau^p \frac{(2n)!}{p!(2l)!}\, \langle x^{2l}(t-\tau) \rangle\, .\quad
\end{eqnarray}
In order to write a recursion relation for $\langle x^{2n}(t) \rangle$, we extract the term corresponding to $p=0$ and $l=n$ in the sum on the right-hand side. This gives
\begin{eqnarray}
   &&\langle x^{2n}(t) \rangle - \int_0^{t} d\tau\, r\, e^{-r \tau}\, \langle x^{2n}(t-\tau) \rangle \nonumber \\
   &&= e^{-rt}\,  \langle D^n \rangle\, t^n\, \frac{(2n)!}{n!}  + \int_0^{t} d\tau\, r\, e^{-r \tau}\, \sum_{\substack{p=1, l=0\\p+l=n}}\, \langle D^p \rangle\,  \tau^p \frac{(2n)!}{p!(2l)!}\, \langle x^{2l}(t-\tau) \rangle\, .
\end{eqnarray}
Here, to treat the integrals over $\tau$ on both sides, it is convenient to use the convolution structure when taking the Laplace transform with respect to the time variable $t$. The equation simplifies to
\begin{eqnarray}
 \frac{s}{r+s} \mathcal{L}_{t\to s}\left[ \langle x^{2n}(t) \rangle\right] = \langle D^n \rangle \, \frac{(2n)!}{(r+s)^{1+n}}  + \sum_{\substack{p=1,l=0\\p+l=n}}\,  \frac{(2n)!}{p!(2l)!}  \langle D^p \rangle\, \mathcal{L}_{t\to s}\left[ r e^{-r t} t^p\right] \mathcal{L}_{t\to s}\left[ \langle x^{2l}(t) \rangle\right] \, ,
\end{eqnarray}
where $\mathcal{L}_{t\to s}\left[ f(t)\right]$ denotes the Laplace transform of $f(t)$. When performing the first Laplace transform on the right hand-side, the expression simplifies to
\begin{eqnarray}
 s\, \mathcal{L}_{t\to s}\left[ \langle x^{2n}(t) \rangle\right] = \langle D^n \rangle \, \frac{(2n)!}{(r+s)^{n}}  +  \sum_{\substack{p=1, l=0\\p+l=n}}\,  \frac{(2n)!}{(2l)!} \langle D^p \rangle \frac{r}{(r+s)^p} \mathcal{L}_{t\to s}\left[ \langle x^{2l}(t) \rangle\right] \, .
\end{eqnarray}
By re-writing the constraint $(p=1,l=0,p+l=n)$ to $l\in[0,n-1]$ with $p=n-l$, the expression now reads
\begin{eqnarray}
 \mathcal{L}_{t\to s}\left[ \langle x^{2n}(t) \rangle\right] =  \langle D^n \rangle \, \frac{(2n)!}{s(r+s)^{n}}  +  \sum_{l=0}^{n-1}\,  \frac{(2n)!}{(2l)!} \langle D^{n-l} \rangle \frac{r}{s(r+s)^{n-l}} \mathcal{L}_{t\to s}\left[ \langle x^{2l}(t) \rangle\right] \, .
\end{eqnarray}
Finally, we re-organize the two sides of the equation such that
\begin{eqnarray}
 \frac{(r+s)^n}{(2n)!}\mathcal{L}_{t\to s}\left[ \langle x^{2n}(t) \rangle\right] =\frac{1}{s}\left[ \langle D^n \rangle + \sum_{l=0}^{n-1}\,  r\,  \langle D^{n-l} \rangle \left(\frac{(r+s)^l}{(2l)!} \mathcal{L}_{t\to s}\left[ \langle x^{2l}(t) \rangle\right]\right)\right] \, ,
\end{eqnarray}
where we now identify a recursive sequence $u_n$ defined as
\begin{eqnarray}
 &&u_n = \frac{\langle D^n \rangle}{s} + \sum_{l=0}^{n-1}\frac{r}{s} \langle D^{n-l} \rangle\, u_l\, .\\
 &&u_n = \frac{(r+s)^n}{(2n)!}\mathcal{L}_{t\to s}\left[ \langle x^{2n}(t) \rangle\right]\, , \quad u_0 = \frac{1}{s}\, .\label{undef}
\end{eqnarray}
Our goal is to find the explicit expression of $u_n$ and then deduce from it $\langle x^{2n}(t) \rangle$ from (\ref{undef}). For this purpose, we can write the first terms of the sequence and try to see if a pattern emerges,
\begin{eqnarray}
u_1 &=& \frac{(r+s)\langle D\rangle}{s^2}, \\
u_2 &=& \frac{(r+s)(r\langle D\rangle^2 + s \langle D^2\rangle)}{s^3}, \\
u_3 &=& \frac{(r+s)(r^2\langle D\rangle^3 + 2\, rs \langle D\rangle \langle D^2\rangle+ s^2\langle D^3\rangle)}{s^4}, \\
u_4 &=& \frac{(r+s)(r^3\langle D\rangle^4 + 3\, r^2s\langle D\rangle^2\langle D^2\rangle + 2\, rs^2\langle D\rangle \langle D^3\rangle + rs^2\langle D^2\rangle^2 + s^3 \langle D^4\rangle)}{s^5}.
\end{eqnarray}
After inspection, for $n\geq1$, we realize that $u_n$ can be written as (see also \cite{OEIS})
\begin{eqnarray}
    u_n=\frac{r+s}{r s^{n+1}}\, [t^n]\left[ \frac{1}{1 - \frac{r}{s} \sum_{i=1}^{\infty} s^i \langle D^i \rangle t^i}\right]\, ,
    \label{un_first}
\end{eqnarray}
where $[t^n][f(t)]$ denotes the coefficient at order $t^n$ of the series expansion of $f(t)$ with respect to $t$. As we will see, we can find a more explicit expression for $u_n$ by first expanding the fraction in squared-brackets. This leads to
\begin{eqnarray}
    u_n=\frac{r+s}{r s^{n+1}}[t^n]\left[\sum_{m=0}^{+\infty} \left(\frac{r}{s}\right)^m\left[\sum_{i=1}^{\infty} s^i \langle D^i \rangle t^i\right]^m\right]\, .
\label{un_expression0}
\end{eqnarray}
It turns out that this can be re-written in terms of partial exponential Bell polynomials $B_{n,k}$ \cite{ComtetL, wikibell}. {\color{black} These polynomials are defined as 
\begin{eqnarray}
B_{n,k}(x_1, x_2, \ldots, x_{n-k+1}) = \sum_{\Vec{j}} \frac{n!}{j_1! j_2! \cdots j_{n-k+1}!} 
\left( \frac{x_1}{1!} \right)^{j_1} 
\left( \frac{x_2}{2!} \right)^{j_2} 
\cdots 
\left( \frac{x_{n-k+1}}{(n-k+1)!} \right)^{j_{n-k+1}}\, ,
\end{eqnarray}
where the summation $\sum_{\Vec{j}}$ over $\Vec{j} = (j_1, j_2, \ldots, j_{n-k+1})$ denotes a sum over all non-negative integers $j_i$ subject to the following constraints
\begin{eqnarray}
j_1 + j_2 + \cdots + j_{n-k+1} &=& k \, , \\
j_1 + 2j_2 + 3j_3 + \cdots + (n-k+1)j_{n-k+1} &=& n \, .
\end{eqnarray}
We have in particular the following identities 
}\begin{eqnarray}
   \left[\sum_{i=1}^{+\infty} \langle D^i \rangle z^i\right]^m &=& m! \sum_{p=m}^{+\infty}  B_{p,m}(a_1,\ldots,a_{p-m+1})\frac{z^p}{p!}\quad \, , \quad a_i = i!  \langle D^i \rangle\, ,\\
   &=& \sum_{p=m}^{+\infty} \hat{B}_{p,m}\left(\langle D \rangle, \ldots,\langle D^{p-m+1} \rangle \right) z^p\, ,\label{bellexpansion}
\end{eqnarray}
where $\hat{B}_{p,m}$ is the ordinary Bell polynomial defined as \cite{wikibell}
\begin{eqnarray}
   \hat{B}_{p,m}\left(\langle D \rangle, \ldots,\langle D^{p-m+1} \rangle \right) = \frac{m!}{p!} \, B_{p,m}\left(\langle D \rangle,2!\langle D^2 \rangle, \ldots,(p-m+1)!\langle D^{p-m+1} \rangle \right)\, .
\end{eqnarray}
Therefore, we can use the identity (\ref{bellexpansion}) with $z=st$ inside the expression for $u_n$ in Eq. (\ref{un_expression0}), and it yields
\begin{eqnarray}
    u_n=\frac{r+s}{r s^{n+1}}[t^n]\left[\sum_{m=0}^{+\infty} \left(\frac{r}{s}\right)^m  \sum_{p=m}^{+\infty} s^p \hat{B}_{p,m}\left(\langle D \rangle, \ldots,\langle D^{p-m+1} \rangle \right) t^p\right]\, .
\end{eqnarray}
To select the term of order $t^n$ we invert the order of the sums such that it gives 
\begin{eqnarray}
    u_n=\frac{r+s}{r s^{n+1}}[t^n]\left[\sum_{p=0}^{+\infty}\sum_{m=0}^{p} \left(\frac{r}{s}\right)^m   s^p \hat{B}_{p,m}\left(\langle D \rangle, \ldots,\langle D^{p-m+1} \rangle \right) t^p\right]\, .
\end{eqnarray}
For $n\geq 1$, it is now possible to write the explicit solution of the sequence
\begin{eqnarray}
    u_n = \frac{r+s}{r\, s}\, \sum_{m=1}^n\, \left(\frac{r}{s}\right)^m \, \hat{B}_{n,m}\left(\langle D \rangle, \ldots ,\langle D^{n-m+1} \rangle \right)\, .
\end{eqnarray}
Using the definition (\ref{undef}) of $u_n$, we can deduce the Laplace transform of the moments of the distribution $p_r(x,t)$, and they read
\begin{eqnarray}
    &&\mathcal{L}_{t\to s}\left[ \langle x^{2n}(t) \rangle\right] = \frac{(2n)!}{r\, s\, (r+s)^{n-1}}\, \sum_{m=1}^n\, \left(\frac{r}{s}\right)^m \, \hat{B}_{n,m}\left(\langle D \rangle, \ldots ,\langle D^{n-m+1} \rangle \right)\, .\label{Laplacemoments}
\end{eqnarray}
Note that we have 
\begin{eqnarray}\label{laplaceformula}
    \mathcal{L}^{-1}_{s\to t}\left[ s^{-(m+1)}(r+s)^{-(n-1)}\right] = t^{m+n-1}\, \frac{M(n-1,m+n,-rt)}{(m+n-1)!}\, ,
\end{eqnarray}
where $M(a,b,x)$ denotes the Kummer's function.
Taking the inverse Laplace transform of Eq.~(\ref{Laplacemoments}) using Eq.~(\ref{laplaceformula}) then leads to the final result
\begin{eqnarray}
    && \langle x^{2n}(t) \rangle = \frac{(2n)!}{r^n} \, \sum_{m=1}^n\, \frac{(r t)^{m+n-1}}{(m+n-1)!}\, M(n-1, m + n, -r t)\, \hat{B}_{n,m}\left(\langle D \rangle, \ldots ,\langle D^{n-m+1} \rangle \right)\, .
    \label{explicitmoments}
\end{eqnarray}
From this formula, and using Eq.~(22) of the End matters, it is possible to compute the cumulants for specific distributions $W(D)$ -- see Fig. \ref{Fig_cumul} for a numerical check in the case of the two-state model (left panel), and for the Wigner semi-circle distribution (right panel).

We present here the first three non-zero moments of the position of the switching diffusion process. These moments are computed directly from the exact expression provided in Eq.~(\ref{explicitmoments}). They are as follows:
\begin{eqnarray}
\langle x^2(t) \rangle &=& 2 \langle D \rangle t, \\
\langle x^4(t) \rangle &=& \frac{e^{-rt}}{r^2}\left[12\left(-2 + e^{rt}\left(2 + rt\left(-2 + rt\right)\right)\right)\langle D \rangle^2 + 24\left(1 + e^{rt}\left(-1 + rt\right)\right)\langle D^2 \rangle\right], \\
\langle x^6(t) \rangle &=& \frac{120\, e^{-rt}}{r^3}\left\{\left[6\left(4 + rt\right) + e^{rt}\left(-24 + rt\left(18 + rt\left(-6 + rt\right)\right)\right)\right]\langle D \rangle^3 \right.  \nonumber\\
&& \left.+ 6\left[-2\left(3 + rt\right) + e^{rt}\left(6 + rt\left(-4 + rt\right)\right)\right]\langle D \rangle \langle D^2 \rangle + 6\left(2 + rt + e^{rt}\left(-2 + rt\right)\right)\langle D^3 \rangle \right\} \, .
\end{eqnarray}
These expressions are valid for any distribution $W(D)$ for which the first three moments are well defined.

\vspace*{0.5cm}

\noindent\textbf{Remark.} It is possible to perform a consistency check and obtain back the expression of $\tilde{p}_r(q,s)$ given in  Eq.~(\ref{explicitLaplaceaveraged}) when summing the Laplace transform of the moments as
\begin{eqnarray}
    \tilde{p}_r(q,s)=\sum_{n=0}^{+\infty} \frac{q^{2 n}}{(2 n)!} \mathcal{L}_{t\to s}\left[ \langle x^{2n}(t) \rangle\right] = \sum_{n=0}^{+\infty} \frac{q^{2 n}}{(r+s)^n}\, u_n\, ,
    \label{laplace_un}
\end{eqnarray}
where $u_n$ is defined in Eq.~(\ref{undef}). To proceed, we use the equivalence 
\begin{eqnarray}
a_n = [t^n]\left( \frac{1}{1 - \frac{r}{s} \sum_{i=1}^{\infty} s^i \langle D^i \rangle t^i}\right) \Longleftrightarrow \frac{1}{1- \frac{r}{s}\sum_{i=1}^{+\infty}s^i \langle D^i \rangle t^i}= \sum_{n=1}^{+\infty}a_n t^n +1\, .
\end{eqnarray}
{\color{black}
Using (\ref{un_first}), we have that
\begin{eqnarray}
    u_n = \frac{r+s}{r s} \frac{a_n}{s^n}\, .
\end{eqnarray}
Hence, the generating function of $u_n$'s is given by
\begin{eqnarray}
    \sum_{n=0}^{\infty} u_n \, z^n= u_0 + \sum_{n=1}^{\infty} a_n\, \left(\frac{z}{s}\right)^n =u_0 + \frac{r+s}{rs}\left(\frac{1}{1-\frac{r}{s}\sum_{i=1}^{+\infty}\langle D^i \rangle z^i } -1\right) \;.
\end{eqnarray}}
Using this relation with $z=q^2/(r+s)$ in Eq.~(\ref{laplace_un}), and using also $u_0 = 1/s$, we find
\begin{eqnarray}
    \tilde{p}_r(q,s)=\frac{r+s}{r s}\, \frac{1}{1-\frac{r}{s}\sum_{i=1}^{+\infty}\left[\frac{q^2}{r+s}\right]^i \, \langle D^i \rangle}-\frac{1}{r}\, ,
\end{eqnarray}
that we can re-write as
\begin{eqnarray}
    \tilde{p}_r(q,s)=\frac{\sum_{i=0}^{+\infty} \left[\frac{q^2}{r+s}\right]^i\,  \langle D^i \rangle}{r +s - r\sum_{i=0}^{+\infty} \left[\frac{q^2}{r+s}\right]^i\,  \langle D^i \rangle} = \frac{J_r(q,s)}{1-r\, J_r(q,s)}\, ,
\end{eqnarray}
{where $J_r(q,s)$ is defined in Eq.~(\ref{explicitLaplaceaveraged}), and the last equality comes from the small $q$ expansion of $J_r(q,s)$. We have indeed
\begin{equation}
   J_r(q,s) = \int_0^{D_{\text{max}}}\hspace*{-0.3cm}dD\, \frac{W(D)}{r+s-D q^2} =\frac{1}{r+s}\int_0^{D_{\text{max}}}\hspace*{-0.3cm}dD\, \frac{W(D)}{1-\frac{D q^2}{r+s}} = \frac{1}{r+s}\, \sum_{i=0}^{+\infty}\left(\frac{ q^2}{r+s}\right)^i\langle D^i\rangle . 
\end{equation}}

\begin{figure}[t]
   \centering
\includegraphics[width=0.4\linewidth]{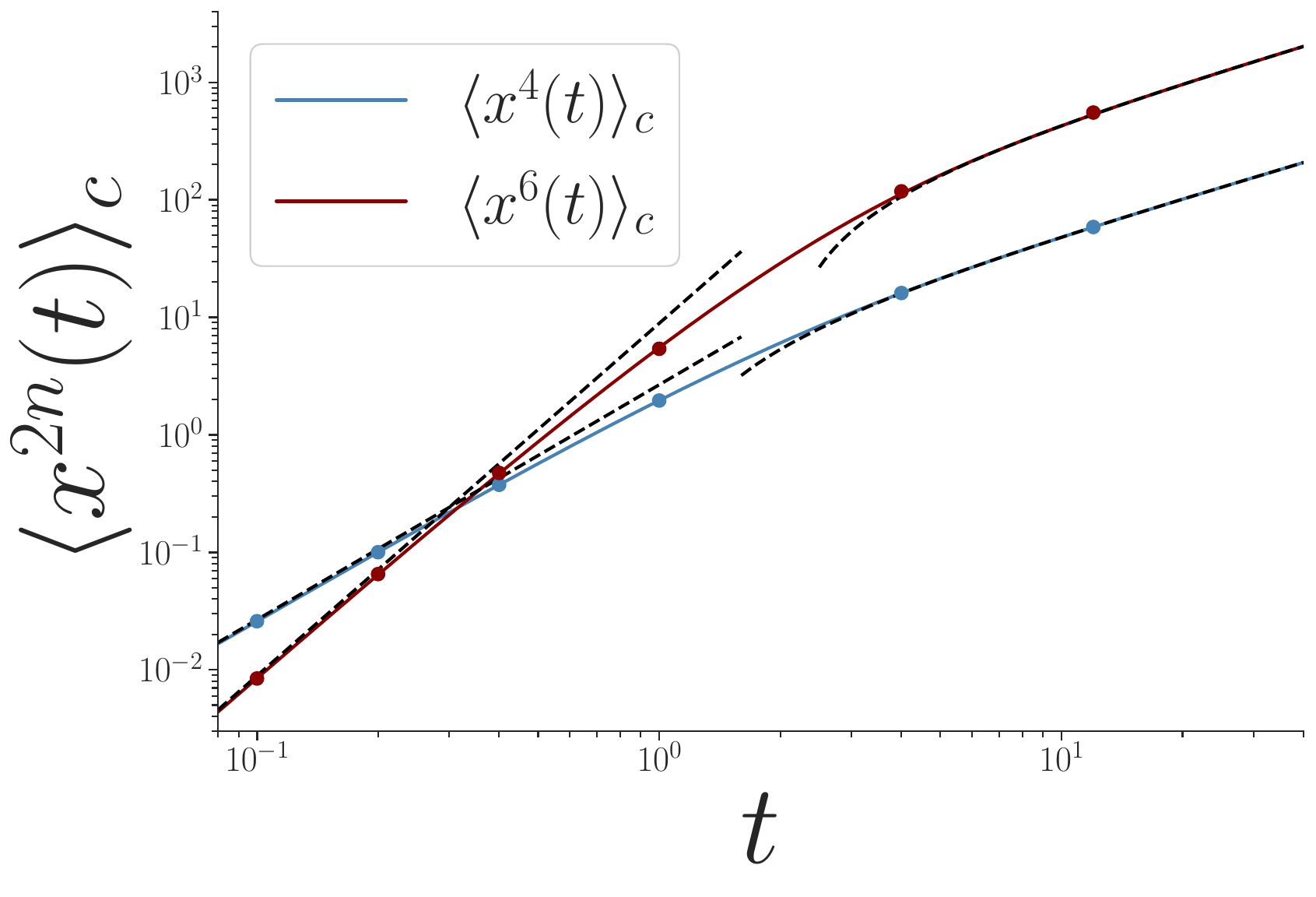}
\includegraphics[width=0.4\linewidth]{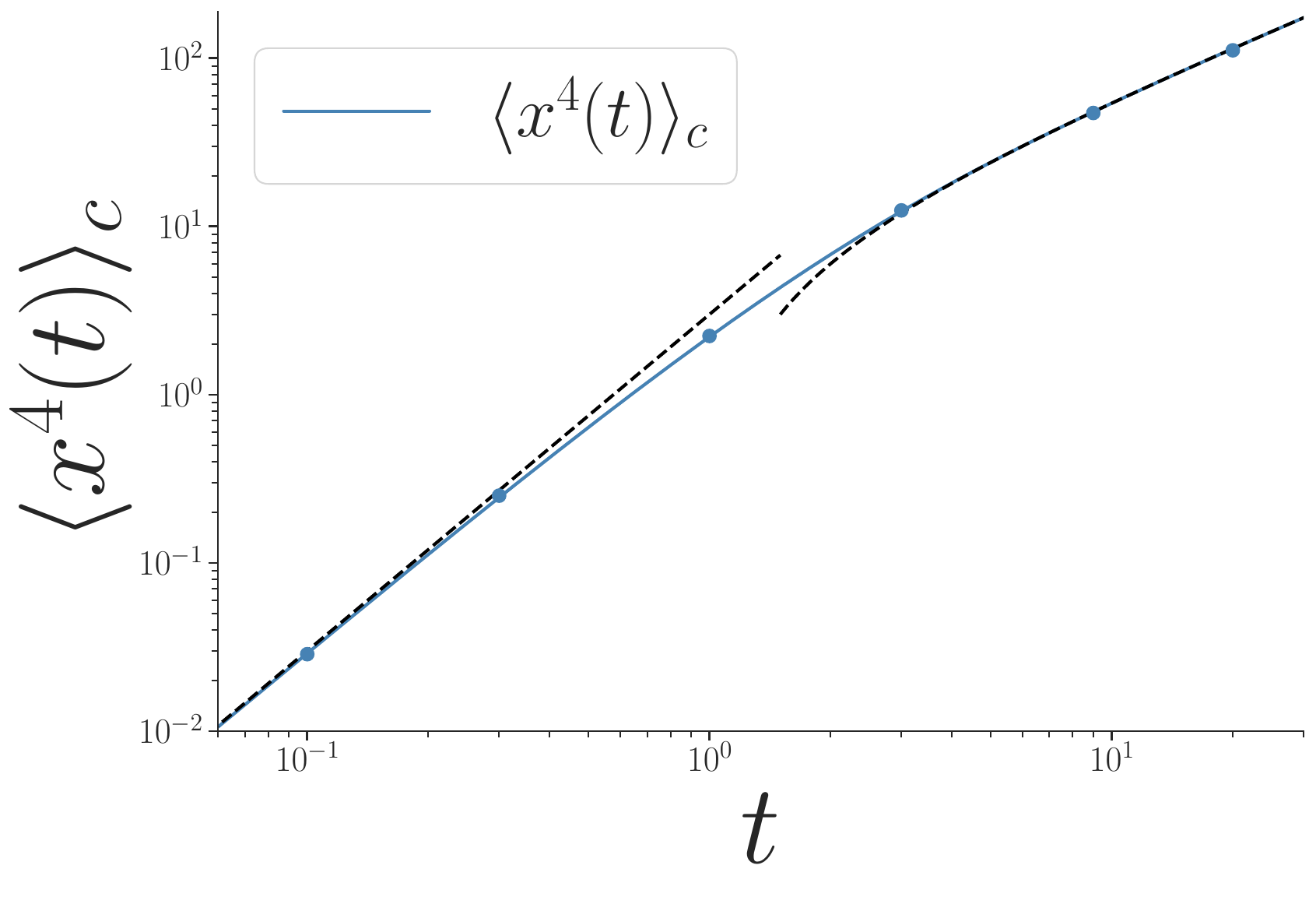}
  \caption{Using the explicit formulas for the moments given in Eq~(\ref{explicitmoments}), we calculate and plot the cumulants (solid lines) and compare them with numerical simulations (dots). The left panel shows results for $W(D)$ for the two-state model -- i.e., $W(D) = p\delta(D-D_1)+(1-p)\delta(D-D_2)$, while the right panel corresponds to the Wigner semi-circle law -- i.e., $ W(D) = 8/(\pi D_{\max}^2) \sqrt{D(D_{\max} - D)}$. The black dotted lines represent the asymptotic predictions given in Eq~(2) of the Letter. Our theoretical predictions show excellent agreement with the simulations for both models. The parameters for the two-state model are: $r = 1$, $p = 1/3$, $D_2 = 1$, $D_1 = 2$. For the Wigner semi-circle law, the parameters are: $r = 1$, $D_{\text{max}}=1$.}
 \label{Fig_cumul} 
\end{figure}

\section{The two-state model $W(D) =  p\, \delta(D-D_1) + (1-p)\delta(D-D_2)$}

In this section, we provide the details of the study of the scaled cumulant generating function (SCGF) $\Psi(q)$ and the rate function $I(y)$ for the two state model corresponding to $W(D) =  p\delta(D-D_1) + (1-p)\delta(D-D_2)$. In this case the function $J_r(q,s)$ in Eq. (\ref{explicitLaplaceaveraged}) reads
\bea \label{J_two_state}
J_r(q,s) = \frac{p}{r+s-D_1 q^2} + \frac{1-p}{r+s-D_2 q^2} \;,
\eea
from which it follows that $\tilde p_r(q,s)$ is given by
\bea \label{ptilde_twost}
\tilde p_r(q,s) = \frac{r+s_+ - q^2(p D_2 + (1-p)D_1)}{s_+-s_-} \frac{1}{s-s_+} + \frac{q^2(pD_2+(1-p)D_1)-r-s_-}{s_+-s_-} \frac{1}{s-s_-} \;.
\eea
where 
 \bea \label{def_spm}
s_{\pm}\equiv s_{\pm}(q) =\frac{1}{2} \left((D_1 + D_2) q^2 - r \pm \Delta(q)\right) \quad \, , \quad \Delta(q) = \sqrt{\left((D_2 - D_1) q^2 + r\right)^2 +4 (D_1 - D_2)p r q^2} \;.
\eea
It is easy to perform the inverse Laplace transform to obtain $\hat p_r(q,t)$, since $\tilde p_r(q,s)$ has two simple pole at $s = s_\pm$ in the complex $s$-plane. This yields
\bea 
\hat p_r(q,t) &=& \int_\Gamma \frac{ds}{2 i \pi} \tilde p_r(q,s)\, e^{st} \label{fouriertransform_twostates} \\
&=& \frac{r+s_+ - q^2(p D_2 + (1-p)D_1)}{s_+-s_-} e^{s_+ \, t} + \frac{q^2(pD_2+(1-p)D_1)-r-s_-}{s_+-s_-} e^{s_- \, t} \nonumber \\
&=& \frac{e^{\frac{1}{2}((D_1+D_2) q^2-r + \Delta(q))t}}{2 \Delta(q)} \left[(D_1-D_2)(2p-1)\, q^2+r+\Delta(q) + e^{-\Delta(q)t}\left(\Delta(q) - (D_1-D_2)(2p-1)q^2-r\right)\right]\;. \nonumber
\eea
At large time, the leading behavior is given by
\bea \label{def_B}
\hat p_r(q,t) = B(q) e^{t s_+(q)} + O(e^{-r\,t}) \quad, \quad B(q) = \left(\frac{(D_1-D_2)(2p-1)q^2 + r + \Delta(q)}{2 \Delta(q)}\right)
\eea
where we have used that $\Delta(q) \geq r$ for all $q$ such that the remainder term in (\ref{fouriertransform_twostates}) is indeed of order $O(e^{-rt})$.

\subsection{The cumulants and the scaled cumulant generating function $\Psi(q)$}

From this exact expression (\ref{fouriertransform_twostates}) we can extract all the information about the cumulants. Indeed, the cumulant generating function is given, to leading order for large $t$ by, 
\bea \label{chi}
\chi_r(q,t) = \ln \hat p_r(q,t) = t\, s_+(q) + \ln \left(\frac{(D_1-D_2)(2p-1)q^2 + r + \Delta(q)}{2 \Delta(q)}\right) + O(e^{-r t}) \;,
\eea
where we have used that $\Delta(q) \geq r$ for all $q$. From this result (\ref{chi}) one can then extract the leading behaviors of the cumulants. We recall indeed that the cumulants $\langle x^{2n}(t)\rangle_c$ are defined as
\bea
\chi_r(q,t) = \ln \hat p_r(q,t) = \sum_{n=1}^\infty \frac{q^{2n}}{(2n)!} \langle x^{2n}(t)\rangle_c \;. 
\eea
Therefore, in principle, the cumulants $\langle x^{2n}(t)\rangle_c$ can be obtained by expanding the expression (\ref{chi}) in powers of $q$. However, given the rather complicated expression of $\Delta(q)$ in (\ref{def_spm}), performing the small $q$ expansion to arbitrary order is a challenging task. Alternatively, the cumulants, to leading order for large $t$ can be computed from the connection to the free cumulants associated to the distribution $W(D) = p \delta(D-D_1)+ (1-p)\delta(D-D_2)$ -- see the second line of Eq. (2) in the main text -- ausing the formula (18), again given in the main text.

\vspace*{0.5cm}
\noindent {\it Cumulants in the case $D_1=D$, $D_2=0$.} We first start by analysing the case $D_1 = D$ and $D_2 = 0$ which is sometimes called randomly flashing diffusion in the literature \cite{Luczka2, BarkaiBurov}. In this case one has
\bea \label{cumul_expl_2st}
\langle x^{2n}(t)\rangle_c &=& \frac{t}{r^{n-1}} (2n)! \, \kappa_{n}(D) + O(1) \;, \nonumber \\ 
\kappa_{n}(D) &=& \langle D^n \rangle + \sum_{j=2}^{n} \frac{(-1)^{j-1}}{j}\binom{n+j-2}{j-1}\sum_{q_1+q_2+\ldots +q_j=n, q_k \geq 1} \langle D^{q_1} \rangle\ldots \langle D^{q_j} \rangle \;.
\eea
The multiple sum over the $q_j$'s in the second term can also be written as
\begin{eqnarray}
    a_{j,n} &&=\sum_{q_1\geq 1}\ldots\sum_{q_j\geq 1}\langle D^{q_1} \rangle\ldots \langle D^{q_j} \rangle \delta_{q_1+\ldots q_j,n} \quad, \quad n \geq j \\
    &&=p^j D^n \sum_{q_1\geq 1}\ldots\sum_{q_j\geq 1}\delta_{q_1+\ldots q_j,n} \quad, \quad n \geq j \;.
\end{eqnarray}
Let us compute the generating function of the sequence $a_{j,n}$, with $n\geq j $
\begin{eqnarray}
    \sum_{n=j}^{+\infty} z^n a_{j,n} &&=p^j  \sum_{q_1\geq 1}\ldots\sum_{q_j\geq 1}(zD)^n\delta_{q_1+\ldots q_j,n}\\
    &&=p^j \left(\sum_{q=1}^{+\infty}(zD)^q\right)^j =p^j \frac{(zD)^j}{(1-zD)^j}= p^j \sum_{k=0}^{+\infty} \frac{\Gamma(j+k)}{k!(j-1)!}(zD)^{k+j}\\
    &&= p^j \sum_{k'=j}^{+\infty} \frac{\Gamma(k')}{(k'-j)!(j-1)!}(zD)^{k'}\, ,
\end{eqnarray}
Therefore, $a_{j,n}= p^j D^n \binom{n-1}{j-1}$ such that
\bea \label{kappa_n_two_st}
\kappa_n(D) = D^n\sum_{j=1}^{n} p^j\frac{(-1)^{j-1}}{j}\binom{n+j-2}{j-1}\binom{n-1}{j-1} \;.
\eea
In fact, for $n \geq 2$ this sum over $j$ can be expressed in terms of an associated Legendre function, leading to the result
\bea \label{kappan_2st}
\kappa_n(D) = 
\begin{cases}
&p\, D \;, \; \hspace*{5.45cm} n=1 \;, \\
&- D^n\, \sqrt{p(1-p) \frac{1}{n(n-1)}}\, P_{n-1}^1(1-2p) \quad, \quad n \geq 2\;,
\end{cases}
\eea
where $P_k^1(x)$ is the associated Legendre function of index $1$ and degree $k$. For instance $P_1^1(x)=-\sqrt{1-x^2}$, $P^1_2(x) =-3 x \sqrt{1-x^2}$. Finally, for in this case $D_1=D$, $D_2=0$, the scaled cumulant generating function can thus be written as
\bea \label{exp_psiq_2st}
\Psi(q) &=& \lim_{t \to \infty} \frac{\chi_r(q,t)}{t} =  \frac{1}{2} \left(D q^2 - r + \Delta(q)\right) \\
&=& p\,q^2 D - \sqrt{p(1-p)} \sum_{n=2}^\infty \frac{q^{2n}}{r^{n-1}} \frac{D^n}{n(n-1)}\,P_{n-1}^1(1-2 p) \;.
\eea

\vspace*{0.5cm}
\noindent {\it Cumulants for arbitrary $D_1$ and $D_2$.} The case of arbitrary $D_1$ and $D_2$ can then easily be deduced from this formula~(\ref{exp_psiq_2st}). Indeed, in this case the scaled cumulant generating function $\Psi(q)$ is given by 
\bea \label{psiq_two_states}
\Psi(q) =  \frac{1}{2} \left((D_1 + D_2) q^2 - r + \Delta(q)\right) \quad \, , \quad \Delta(q) = \sqrt{\left((D_2 - D_1) q^2 + r\right)^2 - 4 (D_2 - D_1)p\, r \,q^2}\;.
\eea
Hence, we see that, except for the second cumulant, the higher order cumulants are only a function of $D_1-D_2$. Therefore we can use the result derived above for $D_1=D$ and $D_2=0$ in Eq. (\ref{exp_psiq_2st}) to expand $\Psi(q)$ as
\bea \label{expansion_Psi_2st}
\Psi(q) = q^2 (p\,D_1 +(1-p)D_2) - \sqrt{p(1-p)} \sum_{n=2}^\infty \frac{q^{2n}}{r^{n-1}} \frac{(D_1-D_2)^n}{n(n-1)}\, P_{n-1}^1(1-2 p) \;,
\eea
from which one can read the cumulants, as given in Eq. (3) in the main text. 

\vspace*{0.5cm}
\noindent {\it The scaled cumulant generating function in the limit $p\to 0^+$.} Interestingly, one finds from the exact expression in (\ref{psiq_two_states}) that the scaled generating function becomes singular in the limit $p \to 0^+$. In this limit, one has indeed $\Delta(q) = |(D_2-D_1)q^2 + r|$, which leads to the singular behavior of $\Psi(q)$ given in Eq. (40) of the End matter, namely
\bea
\lim_{p \to 0^+} \, \Psi (q) = 
\begin{cases}
D_2\, q^2 \;, & \quad q < q_c \;, \\[10pt]
D_1\, q^2 - r \;, & \quad q > q_c \;.
\end{cases} \quad, \quad q_c = \sqrt{\frac{r}{D_1-D_2}} \;.
\label{pzeroPsitwostates_SM}
\eea
One can easily check that $\Psi(q)$ is continuous at $q=q_c$, i.e.,
\bea \label{cont_psi}
\lim_{q\to q_c^-} \lim_{p \to 0^+}\Psi(q) = \lim_{q\to q_c^+} \lim_{p \to 0^+}\Psi(q) = \frac{D_2}{D_1-D_2}\,r \;. 
\eea
However, the first derivative is discontinuous at $q=q_c$, since one has
\bea \label{deriv_psi}
\lim_{q\to q_c^-} \lim_{p \to 0^+}\Psi'(q) = 2\,D_2 \sqrt{\frac{r}{D_1-D_2}} \quad, \quad \lim_{q\to q_c^+} \lim_{p \to 0^+} \Psi'(q) = 2\,D_1 \sqrt{\frac{r}{D_1-D_2}} 
\eea
For small but finite $p$, this transition at $q_c$ is smoothened out over a scale of size $\sqrt{p}$ where $\Psi(q)$ takes the scaling form
\bea \label{interpol}
\Psi(q) - \frac{D_2}{D_1-D_2}\,r \approx \sqrt{p} \, F\left(\frac{q-q_c}{\sqrt{p}} \right)\quad , \quad q\to q_c \; ,
\eea
where $\Psi(q_c) = r D_2/(D_1-D_2)$ the scaling function $F(\tilde q)$ is given by
\bea \label{def_F}
F(\tilde q) = q_c \left(\tilde q(D_1+D_2) + (D_1-D_2) \sqrt{\tilde q^2 + q_c^2} \right) - r \;. 
\eea
One can easily check that this scaling form (\ref{interpol}) interpolates smoothly between the two behaviors in (\ref{pzeroPsitwostates_SM}).

\begin{figure}[t]
    \centering
\includegraphics[width=0.4\linewidth]{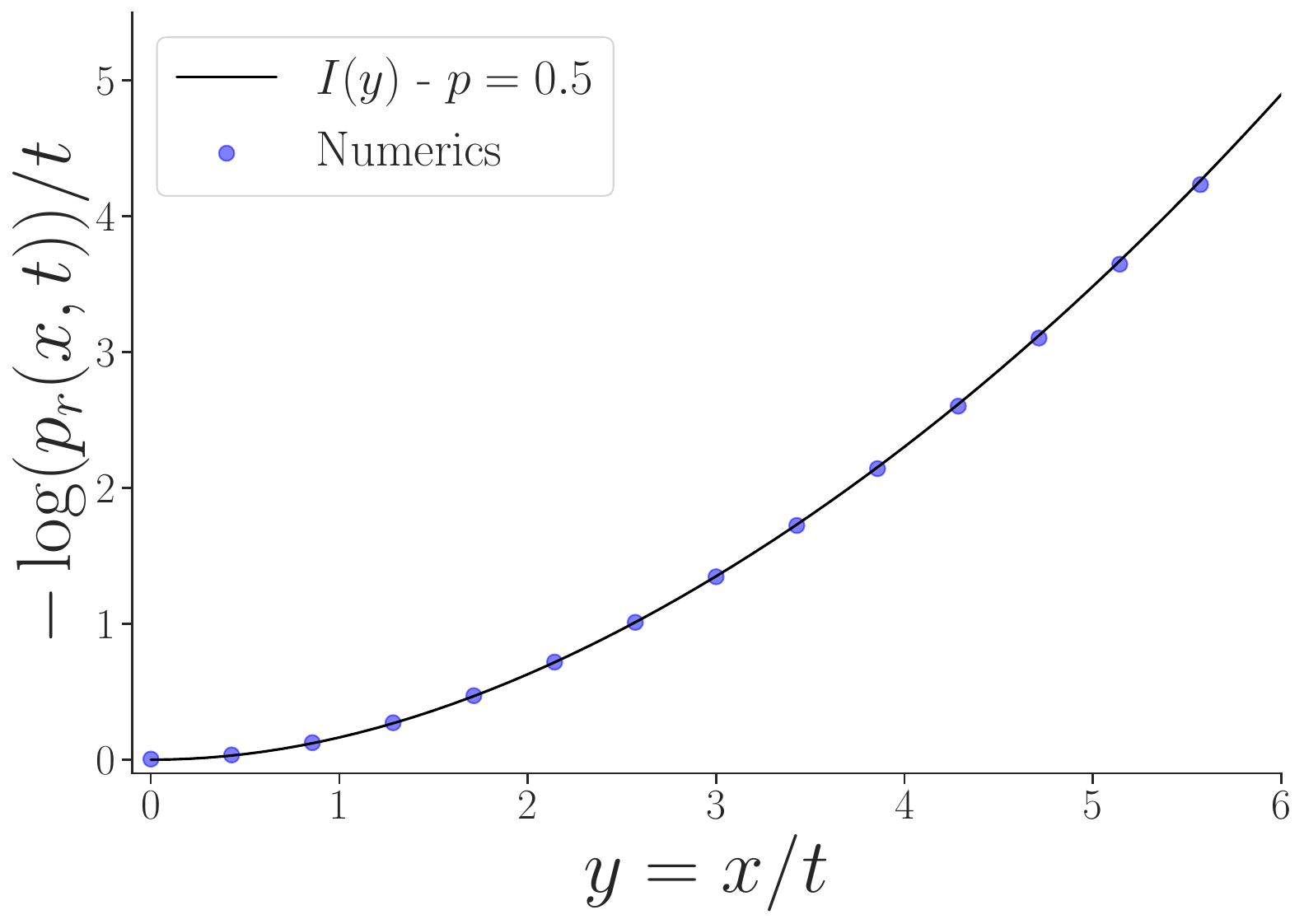}
  \caption{ We show here a plot of the rate function in the two-state model for the parameters $D_1 = 2$, $D_2 = 1$, $r = 1$, $p = 0.5$. The solid line $I(y=x/t)$ has been computed by solving numerically the maximization problem in Eq.~(\ref{maximizationI}). The dots correspond to the numerical resolution of the renewal equation as explained in Section \ref{numericssection} for $t=1000$.} 
  \label{Fig_p05} 
\end{figure}

\subsection{The rate function $I(y)$}

Inverting the exact expression for $\hat p_r(q,t)$ in Eq. (\ref{fouriertransform_twostates}) with respect to $q$ for any finite time $t$ seems quite difficult. However, this inversion can be performed in the limit of large time, using a saddle point computation. Indeed, $p_r(x,t)$ can be formally written as the following contour integral [see Eq. (\ref{inverse_BLT})]
\bea \label{p_x_two_states1}
p_r(x,t) = \int_{\gamma - i \infty}^{\gamma+ i \infty} \frac{dq}{2\pi i}  \, \hat p_r(q,t) \, e^{-q x}\, .
\eea
From the expression of $\hat p_r(q,t)$, one sees that it has a branch cut on the imaginary axis, which we do not specify further, since it will not be useful here. Therefore we can choose some value $\gamma>0$ to define the Browmich contour in Eq. (\ref{p_x_two_states1}). Given the form of $\hat p_r(q,t)$ at large time in Eq. (\ref{def_B}), this Bromwich integral (\ref{p_x_two_states1}) can be evaluated at large time by a saddle-point method, leading to
\begin{eqnarray} \label{saddle_point_2st}
p_r(x=y\,t,t) = \frac{B(q^*)}{\sqrt{2 \pi |\Psi''(q^*)| t}} e^{-t I(y)} \left(1+ O\left(\frac{1}{t}\right) \right), \quad {\rm where} \quad 
\begin{cases}\label{maximizationI}
&I(y) = \underset{{q \in \mathbb{R}}}{\max} (q\,y - \Psi(q)) \\
& \\
& q^* = \underset{{q \in \mathbb{R}}}{\rm argmax}(q\,y - \Psi(q))
\end{cases} \;,
\end{eqnarray}
while the function $B(q)$ is given in Eq.~(\ref{fouriertransform_twostates}). Note that the maximum in Eq. (\ref{saddle_point_2st}) comes from the fact that a minimum along the vertical Bromwich axis becomes a maximum in the horizontal direction along the real axis. Hence, as expected, we see that the rate function $I(y)$ is the Legendre transform of $\Psi(q)$. In principle one can compute $I(y)$ in terms of $q^*$ as
\bea \label{legendre_inv}
I(y) = q^* y - \Psi(q^*) \quad, \quad y = \Psi'(q^*) \;. 
\eea
Except in some special cases, it is difficult to compute explicitly $q^*$ as a function of $y$. However, this form (\ref{legendre_inv}) gives an interesting parametric representation of the function $I(y)$ -- $q^*$ being the parameter -- which can be used to evaluate numerically the function $I(y)$. Alternatively, the function $I(y)$ can also be evaluated by solving numerically the maximization problem. 
In Fig. \ref{Fig_p05}, we plot the function $I(y)$ evaluated by this second method for $p=0.5, r=1, D_1=2$ and $D_2=1$.

We note also the following interesting relation
\bea \label{Iprime}
I'(y) = q^* \;,
\eea
which can be obtained by taking the derivative of the first relation (\ref{legendre_inv}) with respect to $y$ and then using the second relation in (\ref{legendre_inv}). Furthermore, by taking a derivative of the second relation in (\ref{legendre_inv}) with respect to $y$, and using (\ref{Iprime}), one finds the well known identity
\bea
\Psi''(q^*) = \frac{1}{I''(y)} \;,
\eea
with $q^* \equiv q^*(y) = \underset{{q \in \mathbb{R}}}{\rm argmax}(q\,y - \Psi(q))$.

\begin{figure}[t]
    \centering
\includegraphics[width=0.4\linewidth]{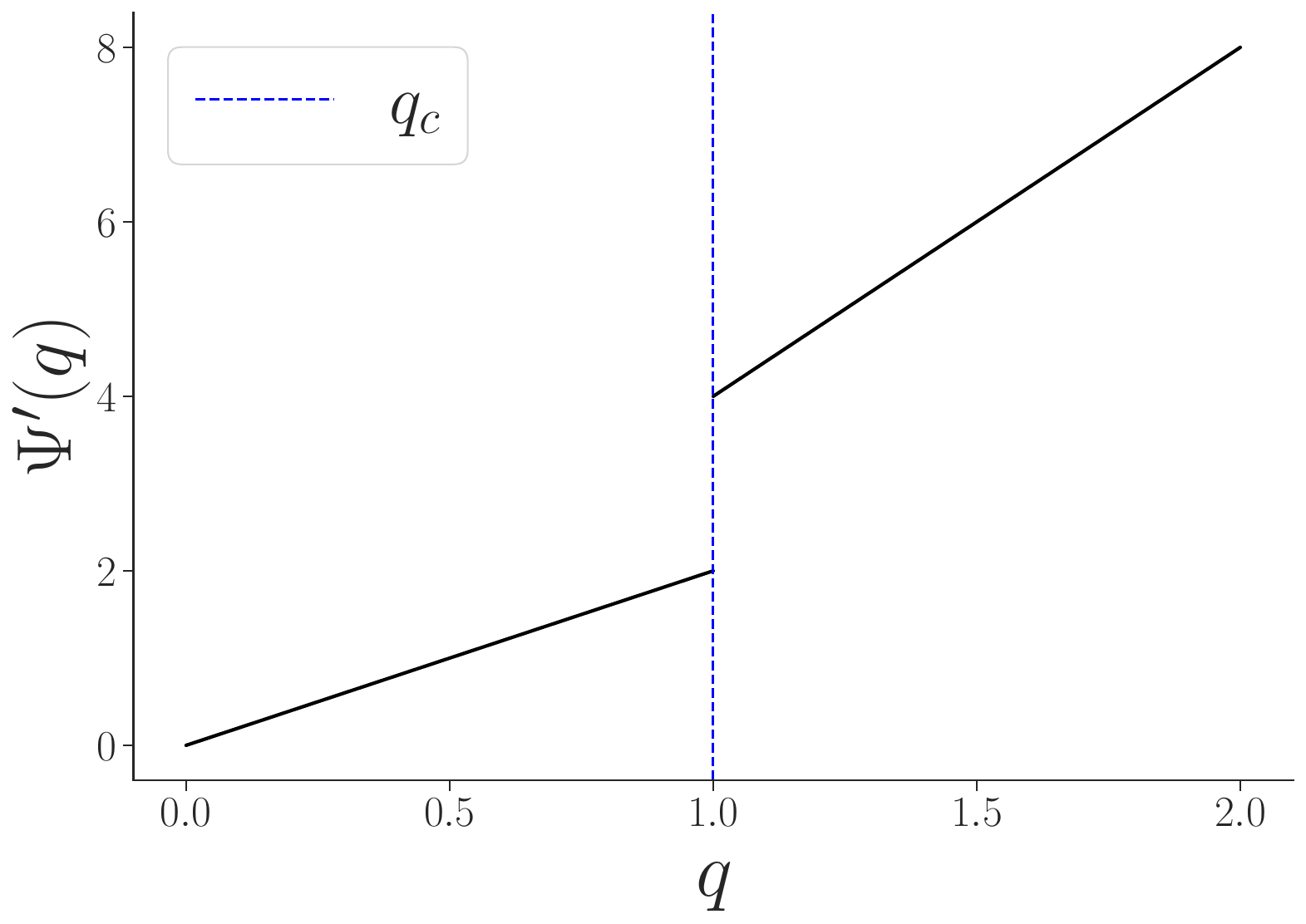}
  \caption{Here, we show a plot of the derivative of $\Psi(q)$ (see Eq.~(\ref{pzeroPsitwostates_SM})) which has a discontinuity at $q_c$ in the limit $p\to 0^+$. Indeed, for $q<q_c$, $\Psi'(q) = 2D_2q$, while for $q>q_c$, $\Psi'(q) = 2D_1q$.} 
  \label{Fig_psiprime} 
\end{figure}

\vspace*{0.5cm}
\noindent{\it The rate function $I(y)$ in the limit $p\to 0$.} Since the rate function $I(y)$ is an even function of $y$, we only consider the range $y \geq 0$. In this case, given the expression of $\Psi(q)$ in Eq. (\ref{pzeroPsitwostates_SM}), one can solve explicitly the equation $y = \Psi'(q)$ in the two regions $0 \leq y \leq 2 D_2 q_c$ and $y \geq  2 D_1 q_c$ since $\Psi'(q)$ has a discontinuity at $q_c = \sqrt{r/(D_1-D_2)}$ -- see Fig. \ref{Fig_psiprime}. One finds
\bea \label{qstar_2st}
q^* = \begin{cases}
& \frac{y}{2 D_2} \;, \; 0 \leq y \leq 2 D_2 q_c\, , \\
& \\
& \frac{y}{2 D_1} \;, \; \hspace*{0.7cm} y \geq 2D_1 q_c \;.
\end{cases}
\eea
On the other hand, for $2 D_2 \, q_c <y< 2 D_1 \, q_c$ one can show, using (\ref{interpol}) that
\bea \label{qstar_freeze}
q^* = q_c + O(\sqrt{p}) \;,
\eea
where the correction of $O(\sqrt{p})$ can in principle be computed explicitly in terms of the function $F(\tilde q)$ in Eq. (\ref{def_F}). Using the expression of $q^*$ in the various regimes of $y$ from Eqs. (\ref{qstar_2st}) and (\ref{qstar_freeze}) one finds the expression of the function $I(y)$ given in Eq. (41) of the End matter, i.e.,
\bea \label{expr_I_2st_sm}
I(y) = 
\begin{cases} 
\frac{y^2}{4 D_{2}} & , \quad |y| \leq v_1 = 2 D_{2} \alpha \\
\alpha\, |y| - D_{2}\, \alpha^2 & , \quad v_1 \leq |y| \leq v_2 \\
r + \frac{y^2}{4 D_{1}} & , \quad |y| \geq v_2 = 2 D_{1} \alpha \;.
\end{cases}
\eea
where $\alpha = \sqrt{r/(D_1-D_2)}$.

\vspace*{0.5cm}
\noindent{\it The pre-exponential factor in the limit $p\to 0$.} We end up this section by evaluating the pre-exponential factor in Eq. (\ref{saddle_point_2st}) in the limit $p\to 0$.

\vspace*{0.5cm}
$\bullet$ For $0\leq y \leq 2 D_2 q_c$, this prefactor is quite easy since $B(q^*) = 1$ while $\Psi''(q^*) = 2 D_2$, leading to
\bea
p_r(x = y t,t) \approx \frac{1}{\sqrt{4 \pi D_2 t}}\, e^{-t \frac{y^2}{4 D_2}} = \frac{1}{\sqrt{4 \pi D_2 t}}\, e^{-\frac{x^2}{4 D_2\,t}} \quad, \quad 0\leq y \leq 2 D_2 q_c \;.
\eea

\vspace*{0.5cm}
$\bullet$ For $y \geq 2 D_1 q_c$, one has $\Psi''(q^*) = 2 D_1$ but $B(q^*) = 0$ to leading order in $p$ and therefore a more careful analysis is required. We first obtain the small $p$ expansion of $q^*$ as
\bea \label{qstar_small_p}
q^* = \frac{y}{2\,D_1} + p (D_1-D_2) r^2 \frac{y}{2 D_1^2} \frac{1}{\left((D_1-D_2)\frac{y^2}{4 D_1^2} - r\right)^2} + O(p^2) \quad, \quad y \geq 2 D_1 q_c \;,
\eea
from which we obtain
\bea \label{Bqstar_2st}
B(q^*) = p \frac{y^4 (D_1-D_2)^2}{(4D_1^2 r + (D_2-D_1)y^2)^2} \quad, \quad y \geq 2 D_1 q_c \;. 
\eea
Note that it has the following asymptotic behaviors
\bea
B(q^*) \approx
\begin{cases}
& p \;, \; \hspace*{2.3cm} y \to \infty \\
& \\
& p\, \frac{D_1^2 r}{(D_1-D_2)} \frac{1}{(y - v_2)^2} \;, \; y \to v_2 = 2 D_1 \alpha \;.
\end{cases}
\eea
Hence in this range we get
\bea \label{p_finale}
p_r(x,t) \approx p \frac{y^4 (D_1-D_2)^2}{(4D_1^2 r + (D_2-D_1)y^2)^2} \frac{1}{\sqrt{4 \pi D_1 t}} e^{-t (r + \frac{y^2}{4 D_1 t})} 
\eea

\vspace*{0.5cm}
$\bullet$ For $2D_2 q_c \leq y \leq 2 D_1 q_c$, the analysis is a bit more complicated and relies on the precise behavior of $\Psi(q)$ around $q_c$ described in Eq. (\ref{interpol}). We find that $p_r(x,t)$ takes the following form (for large $t$ and $p\to 0$) 
\bea \label{p1/4}
p_r(x = y t,t) \approx p^{1/4} \frac{h(y)}{\sqrt{t}}\, e^{-t(\alpha y - D_2 \alpha^2)} \quad, \quad 2 D_2 \alpha < y < 2 D_1 \alpha \;,
\eea
where $\alpha = \sqrt{r/(D_1-D_2)}$ while $h(y)$ is a rather complicated function that can be computed explicitly from $F(\tilde q)$ in Eq. (\ref{def_F}) and its derivatives.

{
\subsection{Fokker-Planck approach and connection to resetting Brownian motion in the limit $p\to 0$}

In this section, we take a different route and provide a Fokker-Planck approach to the two-state model. 
In this case, there are two states with diffusion coefficients $D_1$ and $D_2$, with associated distributions $p(x,t,D_1) \equiv p_1(x,t)\delta(D-D_1)$ and $p(x,t,D_2) \equiv p_2(x,t)\delta(D-D_2)$.  

To derive the Fokker-Planck equation corresponding to $p_1(x,t)$, we first consider that the state at time $t+dt$ is $D_1$. We need to enumerate all possible probability flows from a given state at time $t$ to the state $D_1$ at time $t+dt$. This yields
\begin{eqnarray}
    p_1(x,t+dt)= p\, rdt\left[p_2(x,t) + p_1(x,t)\right]+(1-rdt)\int_{-\infty}^{+\infty}D\xi\, p_1(x-\xi \sqrt{2D_1 dt},t)\, ,
\end{eqnarray}
where $\xi \sim N(0,1)$. The first term on the right-hand side corresponds to the flow of probability arising from a reset in $[t,t+dt]$ with probability $rdt$. There are two possibilities. First, going from $D_2$ at time $t$, reset with probability $rdt$, and going back to $D_1$ with probability $p$, thus the first term inside the brackets. Or, the particle was in state $D_1$ at time $t$, resets with probability $rdt$, and staying at state $D_1$ with probability $p$, hence the second term inside the brackets. The second contribution on the left-hand side simply corresponds to the diffusion when there is no reset with probability $(1-rdt)$. The same reasoning for $p_2(x,t+dt)$ leads to
\begin{eqnarray}
    p_2(x,t+dt)= (1-p)\, rdt\left[p_2(x,t) + p_1(x,t)\right]+(1-rdt)\int_{-\infty}^{+\infty}D\xi\, p_2(x-\xi \sqrt{2D_2 dt},t)\, .
\end{eqnarray}
The Taylor expansion at order $dt$ gives the Fokker-Planck equations for the distributions of both states $D_1$ and $D_2$,
\begin{eqnarray}
    \frac{\partial p_1(x,t)}{\partial t}&&=D_1\frac{\partial^2 p_1(x,t)}{\partial x^2}+ rp\,  p_2(x,t) -r(1-p)\, p_1(x,t)\, ,\label{FP1twostates}\\
    \frac{\partial p_2(x,t)}{\partial t}&&=D_2\frac{\partial^2 p_2(x,t)}{\partial x^2} - rp\,  p_2(x,t) +r(1-p)\,  p_1(x,t) \, .\label{FP2twostates}
\end{eqnarray}
These equations have to be supplemented by the initial conditions
\bea \label{CI}
p_1(x,t=0) = p \delta(x) \quad, \quad p_2(x,t=0) = (1-p) \delta(x) \;.
\eea
The Fokker-Planck equations show that this model is in fact equivalent to a two-state model with different switching rates $k_{12} = r(1-p)$ to switch from the diffusion coefficient $D_1$ to $D_2$ and $k_{21}=rp$ to switch from $D_2$ to $D_1$, as studied in \cite{Grebenkov2019}. In principle, this system of coupled differential equations can be solved exactly via the use of Laplace transform in time and bi-lateral Laplace transform in space (as done in the paper). However, here, to unveil an interesting connection to resetting Brownian motion, we study instead directly the limit $p\to 0$ of this system (\ref{FP1twostates})-(\ref{FP2twostates}).

In the limit $p\to 0$, it is natural to expect $p_2(x,t) = O(1)$, while $p_1(x,t)=O(p)$ -- see Eqs. (\ref{FP1twostates})-(\ref{CI}). In this limit, we thus look for a solution of the form 
\bea \label{smallp_gen}
p_2(x,t) = p_{2,0}(x,t) + O(p) \quad, \quad p_1(x,t) = p \, p_{1,1}(x,t) + O(p^2) \;,
\eea
together with the initial conditions $p_{2,0}(x,0) = p_{1,1}(x,0) = \delta(x)$. By injecting the expansions (\ref{smallp_gen}) in Eqs. (\ref{FP1twostates}) and (\ref{FP2twostates}) one finds
\bea 
\frac{\partial p_{1,1}(x,t)}{\partial t}&&=D_1\frac{\partial^2 p_{1,1}(x,t)}{\partial x^2} - r p_{1,1}(x,t) + r\,  p_{2,0}(x,t) ,\label{FP1twostates_2}\\
    \frac{\partial p_{2,0}(x,t)}{\partial t}&&=D_2\frac{\partial^2 p_{2,0}(x,t)}{\partial x^2}  \, .\label{FP2twostates_2}
\eea
The equation for $p_{2,0}(x,t)$ can easily be solved leading to 
\bea \label{sol_p20}
p_{2,0}(x,t) = \frac{1}{\sqrt{4 \pi D_2 t}}\, e^{-\frac{x^2}{4 D_2 t}} \;.
\eea
Injecting this expression (\ref{sol_p20}) into Eq. (\ref{FP1twostates_2}) leads to
\bea \label{Eq_p11}
\frac{\partial p_{1,1}(x,t)}{\partial t}&&=D_1\frac{\partial^2 p_{1,1}(x,t)}{\partial x^2} - r p_{1,1}(x,t) + \frac{r}{\sqrt{4 \pi D_2 t}}\, e^{-\frac{x^2}{4 D_2 t}} \;.
\eea
Interestingly, this last equation for $p_{1,1}(x,t)$ has a structure which is very similar to the one found for Brownian motion in the presence of stochastic resetting \cite{resettingPRL, resettingReview}. Indeed for the latter, the source term is simply $r \delta(x-X_r)$ (where $X_r$ is the resetting position), instead of the Gaussian in Eq. (\ref{Eq_p11}).  
In fact, in the limit $D_2 \to 0$, this Gaussian term reduces exactly to $r \delta(x)$ -- hence similar to resetting at $X_r=0$. For $D_2 >0$, this delta-function has a certain width $\propto \sqrt{t}$, as described by this Gaussian term in Eq. (\ref{Eq_p11}). It is thus not surprising that the solution found for the two-state model in the limit $p\to  0$ bears some similarity with the resetting Brownian motion. It is also very similar to the model studied in Ref. \cite{Kundureset} where the particle is subjected to a non-instantaneous resetting in the presence of an external linear confining potential. Note that another connection between resetting and a two-state model was also noticed in the context of ``autonomous ratcheting by stochastic resetting'' \cite{Ghosh}. Finally, the interplay between random diffusion models and stochastic resetting was recently studied in \cite{Bressloff2states,Santra}.

Under this form (\ref{Eq_p11}) it is easy to obtain $p_{1,1}(x,t)$ as a convolution of the source term -- the Gaussian in this case -- and the heat kernel with diffusion coefficient $D_1$. This yields
\bea \label{sol_p11}
p_{1,1}(x,t) = e^{-rt} \; \frac{e^{-\frac{x^2}{4 D_1 t}} }{\sqrt{4 \pi D_1 t} }+ r \int_0^t d\tau \int_{-\infty}^\infty dy\, e^{-r(t-\tau)} \frac{e^{-\frac{(x-y)^2}{4 D_1 (t-\tau)}}}{\sqrt{4 \pi D_1 (t-\tau)}} \frac{e^{-\frac{y^2}{4 D_2 \tau}}}{\sqrt{4 \pi D_2 \tau}} \;.
\eea
Note that this equation can also be directly obtained from the small $p$ expansion of the renewal equation (\ref{renewalpropa}). The integral over $y$ can be performed explicitly, leading to (after the change of variable $\tau \to t-\tau$)
\bea \label{sol_p11_2}
p_{1,1}(x,t) = e^{-rt} \; \frac{e^{-\frac{x^2}{4 D_1 t}} }{\sqrt{4 \pi D_1 t}}+ r \int_0^t d\tau e^{-r \tau} \frac{e^{-\frac{x^2}{4D_2(t-\tau)+ 4D_1 \tau}}}{\sqrt{4\pi(D_2(t-\tau)+D_1 \tau)}}  \;.
\eea
This exact expression (\ref{sol_p11}) can now be analysed along the lines of the case of RBM \cite{MMS15} as follows. We first rewrite (\ref{sol_p11_2}) in terms of $y = x/t$ and perform the change of variable $\tau = u t$ in the integral. This gives
\bea \label{sol_p11_3}
p_{1,1}(x=y\,t,t) = \frac{1}{\sqrt{4 \pi D_1 t}} e^{- t \Phi(y)} + r \sqrt{t} \int_0^1 du\, \frac{e^{-t \varphi(u,y)}}{\sqrt{4 \pi(D_2(1-u)+D_1 u )}}
\eea
where we have introduced the two functions
\bea \label{def_phi}
\Phi(y) = r + \frac{y^2}{4 D_1} \quad, \quad \varphi(u,y) = r u + \frac{y^2}{4 D_2(1-u)+4 D_1 u }  \;.
\eea
For large time, the integral in (\ref{sol_p11_3}) can be estimated by a saddle point. As a function of $u$, the function $\phi(u,y)$ admits a single minimum at $u^*$ given by
\bea \label{u_star}
u^* = \frac{1}{D_1-D_2} \left(\frac{y}{2}\sqrt{\frac{D_1-D_2}{r}} - D_2 \right) \;.
\eea
This saddle point $u^*$ occurs within the interval of integration if and only if
\bea \label{condition}
0<u^*<1 \quad \Longleftrightarrow \quad 2 D_2 \sqrt{\frac{r}{D_1-D_2}} \leq y \leq 2 D_1 \sqrt{\frac{r}{D_1-D_2}} \;.
\eea
In this case, evaluating $\varphi(u^*,t)$, one finds
\bea \label{phi_ustar}
\varphi(u^*,y) = - r \frac{D_2}{D_1-D_2} + \sqrt{\frac{r}{D_1-D_2}} y \;.
\eea
Instead, if $u^*<0$ the minimum of $\varphi(u,y)$ is reached at $u=0$, where $\varphi(0,y) = r  + \frac{y^2}{4\, D_2}$ while for $u^*>1$ the minimum is reached at $u^* = 1$ where $\varphi(0,y) = r  + \frac{y^2}{4\, D_1}$. Therefore, this analysis can be summarized as follows
\bea \label{summary_p11}
\lim_{t \to \infty} \frac{\ln p_{1,1}(x=y\,t,y)}{t} = 
\begin{cases}
&r + \frac{y^2}{4\, D_1} \;, \hspace*{3.cm} 0 \leq y \leq 2 D_2 \, q_c \, ,\\
&- r \frac{D_2}{D_1-D_2} + \sqrt{\frac{r}{D_1-D_2}}\,y \quad, \quad  2 D_2 \, q_c \leq y \leq 2 D_1 q_c \, , \\
&r + \frac{y^2}{4\, D_1} \;, \;  \hspace*{3.cm} y \geq 2 D_1 q_c \;.
\end{cases}
\eea
Note that in the first regime $0\leq y \leq 2 D_2 q_c$, the first term in Eq. (\ref{sol_p11_3}) is larger than the integral (since $D_1>D_2$), which explains the first line of Eq. (\ref{summary_p11}).

To analyse $p(x,t,D) = p_1(x,t)\delta(D-D_1)+ p_2(x,t)\delta(D-D_2)$ in the small $p$ limit, one has to analyse $p_2(x,t)$ up to order $O(p)$, i.e., write $p_2(x,t) = p_{2,0}(x,t) + p\, p_{2,1}(x,t) + O(p^2)$. From Eq. (\ref{FP2twostates}), one finds that $p_{2,1}(x,t)$ satisfies
\bea \label{eq_p21}
\frac{\partial p_{2,1}(x,t)}{\partial t} = D_2 \frac{\partial ^2 p_{2,1}(x,t)}{\partial x^2} - r p_{2,0}(x,t)+rp_{1,1}(x,t) \;,
\eea
together with the initial condition $p_{2,1}(x,0) = - \delta(x)$. Its solution reads
\bea \label{sol_p21}
p_{2,1}(x,t) = - \frac{e^{-\frac{x^2}{4 D_2 t}}}{\sqrt{4 \pi D_2 t}} + r \int_0^{t} d\tau \int_{-\infty}^\infty dy \frac{e^{-\frac{y^2}{4 D_2 \tau}}}{\sqrt{4 \pi D_2 \tau}} \left(p_{1,1}(y-x,t-\tau)-p_{2,0}(y-x,t-\tau)\right) \;.
\eea
If one takes the BLT one can easily show that $p_{2,1}(x,t)$ behaves, at leading (exponential) order for large $t$, as $p_{1,1}(x,t)$.  
}

Finally, combining these results for $p_{1,1}(x,t)$ and $p_{2,1}(x,t)$ both given by (\ref{summary_p11}) together with the one for $p_{2,0}(x,t)$ in Eq. (\ref{sol_p20}) one finds that the total probability $p(x,t) = p_{2,0}(x,t)+p(p_{1,1}(x,t)+p_{2,1}(x,t))$ is given by
\begin{align} \label{summary_p11}
\lim_{t \to \infty} -\frac{\ln p(x=y\,t,y)}{t} = 
\begin{cases}
\frac{y^2}{4 D_2} & \quad , \quad 0 \leq y \leq 2 D_2 q_c \;, \\[0.3cm]
- r \frac{D_2}{D_1 - D_2} + \sqrt{\frac{r}{D_1 - D_2}}\, y & \quad , \quad 2 D_2 q_c \leq y \leq 2 D_1 q_c \;, \\[0.3cm]
r + \frac{y^2}{4 D_1} & \quad , \quad y \geq 2 D_1 q_c \;.
\end{cases}
\end{align}
Note that the leading term in the region $ 0 \leq y \leq 2 D_2 \, q_c$ is actually given by $p_{2,0}(x,t)$ in Eq. (\ref{sol_p20}) which is actually dominating over $p_{1,1}(x,t)+p_{2,1}(x,t)$ -- see the first line of Eq. (\ref{summary_p11}) -- 
since $y^2/(4D_2) \leq r + y^2/(4 D_1)$ in this regime. Therefore, we recover the result given in Eq. (41) in the End matter, using a different method which also allows to establish a connection with resetting Brownian motion. Note however that this  small $p$ expansion (which assumes that the exact $p(x,t)$ is an analytic function of $p$) does not allow to recover correctly the $\propto p^{1/4}$ behavior of the pre-exponential factor (see Eq. (\ref{p1/4})).

\section{The case where $W(D)$ has a finite support $[0,D_{\text{max}}]$}

In this section, we provide the details of the study of the scaled cumulant generating function (SCGF) $\Psi(q)$ and the rate function $I(y)$ in the case where $W(D)$ has a finite support $[0,D_{\text{max}}]$ such that $W(D) \sim (D_{\text{max}}-D)^\nu$ when $D\to D_{\text{max}}$, $\nu >-1$. This includes the case where $W(D)$ is given by a semi-circular distribution (corresponding to $\nu =1/2$), or the case of a uniform distribution (for which $\nu = 0$), both studied in Section \ref{specificexamples}.

\subsection{The scaled cumulant generating function $\Psi(q)$}

We recall that $\Psi(q)$ is defined as
\bea \label{def_Psi}
\Psi(q) = \lim_{t \to \infty} \frac{\ln \hat p_r(q,t)}{t} \quad, \quad  \hat{p}_r(q,t) = \int_\Gamma \frac{ds}{2 i \pi} \tilde p_r(q,s)\, e^{st}\, ,
\eea
where we recall that $\tilde p_r(q,s)$ is given by
\bea \label{ptilde_sm}
\tilde p_r(q,s) = \frac{J_r(q,s)}{1-r\,J_r(q,s)} \quad {\rm where} \quad J_r(q,s)= \int_0^{D_{\text{max}}} dD\, \frac{W(D)}{r+s-Dq^2}\, .
\eea
In Eq. (\ref{def_Psi}), $\Gamma$ is a Bromwich contour passing to the right of all the singularities of $\tilde p_r(q,s)$ defined in (\ref{ptilde_sm}) in the complex $s$-plane. In general, $\tilde p_r(q,s)$ admits two types of singularities: (a) the one arising from $J_r(q,s)$ itself -- which is a branch cut -- that exists for all values of $\nu > -1$ and (b) poles -- which are the roots of the denominator in $\tilde p_r(q,s)$ -- which exists only for certain values of $\nu$ and $q$ (see below).

\vspace*{0.5cm}
\noindent {\bf (a) The branch cut}: Since the integral over $D$ defining $J_r(q,s)$ in Eq. (\ref{ptilde_sm}) has a non-integrable singularity for $D = (r+s)/q^2$, the function $J_r(q,s)$, and hence $\tilde p_r(q,s)$ has a branch-cut on the real axis $[-r,-r+D_{\text{max}} q^2]$ (see Fig. \ref{Fig_branchcut}). More precisely, $J_r(q,s)$ can be written as,
\bea \label{CT}
J_r(q,s) = \frac{1}{q^2} g\left( \tilde s = \frac{r+s}{q^2}\right) \quad, \quad g(z) = \int_0^{D_{\max}} \frac{W(D)}{z-D} \;, \; z \notin [0,D_{\max}] \;,
\eea
where $g(z)$ is the  Cauchy-Stieltjes transform of $W(D)$. In particular, for a continuous distribution $W(D)$, $g(z)$ is an analytic function of $z$ which admits a branch cut on $[0,D_{\max}]$, i.e., 
\bea \label{cut}
\lim_{\epsilon \to 0^+} \left[g(D-i \epsilon) - g(D+ i \epsilon) \right] = 2 i \pi W(D) \;, \; D \in [0,D_{\max}] \;. 
\eea
This means that $J_r(q,s)$ is an analytic function of $s$ with a branch cut on the segment $[-r,-r+D_{\text{max}} q^2]$ such that
\bea \label{cut_J}
\lim_{\epsilon \to 0^+} \left[J_r(q,s-i \epsilon) -  J_r(q,s+i \epsilon)\right] = \frac{2 i \pi}{q^2} W\left( \tilde s = \frac{r+s}{q^2}\right) \;, \; s \in [-r,-r+D_{\text{max}} q^2] \;.
\eea

\begin{figure}[t]
    \centering
\includegraphics[width=0.7\linewidth]{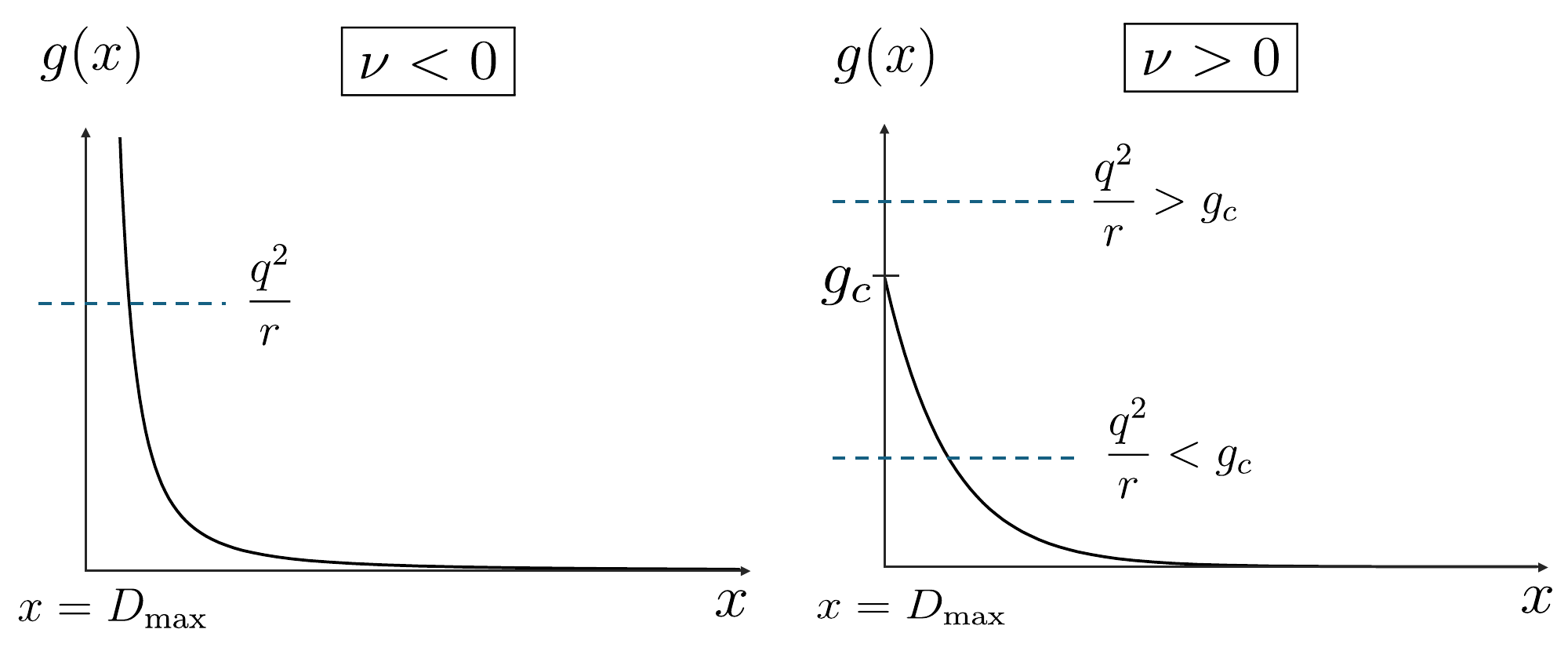}
  \caption{We illustrate here the two situations described in Eqs.~(\ref{gtransition1}) and (\ref{gtransition2}). For a distribution $W(D)\sim (D_{\text{max}}-D)^\nu$, the Cauchy-Stieltjes transform of $W(D)$, denoted by $g(x)$ for $x\in(D_{\text{max}},+\infty)$, exhibits two different behaviors for $x\to D_{\text{max}}$. If $\nu <0$ (left panel), $g(x\to D_{\text{max}})\sim (D_{\text{max}}-x)^\nu$ diverges as $x\to D_{\text{max}}$, while if $\nu >0$ (right panel), $g(x\to D_{\text{max}}) = g_c$ which is finite.}
  \label{Fig_gx} 
\end{figure}

\vspace*{0.5cm}
\noindent {\bf (b) The pole}. To find the pole of $\tilde{p}_r(q,s)$ in the complex $s$-plane, we have to solve (for $s$) the equation
\begin{eqnarray}
  1 - r\, J_r(q,s) =0 \Longleftrightarrow  1=r\int_0^{D_{\text{max}}} dD\, \frac{W(D)}{r+s-D\, q^2} \Longleftrightarrow \frac{q^2}{r}= g\left(\tilde s = \frac{r+s}{q^2}\right) \, ,
    \label{equation_pole}
\end{eqnarray}
where $g(x)$ is defined in (\ref{CT}). Note that the discussion of a similar equation appeared in the study of spherical integrals (i.e., Harish Chandra/Itzykson-Zuber integrals) in random matrix theory \cite{maida_guillonet}. 

To proceed, we analyse the behavior of this function $g(x)$ for real $x \in (D_{\max},+\infty)$. Note that as $g'(x)<0$, it is a decreasing function of $x$, and we have $g(x) \sim 1/x$ when $x\to \infty$. To investigate the behavior of $g(x)$ when $x\to D_{\text{max}}$, we need to distinguish two cases (see Fig. \ref{Fig_gx}):
\begin{eqnarray}\label{gtransition}
    &&(\text{i})\quad  \nu >0\, , \quad  g(x\to D_{\text{max}}) = g_c = \int_0^{D_{\text{max}}} dD \, \frac{W(D)}{D_{\text{max}}-D}\, ,\label{gtransition1}\\
    &&(\text{ii})\quad  \nu < 0\, , \quad  g(x\to D_{\text{max}}) \underset{D\to D_{\text{max}}}{\propto}(D_{\text{max}} - x)^{\nu}\, .\label{gtransition2}
\end{eqnarray}
While, for $\nu>0$ the function $g(x)$ is bounded from above (since $g_c$ is finite), in the case $\nu <0$, the function $g(x)$ diverges as $x \to D_{\text{max}}$ (see Fig.~\ref{Fig_gx}). In the following, we analyse the two situations $-1<\nu \leq 0$ and $\nu  >0$ separately.

\subsubsection{The case $-1<\nu \leq 0$}

Hence, for $\nu < 0$, the equation in (\ref{equation_pole}) admits a solution $s=s^*$ for any $q$ and therefore, $\tilde p_r(q,s)$ has a pole for any $q$ at this value $s=s^*$ (see Fig.~\ref{Fig_branchcut}). Note in addition that $s^* \geq q^2 D_{\text{max}} - r$ and in this case the large time behavior of $\hat p_r(q,t)$ in Eq. (\ref{def_Psi}) is dominated by this pole at $s=s^*$. Therefore, we have
\begin{eqnarray}
    \hat{p_r}(q,t) \isApproxTo{t\to \infty} \text{Res}\left(\frac{J_r(s,q) }{1-r\, J_r(s,q) },s^*(q)\right)\, e^{t s^*(q)}\, .
\end{eqnarray}
To compute the residue, we write the expansion of $J_r(s,q)$ in the vicinity of $s^*(q)$
\begin{eqnarray}
    J_r(s,q) = J_r(s^*(q),q)+\partial_s J_r(s,q)(s-s^*(q))+O((s-s^*(q))^2)\, ,
\end{eqnarray}
and by definition of $s^*$, we have $1-rJ_r(s^*(q),q)=0$ such that $J_r(s^*(q),q)=1/r$. Thus, we obtain
\bea \label{residue}
\hat{p_r}(q,t) \isApproxTo{t\to \infty} \frac{1}{r^2 |\partial_s J_r(s^*(q),q)|} e^{t s^*(q)} \;,
\eea
where we have used that $-\partial_s J_r(s^*(q),q)= |\partial_s J_r(s^*(q),q)|$. In Section \ref{section:preexp}, we give a more explicit expression for the residue and we deduce from it the $O(1)$ corrections for the cumulants for all $W(D)$ with finite moments.

Hence, the scaled cumulant generating function $\Psi(q)$, as defined in Eq. (\ref{def_Psi}), is given by 
\begin{figure}[t]
    \centering
\includegraphics[width=0.49\linewidth]{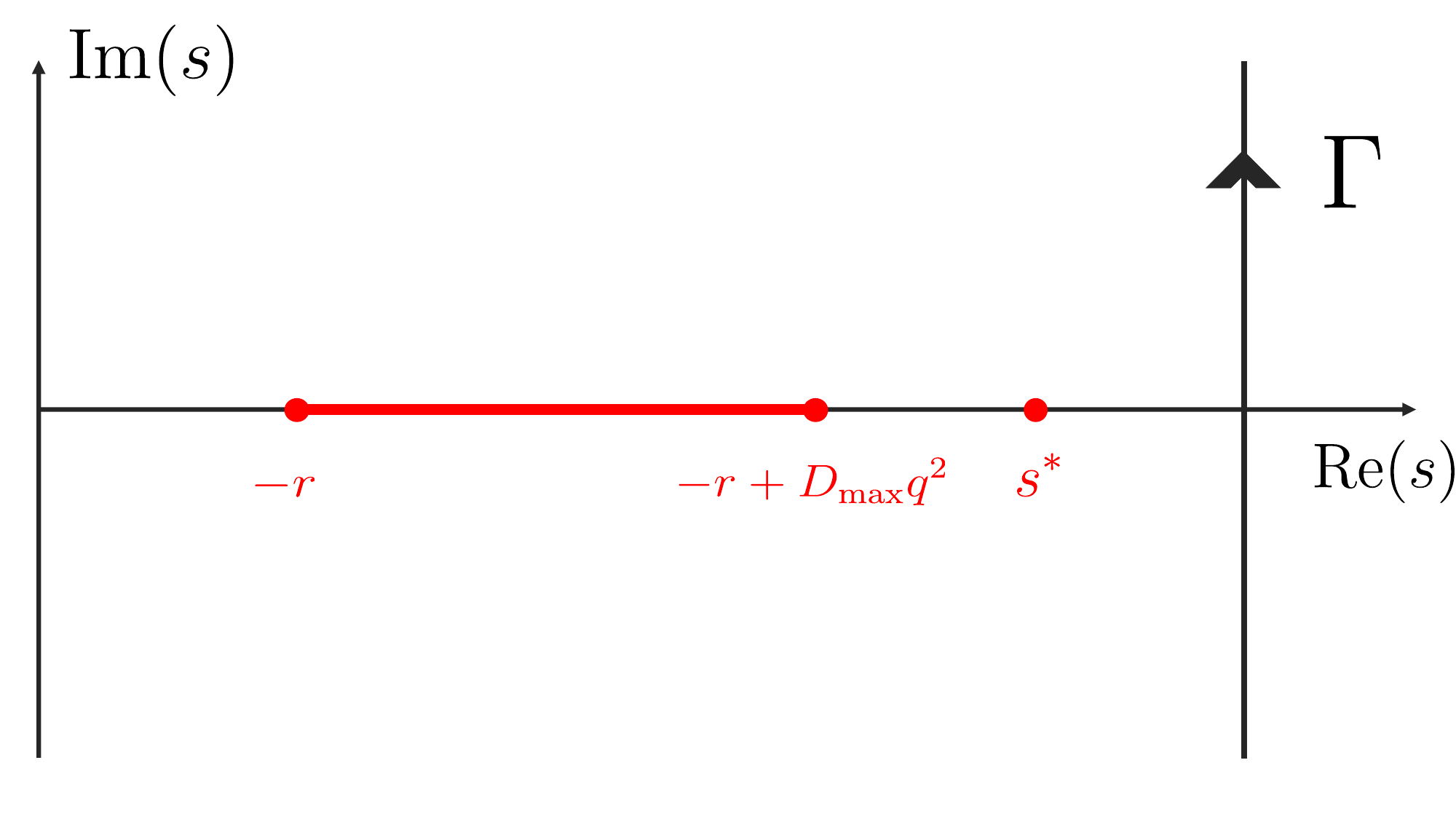}
\includegraphics[width=0.49\linewidth]{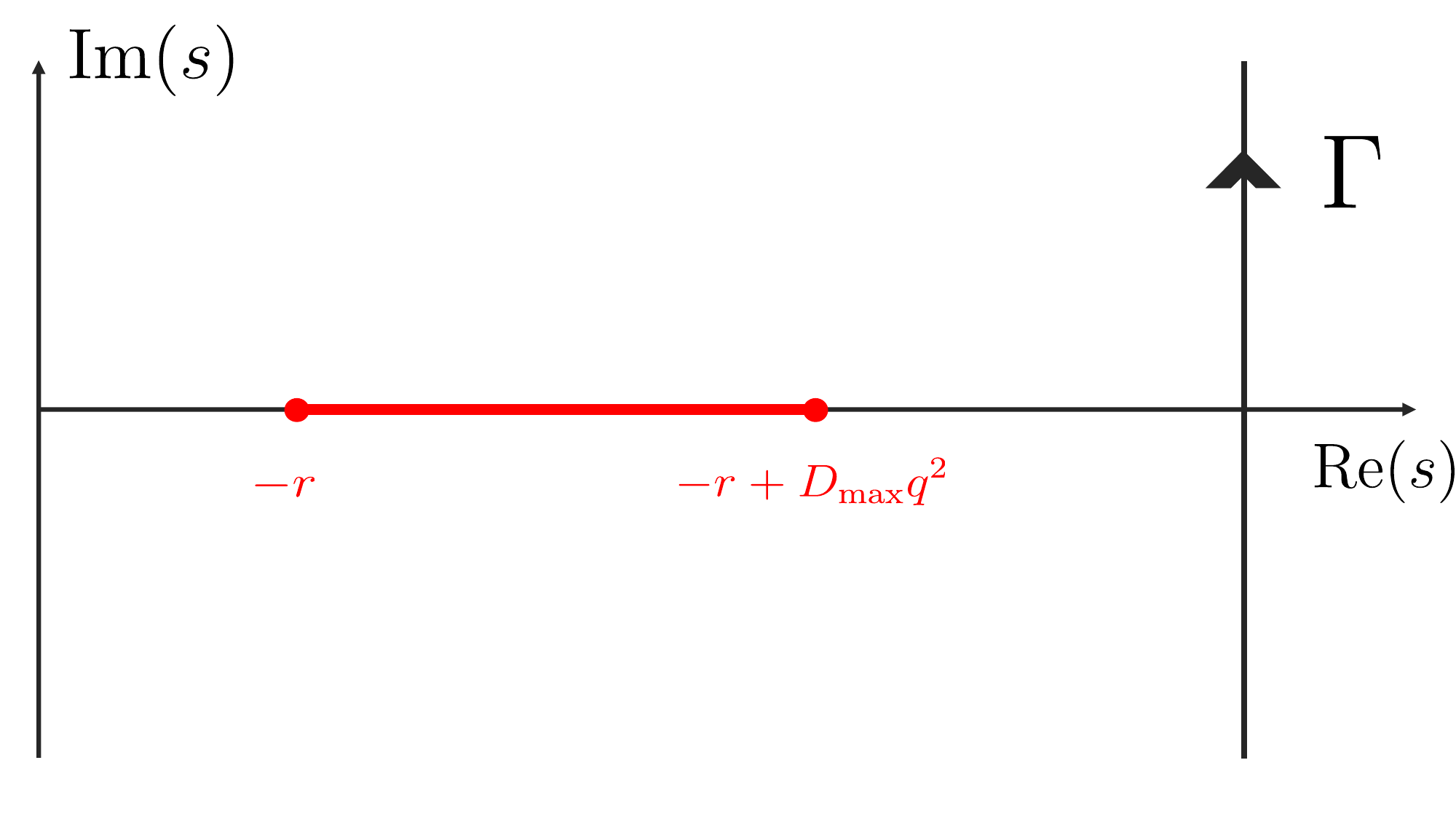}
  \caption{Illustration of the singularities of the function $\tilde p_q(q,s)$ in the complex $s$-plane. The left panel shows the situation corresponding to $-1<\nu < 0$ for all values of $q$ and $\nu > 0$ for $q\leq q_c$ which shows a branch cut on $[-r,-r+D_{\text{max}}q^2]$ as well as a pole at $s^* \equiv s^*(q) \geq -r+D_{\text{max}}q^2$. The right panel shows the situation for $\nu>0$ and $q \geq q_c$ for which there is no pole anymore and the remaining singularity is the branch cut on $[-r,-r+D_{\text{max}}q^2]$. In both figures, the contour $\Gamma$ is the Bromwich contour used in the computation of the generating function $\hat p_r(q,t)$ in Eq. (\ref{def_Psi}).} 
  \label{Fig_branchcut} 
\end{figure}
\bea \label{sstar}
\Psi(q) = s^*(q) \quad, \quad 1 = r\, \int_0^{D_{\text{max}}} dD \frac{W(D)}{r+s^*(q)-Dq^2} \quad, \quad \forall q \in \mathbb{R} \;.
\eea
As discussed in the Letter, this relation shows that $s^*(q)$ can be written in terms of the {\it R-transform} of $W(D)$ namely
\bea \label{series_cum}
\Psi(q) = q^2 R\left( \frac{q^2}{r}\right) = r \sum_{n \geq 1} \left(\frac{q^2}{r} \right)^{n} \kappa_n(D) \;,
\eea
where $\kappa_n(D)$ is the $n$-th free cumulant of $D$. In particular, for small $q$, the leading term is $\Psi(q) \sim \kappa_1(D)\, q^2 = \langle D \rangle q^2$. 

To study the behavior of $\Psi(q)=s^*(q)$ as $q \to \infty$, it is convenient to rewrite the equation satisfied by $s^*(q)$ as
\bea \label{stilde}
\frac{q^2}{r} = \int_0^{D_{\text{max}}} dD\, \frac{W(D)}{\tilde s^*(q) - D} \;, \; \tilde s^*(q) = \frac{r+s^*(q)}{q^2} \;.
\eea
In the large $q$ limit the left hand side is diverging as $q^2/r$ and hence the right hand side must also diverge, implying $\tilde s^*(q) \to D_{\text{max}}$. Therefore, to summarize, for $\nu \leq 0$
\bea \label{asymptpsi_nu_neg}
\Psi(q) = s^*(q) = 
\begin{cases}
&\langle D \rangle \, q^2 \;, \; \hspace{0.4cm} q \to 0 \, ,\\[10pt]
& D_{\text{max}} q^2 - r \;, \; q \to \infty \;.
\end{cases}
\eea
Furthermore, it is easy to check from Eq. (\ref{sstar}) that $\Psi(q)$ and its derivatives are continuous functions of $q$ for all real~$q$. Below, in Section \ref{uniformsection}, we compute $\Psi(q)$ explicitly in the case where $W(D)$ is the uniform distribution over $[0,D_{\max}]$, corresponding to $\nu = 0$.

\subsubsection{The case $\nu > 0$}\label{nug0}

This case turns out to be more interesting. Indeed, for $\nu >0$, the function $\tilde p_r(q,s)$, as a function of $s$, exhibits a pole only for $q^2 \leq q_c^2 = r g_c$ where $g_c$ is given in Eq. (\ref{gtransition}) while there is no pole for $q^2 > q_c^2 = r g_c$ and in that case the large $t$ limit is instead dominated by the branch point at $s_b = -r + D_{\text{max}} q^2$ (see Fig. \ref{Fig_branchcut}). This implies the following behavior of $\Psi(q)$ in this case $\nu > 0$
 \bea
\Psi(q) = 
\begin{cases}
s^*(q)\;, & \quad q < q_c \;, \\[10pt]
D_{\text{max}} q^2-r\;, & \quad q > q_c \;,
\end{cases}
\label{Psi_nupos}
\eea
where $s^*(q)$ is the solution of the implicit equation (\ref{sstar}) and $q_c$ is given by
\bea \label{def_qc}
q_c^2 = r \, \int_0^{D_{\text{max}}} dD \, \frac{W(D)}{D_{\text{max}}-D} \;.
\eea
Let us analyse the properties of $\Psi(q)$ around $q_c$. We first notice that $\Psi(q)$ is continuous at $q=q_c$, which follows straightforwardly from the definition of $q_c$ in (\ref{def_qc}). To analyse its first derivatives, we need to compute ${s^*}'(q)$ as well as ${s^*}{''}(q)$, which can be done by taking derivatives of Eq. (\ref{sstar}) with respect to $q$. This leads to
\bea
{s^*}'(q) = 2q \frac{\int_0^{D_{\text{max}}} dD \frac{D\, W(D)}{(s^*(q)+r-Dq^2)^2}}{\int_0^{D_{\text{max}}} dD \frac{W(D)}{(s^*(q)+r-Dq^2)^2}} \;, \; q \leq q_c \;.
\eea
Since $s^*(q) \to D_{\text{max}} q_c^2 - r$ as $q \to q_c$, we see that if $0<\nu<1$, both the numerator and the denominator are diverging and it is easy to obtain
\bea  \label{nu_less1}
{s^*}'(q) \to 2 q_c D_{\text{max}} \quad \text{as} \quad q \underset{<}{\to} q_c \quad, \quad 0<\nu<1 \;.
\eea
On the other hand, for $\nu>1$ these integrals are converging and ${s^*}'(q)$ converges to a non-trivial value given by
\bea \label{nu_geq1}
{s^*}'(q) \to 2 q_c D_{\rm eff} \quad, \quad   D_{\rm eff} = \frac{\int_0^{D_{\text{max}}} dD \frac{D W(D)}{(D_{\text{max}}-D)^2}}{\int_0^{D_{\text{max}}} dD \frac{W(D)}{(D_{\text{max}}-D)^2}} \leq D_{\text{max}}  \quad {\rm as} \quad q \underset{<}{\to} q_c \quad, \quad \nu>1  \;.
\eea
Using Eq. (\ref{Psi_nupos}) together with (\ref{nu_less1}) and (\ref{nu_geq1}), we thus see that $\Psi'(q)$ is continuous at $q=q_c$ for $0<\nu<1$ while it is discontinuous for $\nu >1$. 

The analysis of $s^*(q)$ near $q_c$ beyond the leading (linear) order requires analyzing the two cases $\nu>1$ and $0<\nu<1$ separately. 

\vspace*{0.5cm}
\noindent {\it The case $\nu>1$.} In this case it is useful to compute ${s^*}''(q)$ from Eq. (\ref{sstar}). It reads
\bea \label{ssec}
{s^*}^{''}(q) = \frac{2}{\int_0^{D_{\text{max}}} dD \frac{W(D)}{(s^*(q)+r-Dq^2)^2}}\left( \int_0^{D_{\text{max}}}dD \frac{W(D)({s^*}'(q)-2 D q)^2}{(s^*(q)+r-Dq^2)^3} + \int_0^{D_{\text{max}}}dD \frac{D W(D)}{(s^*(q)+r-Dq^2)^2}\right)\, .
\eea
On this expression, we see that the behavior of 
${s^*}^{''}(q)$ as $q \to q_c$ is different for $\nu>2$ or $\nu < 2$. For $\nu>2$, one can easily see that ${s^*}^{''}(q)$ remains finite as $q\to q_c$ with the result
\bea \label{ssec2}
{s^*}^{''}(q_c) = 2B_1 = \frac{2}{\int_0^{D_{\text{max}}} dD \frac{W(D)}{(D_{\text{max}}-D)^2}}\left( 4\int_0^{D_{\text{max}}}dD \frac{W(D)(D_{\rm eff}-D)^2}{(D_{\text{max}}-D)^3} + \int_0^{D_{\text{max}}}dD \frac{D W(D)}{(D_{\text{max}}-D)^2}\right) \quad,  \nu > 2 \;.
\eea
However, if $\nu <2$, we see that the first integral in the numerator of (\ref{ssec2}) is diverging, which indicates a singular behavior of ${s^*}^{''}(q)$ as $q \to q_c$. Indeed, in that case one finds
\begin{equation} \label{ssec2}
{s^*}^{''}(q) \approx \nu(\nu-1) B_2 (q_c-q)^{\nu -2} \quad, \quad B_2 = \frac{2 \pi A(D_{\rm eff}-D_{\max})^2}{\int_0^{D_{\text{max}}} dD \frac{W(D)}{(D_{\text{max}}-D)^2}} \frac{1}{\left\vert \sin{(\nu \pi)}\right\vert} \;,
\end{equation}
where the amplitude $A$ is such that $W(D) \approx A(D_{\text{max}}-D)^{\nu}$ as $D \to D_{\text{max}}$. 

%We want to analyse these integrals in the limit $q \to q_c$ and for $0<\nu < 1$ -- since we have already seen that ${s^*}'(q)$ is discontinuous at $q=q_c$. For this purpose, it is useful to use the following asymptotic behaviors of a class of integrals
%\bea \label{asympt1}
%\int_0^{D_{\text{max}}} dD \frac{W(D) f(D)}{(s^*(q)+r-Dq^2)^n}  \approx \frac{A}{q_c^{2n}} \frac{\Gamma(1+\nu) \Gamma(n-1-\nu)}{\Gamma(n)} \left(\frac{{s^*}''(q_c) - 2D_{\text{max}}}{2 q_c^2} \right)^{\nu-n+1} f(D_{\text{max}}) (q_c-q)^{2\nu - 2n  +2} \;, \; q \to q_c \;,
%\eea
%where $f(D)$ is an arbitrary function which is finite as $D \to D_{\text{max}}$ and assuming $W(D) \approx A(D_{\text{max}}-D)^{\nu}$. Of course, in deriving (\ref{asympt1}), we have assumed that ${s^*}''(q_c) \neq 2 D_{\text{max}}$ -- otherwise this leading term is zero and one needs to go beyond this leading order as $q \to q_c$. Using these asympotic behaviors (\ref{asympt1}) in Eq. (\ref{ssec}) one finds the relation
%\bea \label{eq_ssec}
%({s^*}''(q_c) - 2 D_{\text{max}}) (1-2\nu)= 0 \;. 
%\eea

\vspace*{0.5cm}
\noindent {\it The case $0<\nu<1$.} In this case, to analyse the behavior of $s^*(q)$ near $q_c$ beyond the leading order we combine Eqs. (\ref{stilde}) and (\ref{def_qc}) to write
\bea \label{eq_stilde1}
\frac{q_c^2 - q^2}{r} = (\tilde s^*(q)-D_{\text{max}})\int_0^{D_{\text{max}}} dD \frac{W(D)}{(D_{\text{max}}-D)(\tilde s^*(q)-D)} \;.
\eea
Performing the change of variable $u = D_{\text{max}} -D$ one finds that Eq. (\ref{eq_stilde1}) can be re-written as
\bea \label{eq_stilde2}
\frac{q_c^2 - q^2}{r}  = S(q) \int_0^{D_{\text{max}}} du \frac{W(D_{\text{max}}-u)}{u(S(q)+u)} \quad, \quad S(q) = \tilde s^*(q)- D_{\text{max}} \;.
\eea
As $q\to q_c$, $S(q) \to 0$, say as $S(q) \approx B(q_c-q)^\beta$ where $B$ and $\beta$ are yet to be determined. In this limit, the integral over $u$ is thus dominated by its small $u$ behavior, where we can replace $W(D_{\text{max}}-u) \approx A \, u^{\nu}$. To leading order as $q \to q_c$ this equation (\ref{eq_stilde2}) becomes
\bea\label{eq_stilde3}
\frac{q_c^2 - q^2}{r} = A S^\nu \int_0^{D_{\text{max}}/S} dv \frac{v^{\nu}}{v(1+v)}
\eea
For $0<\nu<1$ the upper bound of the integral over $v$ can be sent to $+ \infty$ and the integral can be computed explicitly. We find
\bea \label{eq_stilde3}
S(q) \approx C (q_c-q)^{1/\nu} \;, \; C = \left(2 q_c \frac{\sin(\nu \pi)}{\pi A r} \right)^{1/\nu} \;.
\eea
This leads to the behavior of $s^*(q)$ near $q_c$
\bea \label{eq_stilde4}
s^*(q) = D_{\text{max}} q_c^2-r + 2 D_{\text{max}} q_c(q-q_c)+ D_{\text{max}}(q-q_c)^2 + C q_c^2(q_c-q)^{1/\nu}(1 +o(1)) \;, \; q \to q_c \;.
\eea
Thus we see that, for generic $0<\nu<1$, while the function $\Psi'(q)$ is continuous at $q=q_c$, the function $s^*(q)$ is non-analytic as $q \to q_c$. 

Hence, the leading behavior of $\Psi(q) = s^*(q)$ near $q=q_c$ (with $q<q_c$) can be summarized as follows (see Fig. \ref{Fig_allnu})
\begin{align} \label{summary_ssq}
\Psi(q)- \Psi(q_c) = s^*(q) - s^*(q_c) \approx 
\begin{cases}
& 2 D_{\rm eff}q_c (q-q_c) + B_1(q_c-q)^2 \quad, \quad 2< \nu \\
& 2 D_{\rm eff}q_c(q-q_c) + B_2(q_c-q)^{\nu} \quad, \quad 1 < \nu < 2 \\
& 2D_{\text{max}}q_c(q-q_c)+D_{\text{max}}(q-q_c)^2 + C q_c^2 (q_c-q)^{1/\nu} \quad, \quad 0<\nu<1 \;.
\end{cases}
\end{align}
On the other hand for $q > q_c$, one has from the second line of (\ref{Psi_nupos})
\bea
\Psi(q)- \Psi(q_c) = 2 D_{\max}q_c (q-q_c)+ D_{\max}(q-q_c)^2 \;.
\eea

\begin{figure}[t]
    \centering
\includegraphics[width=0.32\linewidth]{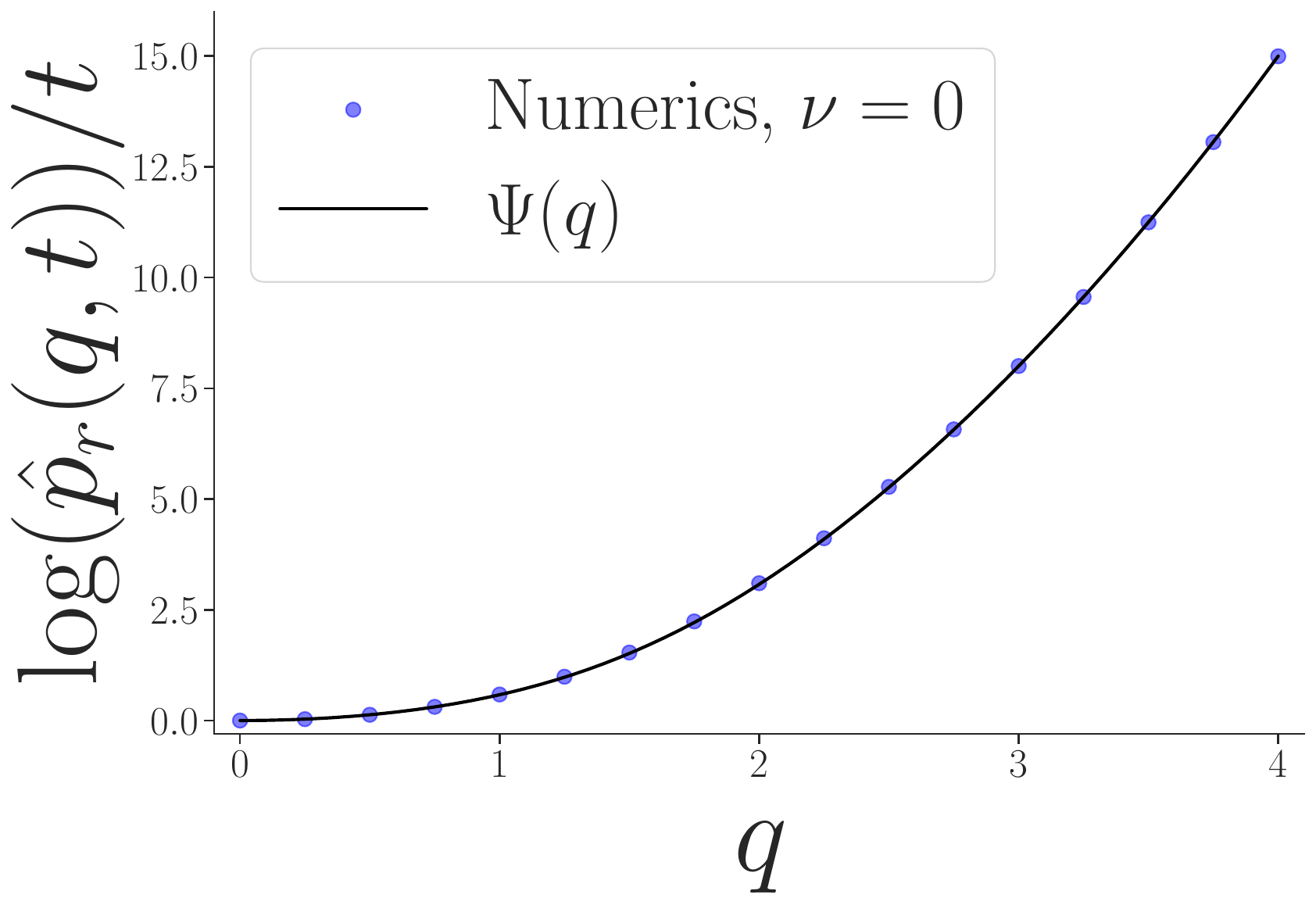}
\includegraphics[width=0.32\linewidth]{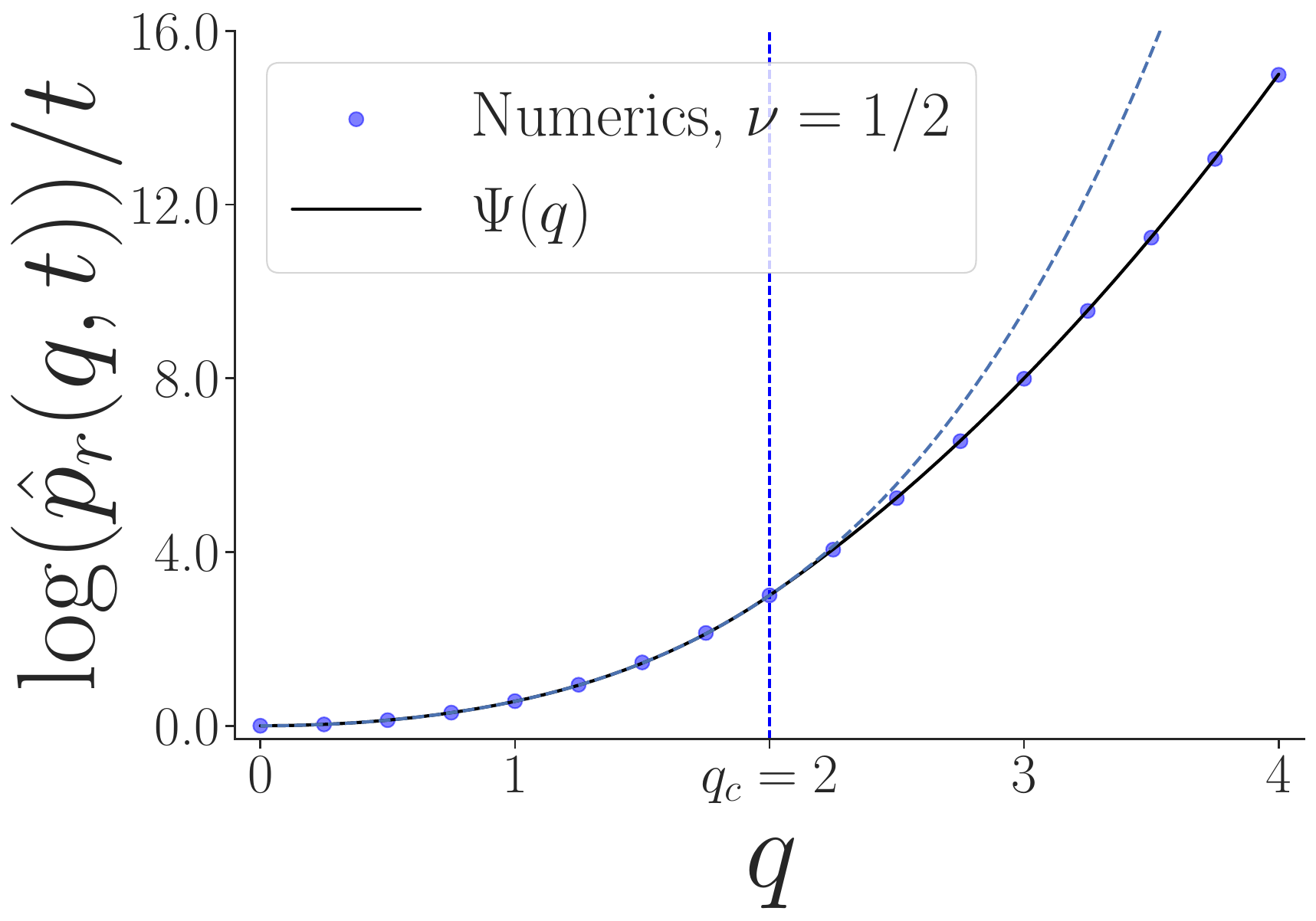}
\includegraphics[width=0.32\linewidth]{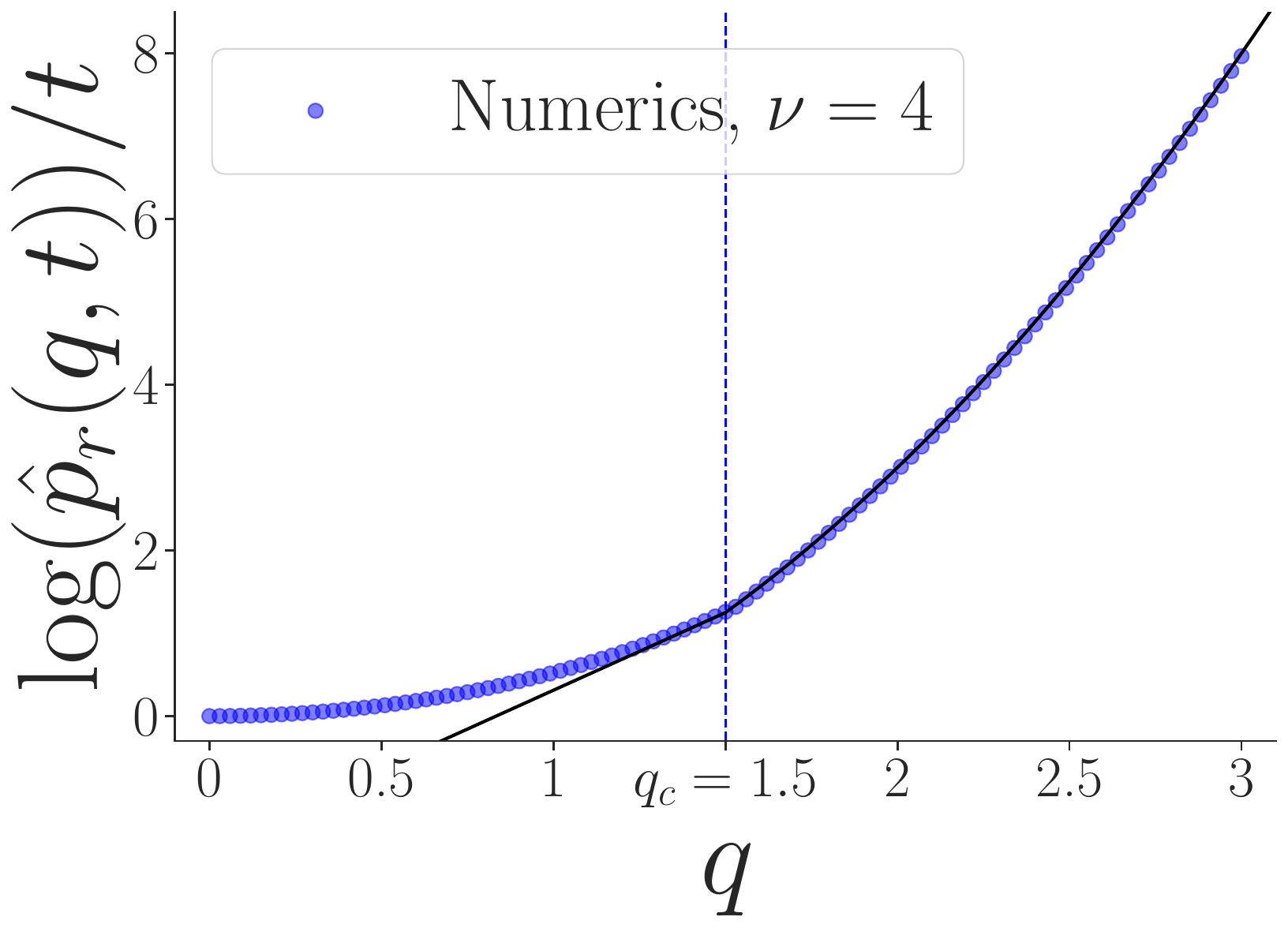}
  \caption{We show here theoretical predictions (solid lines) against numerical results (dots) for $\Psi(q)$ when $W(D) \sim (D_{\text{max}} - D)^{\nu}$ on $[0,1]$ with $r=1$ and $t=1000$. \textbf{Left: $\nu = 0$.} Uniform distribution  (no transition). \textbf{Middle: $\nu = 1/2$.} Wigner distribution (second order transition). The blue dotted line corresponds to the first line of Eq.~(\ref{Psi_nupos}) to emphasize the change of behavior at $q_c$.
  \textbf{Right: $\nu = 4$}. Beta distribution such that $W(D) \sim (1-D)^{4}$ (first order transition). For $q<q_c$, the linear behavior close to $q_c$ is given in Eq.~(\ref{summary_ssq}). For $q>q_c$, we have $\Psi(q) = D_{\max}\, q^2-r$.
  }
  \label{Fig_allnu} 
\end{figure}

\subsection{The rate function $I(y)$}\label{ratefunctionnu}

In this section, we study the behavior of the rate function $I(y)$ which is given by the Legendre transform of $\Psi(q)$, namely
\bea \label{Legendre_sm}
I(y) = \max_{q \in \mathbb{R}} \left(qy - \Psi(q)\right) \;.
\eea
Below we study separately the different cases: (i) $-1 < \nu <0$, (ii) $0<\nu < 1$ and (iii) $\nu > 1$.

\vspace*{0.5cm}
\noindent{\it The case $-1<\nu \leq 0$:} In this case the function $\Psi(q) = s^*(q)$ is regular and its asymptotic behaviors are given in Eq.~(\ref{asymptpsi_nu_neg}). In this case, the Legendre inversion (\ref{Legendre_sm}) can be written as
\bea \label{Legendre2}
I(y) = q_{\rm max} y - s^*(q_{\rm max}) \;, 
\eea
where $ q_{\max} \equiv q_{\max}(y) $ is solution of 
\bea \label{qmax}
y = {s^*}'(q_{\max}) \;.
\eea
Since ${s^*}'(q)$ is a continuous and smooth function of $q$, its inverse function $q_{\max}$ is also a smooth and continuous function of $y$. Hence $I(y)$ is a smooth and continuous function of $y$ on the whole real axis. From the asymptotic behaviors of $s^*(q)$ given in (\ref{asymptpsi_nu_neg}) one immediately gets the asymptotic behaviors of $I(y)$ for small and large arguments as given in the text.}.

\vspace*{0.5cm}
\noindent{\it The case $0<\nu < 1$:} In this case, the rate function has a singularity at $y=y_c$ given by
\bea \label{def_yc}
y_c = 2 D_{\text{max}} q_c \;,
\eea
and to characterize the singularity of $I(y)$, we need to distinguish $0<\nu <1/2$ and $1/2< \nu < 1$: 

For $0<\nu<1/2$ one finds that when $y \to y_c$ from below, i.e. $y<y_c$, the rate function behaves, to leading order as,
\bea \label{nuless1/2}
I(y)-I(y_c) = q_c(y-y_c) + \frac{1}{4D_{\text{max}}}(y-y_c)^2 + o(y-y_c)^2 \;.
\eea
On the other hand for $\nu>1/2$ one finds
\bea \label{nugeq1/2}
I(y) - I(y_c) = q_c(y-y_c) + \left( \frac{\nu}{C\,q_c^2}\right)^{\frac{\nu}{1-\nu}}(1-\nu) |y-y_c|^{\frac{1}{1-\nu}} \;.
\eea
Finally, in the special case $\nu =1/2$ one finds
\bea \label{I_nu1/2}
I(y) - I(y_c) = q_c(y-y_c) + \frac{1}{4(D_{\max}+C\,q_c^2)} (y-y_c)^2  + o(y-y_c)^2 \;.
\eea

\vspace*{0.5cm}
\noindent{\it The case $1<\nu$}. Here, the rate function $I(y)$ has two singular points at $y=y_1$ and $y=y_2=y_c$, namely
\bea \label{def_y1}
y_1 = 2 D_{\rm eff}\, q_c \quad, \quad y_2 = 2 D_{\text{max}} q_c \;.
\eea
On the different intervals $I(y)$ is given for $y\geq 0$ -- note that $I(-y) = I(y)$ -- by
\bea \label{nugeq1}
I(y) = 
\begin{cases}
&q_{\max} y - s^*(q_{\max}) \;, \; 0<y<y_1 \, ,\\
&r + q_c(y-D_{\text{max}}q_c) \;, \; y_1 < y < y_2 \, ,\\
& r + \frac{y^2}{4 D_{\text{max}}} \;, \; y > y_2 \;. 
\end{cases}
\eea
We have explicitly checked that $I(y) > 0$ for all $y >0$. In particular, for the case, $y=y_1$, one can use the explicit expression for $D_{\text{eff}}$ in Eq.~(\ref{nu_geq1}). It is easy to see that, at $y=y_2$, the first derivative of $I(y)$ is continuous while the second is not. However, at $y_1$ the situation is a bit more complicated and depends on $1<\nu<2$ or $\nu >2$:

For $1 <\nu < 2$ one finds
\bea \label{nuless2}
I(y) - I(y_1) = q_c(y-y_1) - \frac{\nu-1}{\nu}\left( \frac{y-y_1}{\nu B_1}\right)^{\frac{\nu}{\nu-1}}  \;, \; y \to y_1 \;\quad {\rm with} \quad \; y<y_1 \;.
\eea
In this case we see that the second derivative is vanishing to the left of $y_1$ and the second derivative is thus continuous. 

On the other hand for $\nu>2$ one finds
\bea \label{nugeq2}
I(y) - I(y_1) = q_c(y-y_1) + \frac{1}{2 s''(q_c)} (y-y_1)^2 \;, \; y \to y_1 \;\quad {\rm with} \quad \; y<y_1 \;.
\eea
In this case we see that the second derivative is finite but discontinuous at $y=y_1$. 

\begin{figure}[t]
    \centering
\includegraphics[width=0.49\linewidth]{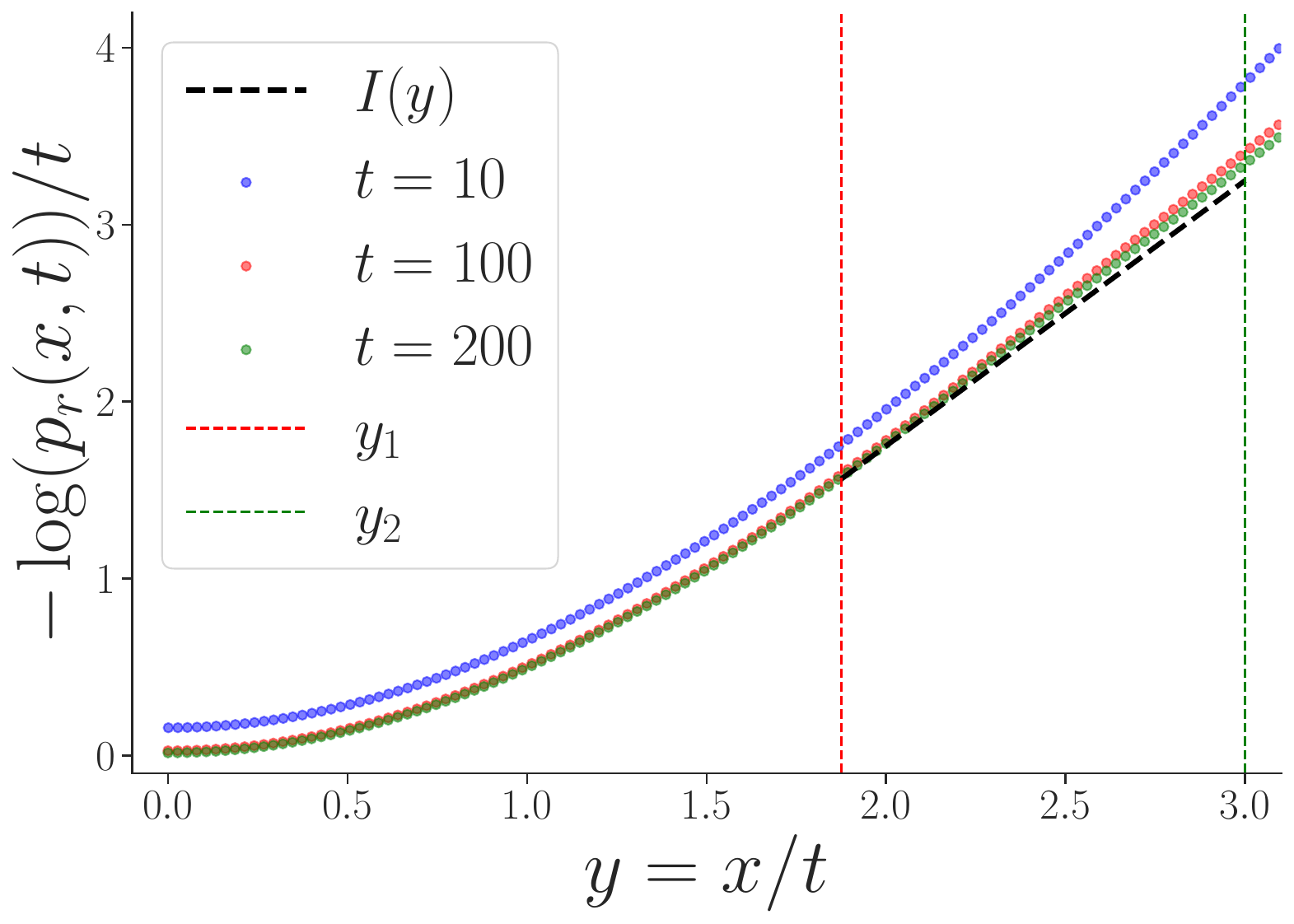}
  \caption{Plot of $-\log(p_r(x,t))/t$ vs $y=x/t$ when $\nu = 4$, i.e. $W(D\to D_{\max}) \sim (D_{\max}-D)^4$. This is a check of the exponential regime given by the second line of Eq.~(\ref{nugeq1}) when $y_1<y<y_2$. The dashed solid line corresponds to the analytic prediction, while the dots are numerical results for different values of $t$. As expected, as $t$ increases, the numerical prediction converges toward $I(y_1<y<y_2)$, although the convergence is seemingly slow. The corresponding values of the probabilities are as small as $10^{-300}$. The parameters are $D_{\text{max}}=1$ and $r=1$.}
  \label{nu4fig} 
\end{figure}

In Fig.~\ref{nu4fig}, we numerically verify convergence to the exponential regime for a beta distribution with $\nu = 4$.

\section{The case where $W(D)$ has an infinite support $[0,+\infty)$}\label{Winfinitesupport}

In this case, it is more convenient to define $\hat p_r(k,t)$ as the Fourier transform $\hat p_r(k,t) = \langle e^{i k x}\rangle$ instead of the generating function (or bilateral Laplace transform) $\hat p_r(k,t) = \langle e^{k x}\rangle$. This is because, when $W(D)$ has an infinite support, the integrals in (\ref{fourierRenewal}) might diverge. In particular, if $W(D) = e^{-D}$ for $D\geq 0$, then the integral
\begin{eqnarray}
    \int_0^{+\infty}dD\, W(D)\, e^{Dq^2t} = \int_0^{+\infty}dD\, e^{D(q^2 t - 1)}
\end{eqnarray}
is diverging beyond a certain time $t$ (for fixed $q$). By working in Fourier space, a global minus sign appears inside the exponential, ensuring the convergence of the integrals. From Eq.~(\ref{renewalpropa3}), we obtain in Fourier space
\begin{eqnarray}
  &&\hat{p}_r(k,t) = e^{-rt}\, \int_0^{+\infty}dD\, W(D)\, e^{-D k^2 t}  + \int_0^{+\infty}dD\, W(D)\, \int_0^{t} d\tau\, r\, e^{-r \tau}\, e^{-D k^2 \tau}\, \hat{p}_r(k,t-\tau)\, ,\\
  &&\hat{p}_r(k,t)= \int_{-\infty}^{+\infty}dx\, e^{ikx}\, p_r(x,t)\, .
  \label{fourier_integral_equation}
\end{eqnarray}
Again, the convolution structure in time can be used to derive an explicit solution via Laplace transformation
\begin{eqnarray}
  &&\hat p_r(k,t) = \int_0^{+\infty}dD\, \frac{W(D)}{D k^2 + r + s}  + r\, \int_0^{+\infty}dD\, \frac{W(D)}{D k^2 + r + s}\, \hat p_r(k,t) \, ,\\
  &&\hat p_r(k,t) = \int_{0}^{+\infty}dt\, e^{-s t}\, \hat{p_r}(k,t)\, .
\end{eqnarray}
Eventually, the explicit solution is given by
\begin{eqnarray}
  \hat p_r(k,t) = \frac{\int_0^{+\infty}dD\, \frac{W(D)}{Dk^2 + r+ s}}{1-r\, \int_0^{+\infty}dD\, \frac{W(D)}{Dk^2 + r+ s}}\, .
  \label{explicitLaplaceaveraged_inf}
\end{eqnarray}
The large time behavior of the Fourier transform can be obtained by taking the inverse Laplace transform as follows
\begin{eqnarray}
    \hat{p}_r(k,t) = \frac{1}{2i\pi}\, \int_{\gamma -i\infty}^{\gamma +i\infty}ds\, e^{st}\, \hat p_r(k,t) = \frac{1}{2i\pi}\, \int_{\gamma -i\infty}^{\gamma +i\infty}ds\, e^{st}\, \frac{\int_0^{+\infty}dD\, \frac{W(D)}{Dk^2 + r+ s}}{1-r\, \int_0^{+\infty}dD\, \frac{W(D)}{Dk^2 + r+ s}} \isApproxTo{t \to \infty} e^{t\tilde \Psi(k)}\, ,
\end{eqnarray}
where $\gamma$ is a real chosen such that all the singularities of $\hat p_r(k,t)$ in the complex $s$-plane are to the left of the Bromwich contour $(\gamma-i\infty, \gamma+i\infty)$, and $\tilde \Psi(k)$ is the singularity of $\hat p_r(k,t)$ with the largest real part. The numerator of $\hat p_r(k,t)$ has a branch cut for $s \in ]-\infty,-r]$ (since here $D_{\max} = + \infty$). Under certain conditions (specified below), the denominator may have a pole $s^* > -r$ located to the right of the branch cut. If such a pole exists, it is the singularity with the largest real part and $\tilde \Psi(k)$ is therefore determined by
\begin{eqnarray}
    1=r\, \int_0^{+\infty}dD\, \frac{W(D)}{Dk^2 + r+ \tilde \Psi(k)}\, ,
    \label{fourierPole}
\end{eqnarray}
which can also be written in terms of the Cauchy-Stieltjes transform $g(x)$ as
\begin{eqnarray}
    \frac{-k^2}{r} = g\left(-\frac{r+\tilde \Psi(k)}{k^2}\right) \quad \text{where} \quad g(x) = \int_0^{+\infty}dD\, \frac{W(D)}{x - D}\, ,
\label{poleFourier}
\end{eqnarray}
with $x\in ]-\infty,0[$. The function $g(x)$ is a strictly decreasing function of $x$ with $g(x\to -\infty)=0$. Whether Eq.~(\ref{poleFourier}) has a solution or not depends on the behavior of the Cauchy-Stieltjes transform $g(x)$ when $x\to 0^-$. As $W(D)$ is a normalizable probability density function, the second integral on the right-hand is well defined (one can indeed set the upper bound of the integral to $+ \infty$). To determine the behavior of the integrand of the first integral when $D\to0$, and when $\epsilon \to 0$, let us assume that $W(D\to0) \approx \alpha\,  D^\nu$, with $\alpha >0$, and $\nu > -1$. It is easy to see that 
\begin{eqnarray}
g(x) \isApproxTo{x\to 0^-}
\begin{cases}
\text{const. } < 0 &\quad , \quad \nu > 0 \;, \\
-\infty &\quad , \quad -1 < \nu \leq 0 \;.
\end{cases}
\end{eqnarray}
%When $\alpha > 0$, the first integrand on the right-hand side causes the integral to diverge. Consequently, in this case, Eq.~(\ref{poleFourier}) always has a solution. However, when $\alpha = 0$, as may occur for a Gamma distribution, the behavior depends on the value of $\nu$. If $\nu < 0$, the integral still diverges, and Eq.~(\ref{poleFourier}) again always has a solution. Conversely, for $\nu > 0$ and $\alpha = 0$, the integral converges. 
Thus, for sufficiently small values of $k$, $\tilde \Psi(k)$ is given by the solution of Eq.~(\ref{fourierPole}) (which exists for all $\nu >-1$).

When $\tilde \Psi(k)$ is given by Eq.~(\ref{fourierPole}), using the fact that for a real probability measure $W(D)$, we have the following relation between the \textit{R-transform} and the Cauchy-Stieltjes transform (see e.g. Theorem 9.23 of \cite{freeBanica})
\begin{eqnarray}
    g\left[R(\xi) +\frac{1}{\xi}\right]=\xi\, ,
\end{eqnarray}
then, it is easy to show that
\begin{eqnarray}
    \tilde \Psi(k) = \lim_{t \to \infty} \frac{\ln \hat p_r(k,t)}{t} = -k^2\, R\left(-\frac{k^2}{r}\right)\, .
\end{eqnarray}

\section{Calculation of the pre-exponential factor of $\hat{p_r}(q,t)$ and the $O(1)$ corrections to the cumulants}\label{section:preexp}
We have shown in Eq.~(\ref{residue}) that the large time behavior of $\hat{p_r}(q,t)$ is given by
\bea \label{residue2}
\hat{p_r}(q,t) \isApproxTo{t\to \infty} \frac{1}{r^2 |\partial_s J_r(s^*(q),q)|} e^{t s^*(q)} \quad {\rm where} \quad J_r(q,s)= \int_0^{D_{\text{max}}} dD\, \frac{W(D)}{r+s-Dq^2}\, .
\eea
We will first compute $\partial_s J_r(s^*(q),q)$, and then deduce from Eq.~(\ref{residue2}) the $O(1)$ corrections to the cumulants.  

{By taking a derivative of Eq. (\ref{sstar}) with respect to $q$, one finds
\bea \label{identity_Jprime}
0 = - {s^*}'(q) \int_0^{D_{\text{max}}} dD \frac{W(D)}{(r+s^*(q)-Dq^2)^2} + 2 q \int_0^{D_{\text{max}}} dD \frac{D\, W(D)}{(r+s^*(q)-Dq^2)^2} \;.
\eea
One can then rewrite the second term as
\bea \label{trick}
2 q \int_0^{D_{\text{max}}} dD \frac{D\, W(D)}{(r+s^*(q)-Dq^2)^2} &=& -\frac{2}{q} \int_0^{D_{\text{max}}} \frac{W(D)}{(r+s^*(q)-Dq^2)} + \frac{2(r+s^*(q))}{q} \int_{0}^{D_{\text{max}}} dD \frac{W(D)}{(r+s^*(q)-Dq^2)^2} \nonumber \\
&=&  -\frac{2}{q\, r} + \frac{2(r+s^*(q))}{q} \int_{0}^{D_{\text{max}}} dD \frac{W(D)}{(r+s^*(q)-Dq^2)^2} \;,
\eea
where we used the fact that $J_r(q,s^*(q))=1/r$.
Finally, using this identity, together with the relation
\bea \label{partial_J}
\partial_s J_r(s^*(q),q) = - \int_{0}^{D_{\text{max}}} dD \frac{W(D)}{(r+s^*(q)-Dq^2)^2}\, , 
\eea
in Eq. (\ref{identity_Jprime}) one finds the following identity
\bea \label{identity}
\frac{1}{r^2 |\partial_s J_r(s^*(q),q)|} = 1  + \frac{1}{r} \left(s^*(q) - \frac{1}{2} q {s^*}'(q) \right) \;.
\eea
We can now inject this identity in Eq. (\ref{residue}) and use the fact that $s^*(q) = q^2 R(q^2/r)$ to write $\hat p_r(q,t)$ at leading order as
\bea\label{residue_2}
\hat p_r(q,t) \approx \left(1 - \frac{q^4}{r^2} R'\left( \frac{q^2}{r}\right) \right) e^{t q^2 R\left(\frac{q^2}{r} \right)} \;.
\eea
This result allows us to calculate the $O(1)$ corrections to the cumulants. To proceed, we write, using (\ref{series_cum})
\bea \label{Rprime}
\frac{q^4}{r^2} R'\left( \frac{q^2}{r}\right) = \sum_{n=1}^\infty \left( \frac{q^2}{r}\right)^{n} (n-1)\kappa_n(D) \;.
\eea
To compute the cumulants, we need to expand the logarithm of $\hat p_r(q,t)$ in Eq. (\ref{residue_2}}). For that purpose, we use the identities
\bea \label{exp_log}
\ln \left( 1 - \frac{q^4}{r^2} R'\left( \frac{q^2}{r}\right)\right) = -\sum_{m=1}^\infty \frac{1}{m} \left(\frac{q^4}{r^2} R'\left(\frac{q^2}{r} \right) \right)^m = - \sum_{m=1}^\infty \frac{1}{m} \left[\sum_{n=1}^\infty \left( \frac{q^2}{r}\right)^{n} (n-1)\kappa_n(D)\right]^m \;.
\eea
We can then finally use the property in (\ref{bellexpansion}) to obtain
\bea \label{order1}
\langle x^{2n}(t)\rangle_c \approx \frac{(2n)!}{r^{n-1}}\, \kappa_n(D)\,t  - \frac{(2n)!}{r^n} \sum_{m=1}^n\frac{1}{m} \hat B_{n,m}\left(\tilde \kappa_1(D), \cdots, \tilde \kappa_{n-m+1}(D) \right) + O(e^{- r t}) \;, \; \tilde \kappa_n(D) = (n-1) \kappa_n(D) \, ,
\eea
where $\kappa_n(D)$ is the $n$-th free cumulant of $D$. Note that the same relation can be derived for $W(D)$ with an infinite support in the framework of the Fourier transform (as done in Section \ref{Winfinitesupport}).

\section{Two specific examples of $W(D)$ with a finite support}\label{specificexamples}

In this section, we derive explicit results for the scaled cumulant generating function (SCGF) $\Psi(q)$, and the rate function $I(y)$, in two specific cases: (i) when $W(D)$ is a uniform distribution ($\nu = 0$), and (ii) when it is a semi-circle distribution ($\nu = 1/2$).

\subsection{Uniform distribution - Case $\nu = 0$}\label{uniformsection}
Let us consider the case where the diffusion coefficients are uniformly distributed such as 
\begin{eqnarray}
W(D) = \frac{1}{D_{\text{max}}}, \quad \text{for} \quad D \in [0, D_{\text{max}}]\, .
\end{eqnarray}
To compute the SCGF, it suffices to determine the \textit{R-transform} associated to the distribution $W(D)$ and use the relation $\Psi(q) = q^2 R\left( \frac{q^2}{r}\right)$. To proceed, we first calculate the Cauchy-Stieltjes transform of $W(D)$ which is given by
\begin{eqnarray}
    g(z)= \frac{1}{D_{\text{max}}}\int_0^{D_{\text{max}}}\, dD \, \frac{1}{z-D}=-\frac{1}{D_{\text{max}}}\log\left(1-\frac{D_{\text{max}}}{z}\right)\, ,
\end{eqnarray}
and we have
\begin{eqnarray}
    g(z) = w \iff z = \frac{D_{\text{max}}}{1-e^{-D_{\text{max}}w}}\, .
\end{eqnarray}
We can then use the identity $R\left(g(z)\right) +\frac{1}{g(z)}=z$ to find that the \textit{R-transform} of a uniform distribution is given by
\begin{eqnarray}
    R(w)= \frac{D_{\text{max}}}{1-e^{-D_{\text{max}} w}}-\frac{1}{w}\, .
\end{eqnarray}
Hence,
\begin{eqnarray}
    \Psi(q) = \frac{D_{\text{max}} q^2}{1-e^{\frac{-D_{\text{max}} q^2}{r}}}-r\, .
\end{eqnarray}
To obtain the rate function, we need to solve the maximization problem
\begin{eqnarray}
    I(y) = \max_{q \in \mathbb{R}} \left(qy - \Psi(q)\right) = q^*y-\Psi(q^*) \;,
\end{eqnarray}
where $q^* \equiv q^*(y)$ is a function of $y$ which is implicitly defined as the solution of
\begin{eqnarray}
    y = -\frac{q^* D_{\text{max}} \left( r - e^{\frac{q^{*^2} D_{\text{max}}}{r}} r + q^{*^2} D_{\text{max}} \right)}{r \left(-1 + \cosh \left(\frac{q^{*^2} D_{\text{max}}}{r}\right)\right)}\, .
\end{eqnarray}
Unfortunately, we cannot solve this equation but we can extract the asymptotic behavior of $q^*(y)$. As the right-hand side is a monotonically increasing function of $q^*$, we can extract the small (resp. large) $y$ behavior of $q^*(y)$ by expanding the right-hand side at small (resp. large) $q^*$ values. Doing so leads to
\begin{eqnarray}
I\left(y\right) =  \begin{cases}
\frac{y^2}{2 {D_{\text{max}}}}+o(y^2)\quad \, , \quad y \to 0 \\
\\
\frac{y^2}{4 {D_{\text{max}}}}+ r + o(y^2)\quad \, , \quad y \to \infty
\end{cases}\, .
\label{LD_unif_asymp}
\end{eqnarray}
The behavior $y\to 0$ in (\ref{LD_unif_asymp}) simply corresponds to the typical fluctuations of the Gaussian where $I(y) = \frac{y^2}{4 \langle D \rangle t }$, with $\langle D \rangle = D_{\text{max}}/2$. On the other hand, the $y \to \infty$ limit corresponds to trajectories that have not experienced any switches (with probability $e^{-rt}$), and have diffused with the maximum diffusion coefficient $D_{\text{max}}$. As predicted in Section \ref{ratefunctionnu}, for $-1<\nu \leq 0$, the function $I\left(y\right)$ interpolates smoothly between the two regimes. Note that the result in Eq.~(\ref{LD_unif_asymp}) obtained for the uniform distribution is in perfect agreement with the general result given in Eq.~(8) of the Letter which is valid for any distribution $W(D)$ with a finite support.

\subsection{Wigner semi-circle distribution - $\nu = 1/2$}
\begin{figure}[t]
    \centering
\includegraphics[width=0.49\linewidth]{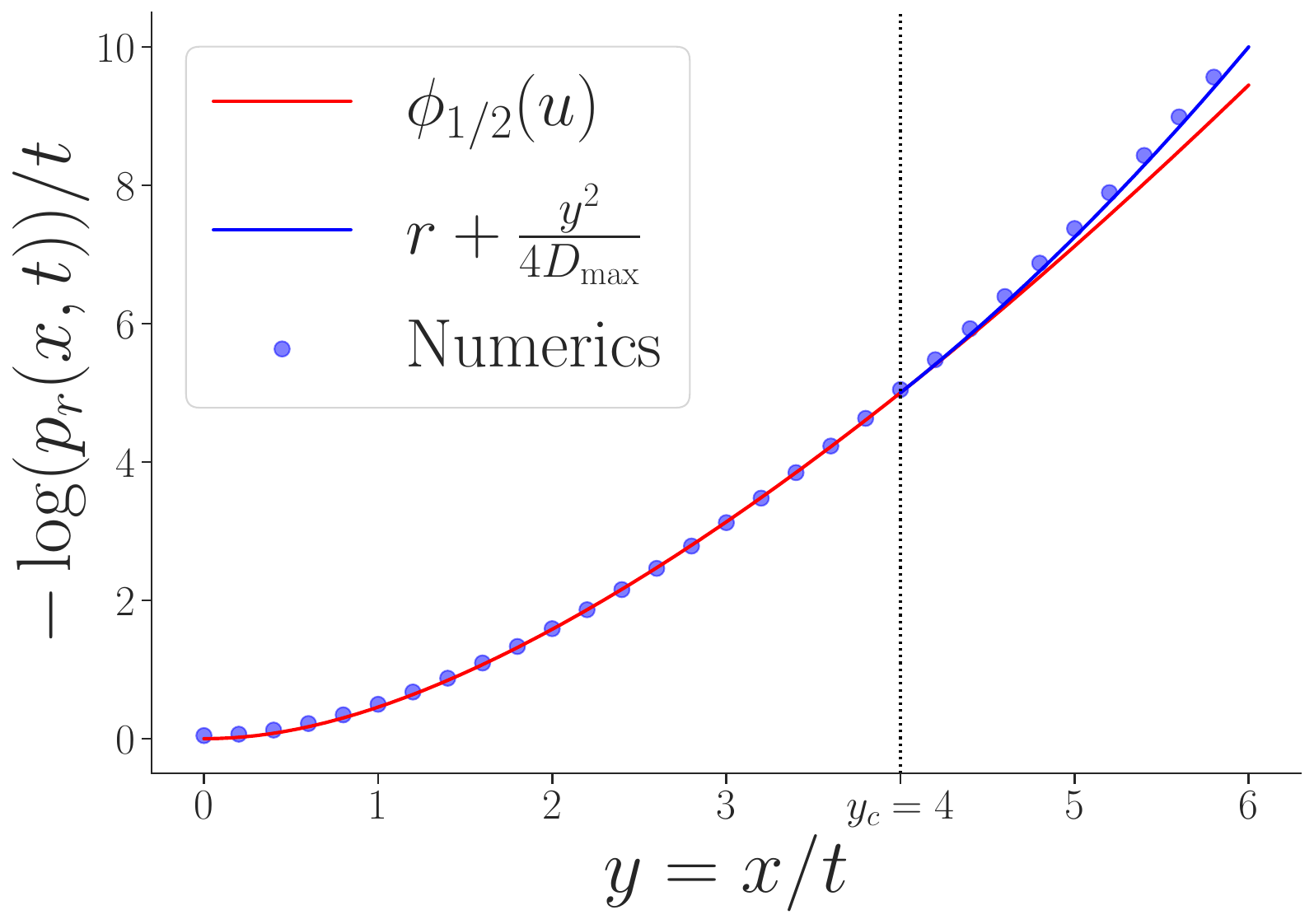}
  \caption{Plot of $-\log(p_r(x,t))/t$ vs $y=x/t$ when $W(D)$ is the Wigner semi-circle defined in Eq.~(\ref{wigner_distri}). For $y<y_c = 4\sqrt{r D_{\text{max}}}$, the rate function is given by $\phi_{1/2}(y)$ -- see Eq.~(\ref{wigner_LDfunction}). For $y>y_c$, it is simply given by $r+y^2/(4D_{\text{max}})$. The solid line corresponds to the analytic prediction, while the dots are numerical results. The agreement with numerics is very good. The corresponding values of the probabilities are as
small as $10^{-200}$. The parameters are $D_{\text{max}}=1$, $r=1$, and $t=50$.}
  \label{wignerratefig} 
\end{figure}

The Wigner distribution corresponds to the case $W(D) \sim (D_{\text{max}}-D)^\nu$, with $\nu =1/2$. The PDF is indeed given by
\begin{eqnarray}
    W(D)= \frac{8}{\pi {D^2_{\text{max}}}} \sqrt{D(D_{\max}-D)} \quad, \quad 0 \leq D \leq D_{\max}
    \label{wigner_distri}
\end{eqnarray}
As demonstrated in section \ref{nug0}, when $\nu = 1/2$, the SCGF has a transition at $q_c$ and it is given by
 \bea
\Psi(q) = 
\begin{cases}
q^2 R\left( \frac{q^2}{r}\right) = r \sum_{n \geq 1} \left(\frac{q^2}{r} \right)^{n} \kappa_n(D)\;, & \quad q < q_c \;, \\[10pt]
D_{\text{max}} q^2-r\;, & \quad q > q_c \;,
\end{cases}
\label{Psi_nupos}
\eea
where $q_c = \sqrt{4r/D_{\text{max}}}$ is determined by Eq.~(\ref{def_qc}). It is well known that in free probability theory, the Wigner semi-circle distribution plays the same role as the Gaussian distribution in classical probability theory in the sense that all its free cumulants $\kappa_n(D)$ vanish for $n>2$. It is easy to compute the first free cumulants, for instance using the formulae given in the End Matter of the letter. They are given by $\kappa_1(D)=D_{\text{max}}/2$ and $\kappa_2(D) =D_{\text{max}}^2/16 $ such that
 \bea
R(z) = \frac{D_{\max}}{2} + z \frac{D^2_{\max}}{16} 
\Longrightarrow
\Psi(q) = 
\begin{cases}
\frac{D_{\text{max}}}{2}q^2 + \frac{D_{\text{max}}^2}{16 }\frac{q^4}{r}\;, & \quad q < q_c \;, \\[10pt]
D_{\text{max}} q^2-r\;, & \quad q > q_c \;.
\end{cases}
\label{Psi_nupos}
\eea
In the middle panel of Fig. \ref{Fig_allnu}, we have checked the equation above numerically. As shown in Section \ref{ratefunctionnu} for $\nu =1/2$, the rate function $I(y)$ exhibits a second order transition at $y_c = 2 D_{\text{max}}q_c = 4\sqrt{r D_{\text{max}}}$. We have
\bea
I(y) = 
\begin{cases}\label{IrateWigner}
\phi_{1/2}(y) &\quad, \quad y \leq y_c \\
 r + \frac{y^2}{4 D_{\text{max}}} &\quad, \quad y \geq y_c \;,
\end{cases}
\eea
where 
\bea 
\phi_{1/2}(y) = \max_{q \in \mathbb{R}} \left(qy - \frac{D_{\text{max}}}{2}q^2 - \frac{D_{\text{max}}^2}{16 }\frac{q^4}{r}\right) \;.
\eea
Hence, one needs to solve the following equation for $q^*$
\bea
y = D_{\text{max}} q +\frac{D_{\text{max}}^2}{4}\frac{{q^*}^3}{r}\, .
\eea
This equation has two complex roots and one real root. The large deviation function is real only for the real root, which is given by
\begin{eqnarray} 
q^*(y) = \frac{2 \cdot 6^{\frac{2}{3}}\,  D_{\text{max}}^3 r - 6^{\frac{1}{3}} \left(rD_{\text{max}}^4 \Delta\right)^{\frac{2}{3}}}{3 D_{\text{max}}^2 \left(r D_{\text{max}}^4 \Delta\right)^{\frac{1}{3}}}\quad , \quad \Delta = -9y + \sqrt{48 D_{\max} r + 81 y^2}\, .
\end{eqnarray}
such that we obtain
\begin{eqnarray}
\phi_{1/2}(y) &=& \frac{r^{\frac{1}{3}} \left(-2 \cdot 6^{\frac{2}{3}} \cdot (D_{\text{max}} r)^{\frac{1}{3}} + 6^{\frac{1}{3}} \Delta^{\frac{2}{3}}\right)}{36 \cdot D_{\text{max}}^{\frac{2}{3}} \Delta^{\frac{4}{3}}} \times \left[81 y^2 - 9y \sqrt{48 D_{\text{max}} r + 81 y^2} + 9 \cdot 6^{\frac{1}{3}} y \left(D_{\text{max}} r \Delta \right)^{\frac{1}{3}} \right.\nonumber\\
&&\left.- 2^{\frac{1}{3}} \cdot 3^{\frac{5}{6}} \sqrt{16 D_{\text{max}} r + 27 y^2} \left(D_{\text{max}} r  \Delta \right)^{\frac{1}{3}} + 2 \cdot 6^{\frac{2}{3}} \left(D_{\text{max}} r  \Delta \right)^{\frac{2}{3}}\right]\, .
\label{wigner_LDfunction}
\end{eqnarray}
As expected, one can check that at small argument, we retrieve the Gaussian fluctuations $\phi_{1/2}(y) = \frac{y^2}{4\langle D \rangle}+o(y^2)$, where $\langle D \rangle = D_{\text{max}}/2 $. In Fig. \ref{wignerratefig}, we numerically verify our prediction for the rate function in Eq.~(\ref{IrateWigner}) with an accuracy up to $10^{-200}$. From these explicit expressions, one can check that the rate function $I(y)$ as well as its derivative are continuous at $y=y_c$. However, $\phi_{1/2}''(y_c) = 1 /(4D_{\max})$ while $I''(y\to y_c^-)=1/(2D_{\max})$ (with $y>y_c$), hence clearly the second derivative is discontinuous at $y_c$. This is consistent with the relation $I''(y_c)=1/\Psi''(q_c)$ where $\Psi''(q\to q_c)=4D_{\max}$ can be computed using the third line of Eq.~(\ref{summary_ssq}) for $q<q_c$.

\subsubsection{Correction of order $O(1)$ to the cumulants}

We can use the formula (\ref{residue_2}) to compute the $O(1)$ corrections to the cumulants in the large time limit. Indeed using that the $R$-transform of the Wigner distribution on $[0,D_{\max}]$ is given by
\bea
R(z) = \frac{D_{\max}}{2} + z \frac{D^2_{\max}}{16} \;,
\eea
one obtains from (\ref{residue_2})
\bea \label{ptilde_Wigner}
\tilde p_r(q,t) \approx \left(1 - \frac{q^4}{r^2} \frac{D^2_{\max}}{16} \right) e^{t \left(q^2 \frac{D_{\max}}{2} + q^4 \frac{D^2_{\max}}{16\,r} \right)} \;.
\eea
This allows us to compute the cumulants up to order $O(1)$ in the large $t$ limit, leading to
\bea \label{cumul_Wigner}
\langle x^2(t) \rangle_c = D_{\max}\,t \quad, \quad  \langle x^{4n}(t) \rangle_c \approx \begin{cases} 
&\frac{3}{2} \frac{D^2_{\max}}{r}(t-1/r) + O(e^{-rt}) \;, \; n=1 \\
& - \frac{(4n)!}{n} \left(\frac{D^2_{\max}}{16 r^2} \right)^n + O(e^{-rt})  \;, \; n>1
\end{cases}
\eea
while the other cumulants are exponentially small, i.e., of order $O(e^{-r t})$ or smaller.

\section{Discussion of the dynamical transitions for bounded $W(D)$}
We consider a distribution $W(D)$ supported over a finite interval $[0,D_{\max}]$ and we assume that, near $D = D_{\max}$, the distribution $W(D)$ behaves as 
\bea \label{def_nu}
W(D) \sim (D_{\max}-D)^{\nu} \;, \; \nu > - 1 \;.
\eea
For $-1<\nu < 0$, the distribution $W(D)$ has an integrable divergence as $D \to D_{\max}$, while at $\nu = 0$, it approaches a constant as $D \to D_{\max}$. In contrast, for $\nu >0$, the distribution $W(D)$ vanishes as $D \to D_{\max}$. For $\nu > 1$, not only $W(D)$ but also its derivative $W'(D)$ vanishes at the upper edge $D_{\max}$. As discussed in the main text, we find different behaviors for the position distribution $p_r(x,t)$, depending on the exponent $\nu>-1$ summarized in Eqs. (14) and (15) in the main text. 
\begin{itemize}

\item[$\bullet$]{$-1 < \nu \leq 0$: in this case we find $p_r(x,t) \sim e^{- t I(y=x/t)}$ where the function $I(y)$ is a smooth function, interpolating smoothly between the two limiting behaviors 
\bea \label{asympt_I_sm}
I(y) \approx
\begin{cases}
& \frac{y^2}{4 \langle D \rangle} \quad \quad \quad \;, \quad y \to 0 \;,
 \\
& r + \frac{y^2}{4 D_{\max}} \quad, \quad y \to \infty \;.
 \end{cases}
\eea
The limit $y \to 0$ corresponds to typical trajectories that have undergone a lots of switchings till time $t$. In contrast, the limit $y \to \infty$ corresponds to extremely rare trajectories that start with a diffusion coefficient close to $D_{\max}$ and undergo essentially no switching, till time $t$. The latter occurs with probability $e^{-rt}$ (thus explaining the shift by a constant $r$ in the second line of Eq. (\ref{asympt_I_sm})). The quadratic part in the second line corresponds to standard diffusion with a diffusion coefficient $D_{\max}$. In this case, since $W(D)$ does not vanish close to $D_{\max}$, one can have several trajectories where $D$ is not necessarily $D_{\max}$ but quite close to it, undergoing few switchings. The probability weight coming from such configurations is close to the extreme trajectories. The existence of such ``near-extreme'' trajectories indicates that the position distribution gets smoothly interpolated between contributions from the extreme and the typical trajectories. This is the physical reason behind the absence of any sharp transition in the rate function $I(y)$.  
}
\item[$\bullet$]{$0 < \nu < 1 $:
In contrast to the case discussed above, the rate function $I(y)$ undergoes a sharp transition at $y=y_c$, with the following behaviors
\bea 
\label{ratenu01_sm}
I(y) = 
\begin{cases}
\phi_\nu(y) &\quad, \quad y \leq y_c = 2 D_{\max} q_c \, ,\\
 r + \frac{y^2}{4 D_{\text{max}}} &\quad, \quad y \geq y_c \;,
\end{cases}
\eea
where $\phi_\nu(y)$ is a smooth function and $y_c$ is a constant given in the main text. Around $y=y_c$, the function $I(y)$ is continuous but nonanalytic -- see Section \ref{ratefunctionnu}. Thus in this case the lines $x = \pm y_c t$ in the space-time plane act like a light cone that separates the extremely rare trajectories from the typical ones (see the left panel of Fig. 3 in the main text). This happens when $W(D)$ vanishes as $D \to D_{\max}$, leading to the absence of the intermediate trajectories interpolating between the rare and the typical ones, as in the previous case $-1 < \nu \leq 0$. 
}
\item[$\bullet$]{$\nu \geq 1$:
In this case, the behavior of the rate function is even richer, displaying three different regimes
\bea
\label{nugeq1_sm}
I(y) = 
\begin{cases}
\phi_\nu(y) &, \quad 0 < y < 2 D_{\rm eff} q_c \;, \\
q_c y -\gamma &, \quad 2 D_{\rm eff} q_c < y < y_c \;, \\
r + \frac{y^2}{4 D_{\text{max}}} &, \quad y > y_c \;,
\end{cases}
\eea
where $D_{\rm eff} < D_{\max}$ and $\gamma =D_{\max}q_c^2 - r >0$. Thus, compared to the case $0<\nu < 1$ in (\ref{ratenu01_sm}), there is an additional regime sandwiched between the typical ($y<2 D_{\rm eff} q_c$) and extremely rare trajectories ($y > y_c = 2 D_{\max}q_c$). In this intermediate regime, the position distribution reads 
\bea \label{p_exp}
p_r(x,t) \approx e^{- q_c(x - v\,t)} \quad, \quad 
\eea
where $v = y_c - r D_{\max}/y_c > 0$. In addition, one finds $v<y_c$. Thus the position distribution has the shape of a traveling front with an exponential tail that moves forward with a nontrivial speed $v$. In the space-time plane, we now have two light cones at $x = \pm 2 D_{\rm eff} q_c t$ and $x = \pm y_c t$ (see the right panel of Fig. 3 in the main text). The existence of this new intermediate regime can be traced back to the fact that for $\nu >1$, the derivative $W'(D)$ vanishes as $D \to D_{\max}$: this corresponds to trajectories with a typical $D$ close to $D_{\max}$ which however undergo many switchings compared to the extreme trajectories. As a result of that, the factor $e^{-rt}$ that appears in the extreme tail is absent in this intermediate regime. Thus in the typical regime $|x|< 2 D_{\rm eff} q_c t$, the trajectories are associated with ``small'' values of $D$ but with a large number of switchings. In the rare regime, when $|x|> y_c t$ the trajectories up to time $t$ have a $D$ close to $D_{\max}$ and undergo almost no switchings. In the intermediate regime, when $2 D_{\rm eff} q_c t<|x|<y_c t$, the trajectories up to time $t$ have typically $D$ close to $D_{\max}$ but undergo a large number of switchings.  
}

\end{itemize}

\section{Switching diffusion in higher dimensions}

Let us consider a switching diffusion process in $d$ dimension. Each coordinate $x_i$ follows a switching dynamics
\begin{eqnarray}
    \dot{x}_i(t) = \sqrt{2D(t)}\eta(t)\quad \, , \quad i\in\{1,\ldots,d\}\, .
\end{eqnarray}
We consider the case where the particle starts its motion at the origin such that for all $i$, $x_i(0)=0$. The important point is that the dynamics of the different components $x_i$'s are correlated because they share the same diffusion coefficient and the same switching events.

\subsection{Renewal equation and explicit solution in $d$ dimension}

We define $P_r[\{ x_i\},D,t| {\sf D}_1]$ to be the joint PDF (JPDF) of the $x_i$'s with initial diffusion coefficient ${\sf D}_1$, with final positions $\{x_i\} = (x_1,x_2,\ldots,x_d)$ and final diffusion coefficient $D$ at time $t$. We can write a renewal equation which reads
\begin{eqnarray}
  P_r[\{ x_i\},D,t| {\sf D}_1] &&= e^{-rt}\, \prod_{i=1}^d\frac{e^{-\frac{x_i^2}{4{\sf D}_1t}}}{\sqrt{4\pi {\sf D}_1 t}}\, \delta(D-{\sf D}_1)\,  + \nonumber \\
  &&\int_0^{t} d\tau\, r\, e^{-r \tau} \int_{-\infty}^{+\infty}dy_1\, \ldots \, dy_d \int_0^{+\infty}dD'\, P_r[\{ y_i\},D',t-\tau| {\sf D}_1]\, W(D)\, \prod_{i=1}^d\frac{e^{-\frac{(x_i-y_i)^2}{4D \tau}}}{\sqrt{4\pi D \tau}} \, .
  \label{renewalpropa}
\end{eqnarray}
The first contribution comes from the event, that occurs with probability $e^{-r t}$, where there is no switch up to time $t$ and each component follows a simple Brownian motion with diffusion coefficient ${\sf D}_1$. The second term accounts for the event where the last switch occurred at time $t - \tau$, at which point the component $i$ was at position $y_i$ with diffusion coefficient $D'$. The probability that no reset occurred between $t - \tau$ and $t$ is $e^{-r \tau}$, while the probability of a reset occurring within the small time interval $[t - \tau, t - \tau + d\tau]$ is $r d\tau$. To account for all possible switch times, we integrate over $\tau$. Next, we integrate over $y_i$'s and $D'$, taking into account the propagator $P_r[\{ y_i\},D',t-\tau| {\sf D}_1]$ that describes the paths from the origin $x_i = 0$ at $t = 0$ to position $y_i$ at time $t - \tau$. We also include the Gaussian propagator that governs the motion from $y_i$ at time $t - \tau$ to $x_i$ at time $t$. Finally, we need to account for the transition probability of the diffusion coefficient changing from $D'$ (the value just before the reset at time $t - \tau$) to $D$ (the value immediately after the reset). Since the diffusion coefficients are i.i.d., this is simply given by the distribution $W(D)$. 

The equation (\ref{renewalpropa}) can be simplified by averaging over ${\sf D}_1$, and integrating over $D$. It gives the simpler equation
\begin{eqnarray}
  P_r[\{ x_i\},t] &&= e^{-rt}\, \int_0^{+\infty}dD\, W(D)\,  \prod_{i=1}^d\frac{e^{-\frac{x_i^2}{4Dt}}}{\sqrt{4\pi  D t}} \\ \nonumber
  &&+ \int_0^{t} d\tau\, r\, e^{-r \tau} \int_{-\infty}^{+\infty}dy_1\, \ldots \, dy_d \, P_r[\{ y_i\},t-\tau]\int_0^{+\infty}dD\, W(D)\, \prod_{i=1}^d\frac{e^{-\frac{(x_i-y_i)^2}{4D \tau}}}{\sqrt{4\pi D \tau}} \, .
  \label{renewalpropa2}
\end{eqnarray}
Using the convolution structure in space, one can use the BLT with respect to the positions $\{x_i\}$ to obtain
\begin{eqnarray}
  \hat{P}_r[\{ q_i\},t] &&= e^{-rt}\, \int_0^{+\infty}dD\, W(D)\, e^{|\textbf{q}|^2 D t} + \int_0^{t} d\tau\, r\, e^{-r \tau}  \hat{P}_r[\{ q_i\},t-\tau]\, \int_0^{+\infty}dD\, W(D)\, e^{|\textbf{q}|^2 D \tau} \, ,
  \label{renewalpropafourier}
\end{eqnarray}
where $|\textbf{q}|^2 = \sum_{i=1}^dq_i^2$ and
\begin{eqnarray}
   \hat{P}_r[\{ q_i\},t] =\int_{-\infty}^{+\infty}dx_1\ldots dx_d\, 
   e^{\sum_{i=1}^dq_i x_i}\, P_r[\{ x_i\},t] \;.
\end{eqnarray}
In addition, going in Laplace space with respect to time allows us to exploit the convolution structure in time such that
\begin{eqnarray}
  \tilde{P}_r[\{ q_i\},s] &&= \int_0^{+\infty}dD\, W(D)\, \mathcal{L}_{t\to s} \left[e^{-rt}\, e^{|\textbf{q}|^2 D t}\right] +r\,  \tilde{P}_r[\{ q_i\},s]\, \int_0^{+\infty}dD\, W(D)\, \mathcal{L}_{t\to s}\left[e^{-r t}e^{|\textbf{q}|^2 D t}\right] \, ,
  \label{renewalpropaLaplace}
\end{eqnarray}
where $\mathcal{L}_{t\to s}$ denotes the Laplace transform and
\begin{eqnarray}
   \tilde{P}_r[\{ q_i\},s] = \int_{0}^{+\infty}dt\, e^{-s t}\, \hat{P}_r[\{ q_i\},t]\, . 
\end{eqnarray}
Finally, we obtain a close expression for the JPDF which is given by
\begin{eqnarray}\label{closedpropa}
    \tilde{P}_r[\{ q_i\},s] = \frac{\int_0^{+\infty}dD\, \frac{W(D)}{r+ s-D|\vec{q}|^2}}{1-r\, \int_0^{+\infty}dD\, \frac{W(D)}{r+ s -D|\vec{q}|^2 }}\, .
\end{eqnarray}
This is exactly the solution for the one-dimensional process, Eq.~(\ref{explicitLaplaceaveraged}), with $q$ replaced by its norm. This shows that the distance from the origin in the $d$-dimensional process has the same properties as the one-dimensional process.

\subsection{Mixed cumulants: example in the case $d=2$}
As a consequence of the explicit solution (\ref{closedpropa}), we have the following relation
\begin{eqnarray}
   \hat{P}_r[\{ q_i\},t] = \langle e^{\vec{q}.\vec{x}}\rangle\isApproxTo{t\to \infty} e^{t\Psi(|\vec{q}|)} \quad \, , \quad \Psi(|\vec{q}|) = |\vec{q}|^2R\left(\frac{|\vec{q}|^2}{r}\right)\quad \text{when } \, q\to 0\, .
\end{eqnarray}

For a simple illustration, let us compute the connected two-point function $ \langle x_1^2 x_2^2\rangle - \langle x_1^2\rangle\langle x_2^2 \rangle $. In two dimensions, we have
\begin{eqnarray}
    \langle e^{\vec{q}.\vec{x}}\rangle &=& 1+ \frac{1}{2}\langle  (\vec{q}.\vec{x})^2 \rangle + \frac{1}{4!}\langle  (\vec{q}.\vec{x})^4 \rangle +\cdots\\
    &=& 1+\frac{1}{2}\langle (q_1 x_1 + q_2 x_2)^2 \rangle + \frac{1}{4!} \langle (q_1 x_1 + q_2 x_2)^4)\rangle +\cdots\\
    &=& 1+\frac{1}{2} q_1^2 \langle x_1^2\rangle + \frac{1}{2} q_2^2\langle x_2^2\rangle + \frac{1}{4!} \left(q_1^4\langle x_1^4\rangle + 6 q_1^2 q_2^2 \langle x_1^2 x_2^2\rangle + q_2^4 \langle x_2^4\rangle  \right) +\cdots\, ,
\end{eqnarray}
where one can show that the odd terms vanish by using the fact that the joint distribution of the components is symmetric in each component, and where higher-order terms are neglected.
Taking the logarithm of the generating function leads to
\begin{eqnarray}
    \ln \langle e^{\vec{q}.\vec{x}}\rangle  &=& \ln\left[1+\frac{1}{2} q_1^2 \langle x_1^2\rangle + \frac{1}{2} q_2^2\langle x_2^2\rangle + \frac{1}{4!} \left(q_1^4\langle x_1^4\rangle + 6 q_1^2 q_2^2 \langle x_1^2 x_2^2\rangle + q_2^4 \langle x_2^4\rangle  \right) +\cdots\right]\\
    &=& \frac{1}{2} q_1^2 \langle x_1^2\rangle + \frac{1}{2} q_2^2\langle x_2^2\rangle + \frac{1}{4!} \left(q_1^4\langle x_1^4\rangle + 6 q_1^2 q_2^2 \langle x_1^2 x_2^2\rangle + q_2^4 \langle x_2^4\rangle  \right)\\& -&\frac{1}{2}\left(\frac{1}{4} q_1^4 \langle x_1^2\rangle^2 + \frac{1}{4} q_2^4 \langle x_2^2\rangle^2 +\frac{1}{2} \langle x_1^2\rangle \langle x_2^2\rangle q_1^2 q_2^2\right) +\cdots\, .
\end{eqnarray}
Therefore, the connected two-point function is given by
\begin{eqnarray}
    \frac{\partial^2}{\partial q_1^2}\frac{\partial^2}{\partial q_2^2}\ln \langle e^{\vec{q}.\vec{x}}\rangle \bigg|_{q_i=0} = \langle x_1^2 x_2^2\rangle - \langle x_1^2\rangle\langle x_2^2 \rangle\, .
\end{eqnarray}
On the other hand, we also have
\begin{eqnarray}
    \ln \langle e^{\vec{q}.\vec{x}}\rangle  \isApproxTo{t\to \infty} t \Psi(|\vec{q}|) = t \sum_{n=1}^{+\infty} \kappa_n(D)(q_1^2+q_2^2)^n\, r^{1-n}\, ,
\end{eqnarray}
such that 
\begin{eqnarray}
    \frac{\partial^2}{\partial q_1^2}\frac{\partial^2}{\partial q_2^2}\ln \langle e^{\vec{q}.\vec{x}}\rangle \bigg|_{q_i=0} \isApproxTo{t\to \infty} \frac{2}{r}\, \kappa_2(D) t\, .
\end{eqnarray}
Hence,
\begin{eqnarray}
    \langle x_1^2 x_2^2\rangle - \langle x_1^2\rangle\langle x_2^2 \rangle \isApproxTo{t\to \infty}  \frac{2}{r}\, \kappa_2(D) t\, .
\end{eqnarray}
This computation can straightforwardly be generalized to any mixed cumulants in $d$ dimension.

\section{Numerical method}\label{numericssection}

In this section, we explain the algorithm used to compute numerically the scaled cumulant generating function $\Psi(q)$ and the large deviation function $I(y)$. To obtain high numerical accuracy, we implemented the code in \texttt{Julia}, utilizing the \texttt{BigFloat()} type for arbitrary precision arithmetic.

\subsection{Numerical evaluation of $\Psi(q)$}\label{numericalrenewal}
Recall that the scaled cumulant generating function (SCGF) is defined as $\Psi(q) = \lim_{t \to \infty} \ln \hat{p}_r(q,t)/t$. We will directly compute  numerically $\hat p_r(q,t)$. As shown in Eq.~(\ref{fourierRenewal}), $\hat{p}_r(q,t)$ satisfies an integral equation, which we solve numerically. The equation is given by
\begin{eqnarray}
  &&\hat{p}_r(q,t) = e^{-rt}\,  \hat{G}_0(q,t) + \int_0^{t} d\tau\, r\, e^{-r \tau}\, \hat{G}_0(q,\tau) \, \hat{p}_r(q,t-\tau)\quad \ , \quad \hat{G}_0(q,t) = \int_0^{D_{\text{max}}}dD\, W(D)\, e^{D q^2 t}\, ,\label{numerical_fourier}\\
  &&\hat{p}_r(q,t)= \langle e^{qx} \rangle =\int_{-\infty}^{+\infty}dx\, e^{q x}\, p_r(x,t) \quad \, , \quad \hat{p}_r(0,t)=1\, ,
\end{eqnarray}
where $\hat{p}_r(0,t)=1$ is just the normalization. Here, $\hat{G}_0(q,t)$ represents the bilateral Laplace transform of the Brownian motion propagator with diffusion coefficient $D$, averaged over the distribution $W(D)$. Interestingly, this integral equation resembles that of resetting Brownian motion (rBM) \cite{resettingPRL, resettingReview}. Indeed, the cumulative distribution $Q_r(M,t)$ of the maximum of a rBM starting at the origin up to time $t$ obeys the same integral equation (with the identification $q\to M$), but with a different function $\hat{G}_0(q,t)$. We solve this equation numerically using a recursive method, discretizing time into small intervals $\Delta t$, as explained in Section IV of~\cite{hartmannreset}.

\subsection{Numerical evaluation of $I(y)$}

The rate function is defined as $I(y= x/t) = \lim_{t \to \infty} -\ln p_r(x,t)/t$. To compare with our analytical prediction, we will compute the Fourier transform of the distribution of the position of the particle $\hat p_r(k,t)$ defined in Eq.~(\ref{fourier_integral_equation}), and then invert it. We choose to work in Fourier space because it is easier to numerically invert than the bilateral Laplace transform. We first compute numerically $\hat p_r(k,t)$ as explained in the previous section by replacing $q^2 \to -k^2$. The inverse Fourier transform is approximated as
\begin{eqnarray}\label{invertfouriernumeric}
    p_r(x,t)=\frac{1}{2\pi}\, \int_{-\infty}^{+\infty} dk\, e^{-ikx}\, \hat{p}_r(k,t) = \frac{1}{\pi}\, \int_0^{+\infty} dk\, \cos(kx)\, \hat{p}_r(k,t) \approx \frac{1}{\pi}\, \int_0^{k_{\rm Max}} dk\, \cos(kx)\, \hat{p}_r(k,t)\, ,
\end{eqnarray}
where the second equality we have used the fact that $p_r(x,t)$ is symmetric and real. The last approximation comes from the fact that we cannot integrate up to $+\infty$ as we evaluate $\hat{p}_r(k,t)$ numerically. Therefore, we need to specify an upper bound for the integral which we call $k_{\rm Max}$. We choose the value $k_{\rm Max}$ to achieve the desired precision in our evaluation. As we have a prediction for the tail of $I(y)$ (see Eq.~(10) of the letter), we can estimate the precision required to compute $p_r(x = yt,t)$ for a given $x$ as
\begin{eqnarray}\label{gaussnum}
    p_r(x = yt,t) \isApproxTo{t \to \infty} \exp\left(-\frac{y}{4D_{\text{max}}}\right)\, .
\end{eqnarray}
The value of $k_{\rm Max}$ is then chosen such that $\hat{p}_r(k_{\rm Max},t)$ is of the same order as  the right hand side of Eq.~(\ref{gaussnum}) in order to estimate the integral $\int_0^{k_{\rm Max}} dk\, \cos(kx)\, \hat{p}_r(k,t)$ with the required precision.
%to also be of the same order as $\hat{p}_r(k_{\rm Max},t)$ (as $\cos(kx) \sim O(1)$ and $\hat{p}_r(k,t)$ is a strictly decreasing function of $k$ {\color{red} is this true?}). 
To probe the large deviations, one needs to go at high values of $t$ and $x$ (typically, $x$ is of the order of $10^3$ to $10^4$). Therefore, the cosine in the integral has a really small period and the integrand highly oscillates. To numerically compute the integral in Eq.~(\ref{invertfouriernumeric}), we employ Filon’s method \cite{Filon}, which is effective for oscillatory integrals.

\section{Mapping to a random growth model}

A recent study~\cite{BBLD2025} analyzed a population growth model with broad applications, including ecology, directed polymers, and immunology. In this framework, $x_i(t)$ can, for instance, represent the population of city $i \in [1, N]$. In the mean-field, fully connected limit, the random multiplicative growth model has the following dynamics~\cite{BBLD2025}
\begin{equation}
\frac{dx_i(t)}{dt} = \left(m_i + \sigma \xi_i(t) - \varphi\right)x_i(t) + \varphi \bar{x}(t) \quad ; \quad \bar{x}(t) := \frac{1}{N} \sum_i x_i(t)\, ,
\label{popdyn}
\end{equation}
The mean growth rates $m_i$'s are drawn from a distribution $\rho(m)$, $\varphi$ denotes the migration rate, and $\xi(t)$'s are  Gaussian white noise. In the case $\sigma = 0$ and at large times, the average population $\bar{x}(t)$ grows as $\bar{x}(t) \propto e^{\gamma t}$, defining the asymptotic growth rate $\gamma$. Interestingly, when $\rho(m)$ has finite support, the authors show in this limiting case $\sigma = 0$ that there exists a critical migration rate $\varphi_c$ such that for $\varphi > \varphi_c$, the growth rate $\gamma$ satisfies $\gamma = R(1/\varphi)$, where $R$ is the \emph{R-transform} of $\rho$. For $\varphi < \varphi_c$, the growth rate becomes $\gamma = m_> - \varphi$, where $m_>$ is the upper bound of the support of $\rho(m)$. As explicitly stated by the authors, this transition exactly matches the one we found in Eq.~(12) in the main text.

The mapping between the two models becomes even more striking when, in our case, we consider the $N$-state model with $W(D) = \frac{1}{N} \sum_{i=1}^N \delta(D - D_i)$. As observed in~\cite{BBLD2025} (see Appendix B there), the growth rate $\gamma$ is the analogous to the function $\Psi(q)$, which characterizes the exponential growth of $\langle e^{qx}\rangle = \hat{p}_r(q,t)$. To understand better this mapping, let us write the Fokker-Planck equation for the BLT of the joint density $p_r(x, t, D_i)$ denoted $\hat{p}_r(q,t,D_i)$ -- where $p_r(x, t, D_i) \, dx \, dt$ is the probability that the particle is in state $D_i$ within the interval $[x, x + dx]$ at a time between $t$ and $t + dt$ -- we obtain
\begin{equation}
    \frac{\partial \hat{p}_r(q,t,D_i)}{\partial t} = (D_i q^2 -r)\hat{p}_r(q,t,D_i) + r \hat{p}_r(q,t)\quad ; \quad \hat{p}_r(q,t)= \frac{1}{N}\sum_{i=1}^N\hat{p}_r(q,t,D_i)\, .
\end{equation}
This is exactly the same equation satisfied by the dynamics of the $x_i$'s given in Eq.~(\ref{popdyn}), when $\sigma=0$. Therefore, at a fixed value of $q$, for instance $q=1$, there is a mapping between the two models: $x_i \equiv \hat{p}_r(q,t,D_i)$, $\bar{x}(t)\equiv \hat{p}_r(q,t)$, $\gamma \equiv \Psi(q)$, $m_i \equiv D_i\, q^2$, $\rho(m) \equiv W(D)$ and $\varphi \equiv r$~\cite{BBLD2025}.